\documentclass[12pt]{article}

\usepackage{amssymb,amsmath,amsthm,bbm,enumitem,xcolor}
\usepackage{dsfont}
\usepackage[a4paper, total={6.3in, 9.2in}]{geometry}
\usepackage{setspace}
\usepackage{authblk}
\usepackage{graphicx}
\usepackage{subfigure}
\usepackage[title]{appendix}
 \usepackage{mathtools}
\usepackage{algorithm,algpseudocode}
\usepackage{tikz}
\usepackage{multirow}
\usepackage{makecell}
\usepackage[normalem]{ulem}
\usepackage{comment} 
\usepackage{placeins} 
\PassOptionsToPackage{hyphens}{url}

\usepackage[colorlinks = true,
            linkcolor = blue,
            urlcolor  = blue,
            citecolor = blue,
            anchorcolor = blue]{hyperref}
\hypersetup{breaklinks=true}
 
\usetikzlibrary{shapes.geometric, arrows, decorations.markings,shapes.multipart,cd,positioning,patterns,snakes}
\tikzstyle{startstop} = [rectangle, rounded corners, minimum width=2cm, minimum height=1cm,text centered, draw=black]
\tikzstyle{vecArrow} = [thick, decoration={markings,mark=at position
   1 with {\arrow[semithick]{open triangle 60}}},
   double distance=1.4pt, shorten >= 5.5pt,
   preaction = {decorate},
   postaction = {draw,line width=1.4pt, white,shorten >= 4.5pt}]
\tikzstyle{innerWhite} = [semithick, white,line width=1.4pt, shorten >= 4.5pt]
   \tikzset{|/.tip={Bar[width=.8ex,round]}}

\usepackage{apacite} 
\AtBeginDocument{%

   \urlstyle{APACsame}%
}

\usepackage[round]{natbib}    

\newtheorem{theorem}{Theorem}[section]
\theoremstyle{definition}

\theoremstyle{definition}

\theoremstyle{definition}
\newtheorem{assumption}{Assumption}[section]
\theoremstyle{definition}
\newtheorem{proposition}{Proposition}[section]
\theoremstyle{definition}
\newtheorem{corollary}{Corollary}[section]
\theoremstyle{remark}
\newtheorem{remark}{Remark}[section]
\theoremstyle{definition}
\newtheorem{lemma}{Lemma}[section]

\theoremstyle{definition}
\newtheorem{problem}{Problem}

\DeclareMathOperator{\E}{\mathbb{E}}

\newcommand{\hide}[1]{}

\allowdisplaybreaks
\setlist[itemize]{leftmargin=0.4cm,labelindent=\parindent}

\title{Mean Field Analysis of Mutual Insurance Market}
\date{\today}

 \author[$1$]{\small{Bohan Li}}
\author[$2$]{\small{Wenyuan Li}}
\author[$3$]{Kenneth Tsz Hin Ng}
\author[$4$]{Sheung Chi Phillip Yam}

\affil[$1$]{Center for Financial Engineering, Soochow University, Suzhou, Jiangsu, China. Email: 	
bhli@suda.edu.cn}
\affil[$2$]{Department of Statistics and Actuarial Science, The University of Hong Kong, Pokfulam, Hong Kong. Email: 	
wylsaas@hku.hk}
\affil[$3$]{Department of Mathematics, The Ohio State University, Columbus, Ohio, US. Email: ng.499@osu.edu}
\affil[$4$]{Department of Statistics and Data Science, The Chinese University of Hong Kong, Shatin, Hong Kong. Email: scpyam@sta.cuhk.edu.hk}

\newcommand{\bcm}[1]{{\color{blue}#1}}

\begin{document}
\maketitle




\begin{abstract}
   A mutual insurance company (MIC) is a type of consumer cooperative owned by its policyholders. By purchasing insurance from an MIC, policyholders effectively become member-owners of the company and are entitled to a share of the surplus,  which is determined by their own collective claims and premium contributions. This sharing mechanism creates an  interactive environment in which individual insurance strategies are influenced by the actions of others. Given that mutual insurers account for nearly one-third of the global insurance market, the analysis of members' behavior under such a sharing mechanism is of both practical and theoretical importance. This article presents a first dynamic study of members' behavior in the prevalent mutual insurance market under the large-population limit. With members' wealth processes depending on the law of the insurance strategies, we model the surplus-sharing mechanism using an extended mean field game (MFG) framework and address the fundamental question of how strategic interactions in this setting influence individual decisions. Mathematically, we establish the global-in-time existence and uniqueness of the mean field forward-backward stochastic differential equation (MF-FBSDE) characterizing the Nash equilibrium strategy, employing techniques to accommodate realistic insurance constraints. 
    Computationally, we develop a modified deep BSDE algorithm capable of solving the extended MFG problem with an additional fixed-point structure on the control. Utilizing this scheme, we examine how structural features of the MIC’s design, such as the composition of risk classes and surplus-sharing proportions, reshape members’ decisions and wealth through collective interactions, underscoring the central role of these mechanisms in MICs.
 \end{abstract}


 
\section{Introduction}

Mutual insurance companies (MICs) are one of the two most prevalent forms of centralized insurance providers in the industry, with a history dating back to the 18th century. Originating as community-based risk-sharing arrangements, early MICs gained traction in response to emerging urban risks, particularly frequent house fires. During the 19th century, industrialization introduced new hazards to high-risk occupations such as railroad workers. In response, U.S. railroad workers formed mutual benefit societies like the Brotherhood of Railroad Trainmen, which pooled member dues to provide life and disability benefits, reflecting the same mutual aid principles found in modern MICs. Today, the global mutual insurance market remains stable, accounting for 26.37\% of the global insurance industry and generating approximately USD 1.41 trillion in premiums worldwide.\footnote{According to the International Cooperative and Mutual Insurance Federation’s \href{https://www.icmif.org/mms-2024/}{global mutual market share report in 2024}} 

Unlike shareholder-owned insurance companies (SICs), the other major form of insurance providers, an MIC is owned entirely by its policyholders or members  (\cite{vaughan2007fundamentals,RejdaMcNamara2016}). Consequently, the surplus (or deficit) in an MIC, calculated as the premium income minus claims paid, reserves, and operating expenses, is shared among the members. This surplus may be distributed as dividends, premium adjustments, or other benefits, depending on the practice of the company. Hence, the net price of a policy is known \textit{ex post}, which is defined as the premium paid minus the shared surplus received. Such a sharing mechanism is absent in SICs, as policyholders are not necessarily the owners of the company. The following table compares these two types of insurance companies. 
\begin{table}[htbp]
	          
	\centering 
    \caption{Comparisons between MICs and SICs}
    \smallskip
	\begin{tabular}{c|c|c}
        \hline
            &  MIC  &  SIC \\ \hline\hline
               Ownership & Policyholders      & Shareholders \\ \hline
               Capital Required? & No & Yes \\ \hline 
               Net Price of Policy & Known \textit{ex post} & Known \textit{ex ante} \\ \hline 
               Manager's Earnings & Expense Saving & Investment Profit \\ \hline 
	\end{tabular}
                 \label{tab:mf:SIC:MIC}
\end{table}

MICs offer several advantages over SICs, with one of the most prominent being the mitigation of the policyholder-agent conflict. This comes as no surprise: employees often work for the best interest of the owners of the company, who, in the case of an MIC, are the members themselves. In addition, the risk and surplus sharing mechanism between members is found to be efficient in diversifying idiosyncratic risks (see \cite{cass:individual:mutual:pareto:1996}). Evidently, MICs are not always superior to SICs. In particular, the ability to raise capital from the financial market enables SICs to  enhance their liquidity and financial flexibility, which therefore allows SICs to expand their operations and innovate more readily compared to MICs. The relative merits of MICs versus SICs constitute an important and long-standing debate in the literature. Over centuries of development, both forms of insurance have evolved and now coexist with significant and enduring presence in the market. Our study takes the relevance of MICs as given and does not further explore this comparative aspect; interested readers are referred to \cite{mcnamara1992ownership,cummins:coexistence:1999,BIENER2012454,BRAUN2015875,SCHMEISER202192}. 

The objective of this article is to provide a quantitative and dynamic analysis of the members' behavior under the surplus-sharing mechanism of an MIC. To name a few representative studies in the literature, from the   perspective of an MIC or a mutual-aid platform, \cite{TAPIERO1984241}  addressed the problem of determining optimal premium rates.  
Regarding individual members' viewpoints, valuation problems were proposed using expected utility, Choquet expected utility and distortion risk measures in \cite{ALBRECHT2017180}; and mean-variance objective in  \cite{gatzert2012merits}. From a community perspective, \cite{BAUERLE201837} considered socially optimal reinsurance treaties among insurers and a reinsurance company, \cite{chen:optimal:2021} formulated the optimal risk-sharing to achieve Pareto optimality without a surplus/loss-sharing mechanism. More recently, peer-to-peer (P2P) insurance models, which are built on the principle of mutuality in a decentralized structure, have drawn attention in the study of optimal risk-sharing; see e.g., 
\cite{dhaene,dhaene:2021,denuit:comonotonicity:2023}. 

Despite the rich landscape of inspiring work on mutual and P2P insurances, a fundamental and still underexplored question is how much risk each participant optimally chooses to transfer to the platform in this interactive environment due to the sharing mechanism, especially under a dynamical continuous-time model. Indeed, these individual decisions and loss experiences directly impact the platform's stability and efficiency, while the distribution of surplus or deficit not only shapes their incentives but also couples their decisions, creating complex interdependencies that deserve careful study. For instance, a member's behavior may vary across platforms with different compositions of risk class, or when entitled to a larger or smaller share of the surplus/deficit. The major technical challenge arises from the interactions created by the sharing mechanism, which couples members and results in a less mathematically tractable optimal decision problem, especially when the number of members is large. Additionally, since members may incur claims at different times and adjust their insurance choices in response to the evolving collective experience, a continuous-time framework is naturally suited to capturing such dynamic feedback in contrast to discrete-time models, which are less equipped to capture this level of temporal heterogeneity. These complexities call for a modeling framework capable of handling large-population strategic interactions. Recent advances in \textit{extended mean field game} (MFG) (\cite{carmona:extended:2019,alasseur2020extended,carmona2021probabilistic,munoz2023classical,li:2024:cryptocurrency,bensoussan2025linear}) have emerged to meet this need, offering a powerful approach for modeling the optimal control problem from the members’ perspective.

In this article, we formulate the optimal insurance problems for an MIC under an extended MFG framework. Instead of modeling direct interactions among participants, MFGs capture their behavior through interaction with a common macroscopic factor, known as the mean field term, providing an asymptotic approach to solving optimal decision problems involving a large population. Due to their mathematical tractability and practical relevance, MFGs have been applied across various domains, including finance (\cite{casgrain2019algorithmic,HAN2022,bensoussan2022dynamic}), machine learning (\cite{ruthotto2020machine}), and cryptocurrency mining (\cite{lion:bitcoin:2024,li:2024:cryptocurrency}).  Recently, MFGs have begun to gain traction in the insurance and actuarial context. For example, \cite{BO2024202} analyzed the behavior of competitive insurers that interact through relative performance in their objective functions, while their wealth processes evolve independently. In contrast, our work incorporates explicit interactions in participants' wealth processes, making it one of the first in the actuarial domain to do so. Although our primary focus is on MICs, the model introduced herein can be readily applied to other mutual-aid platforms that share this mutuality and risk/surplus sharing mechanism.

Our model consists of members classified into $H$ different membership or risk classes, where members are homogeneous within class, and heterogeneous between classes.  This classification structure is crucial in insurance pricing and underwriting, as members are often grouped based on various risk and demographic factors, such as age, region of residence, smoking status, and other relevant characteristics. Our model stands out by encompassing the surplus-sharing mechanism in a pro-rata basis, which depends on the insurance strategies and claim experience of all other members within the MIC. Consequently, the wealth process of a member is influenced not only by their own actions but also by the collective strategies of other members within the system. The MFG is termed \textit{extended} here because it explicitly captures this additional layer of interaction arising from the direct impact of collective strategies within the company. Our model yields important insights into how the surplus-sharing mechanism within an MIC impacts the proportional insurance strategies of individual members, particularly in terms of reaching a Nash equilibrium.


The contributions of the present article are highlighted below. From a mathematical perspective, our work contributes to providing the solution of an extended mean field game characterized by a system of mean field forward-backward stochastic different equations (MF-FBSDEs) associated with games, and establishes 
a result of global-in-time existence and uniqueness of the solution. When a practical constraint on the insurance strategy is imposed, the strict monotonicity condition (see e.g.~\cite{pardoux2014stochastic}) for FBSDEs no longer holds due to the non-expansive property of a projection map. To address this, we derive a weaker form of monotonicity by utilizing the properties of the projection map, and employ an adaptation of the celebrated continuation approach to bypass the standard condition to establish a global existence result. Our sufficient condition merely requires a small mean field effect on each member, which is in line with the finding in the literature (see e.g.~\cite{CHU2025112028}).

 From a numerical perspective, to address the fact that the MF-FBSDE lacks a closed-form solution under the insurance constraint, we adopt a deep neural network (DNN) approach to solve the equation and implement the resulting optimal insurance strategies. Due to the presence of the mean field terms, standard Monte-Carlo methods are not directly applicable. To address this, we adapt and modify the forward method introduced in \eqref{eq:FBSDE:lq:insurance:constrained} (see \cite{germain2022numerical, carmona2022convergence, han2024learning}). Our proposed method includes an additional penalty term to match the output of the network with the mean field equilibrium strategy under the extended mean field game framework.In the absence of insurance constraints, the proposed method aligns with the known closed-form solution in the linear-quadratic setting, which demonstrates the accuracy of the algorithm. 
 
 From an economic perspective, we conduct a series of sensitivity analyzes to examine how the risk characteristics of members and the surplus-sharing mechanism influence their wealth and insurance strategies. First, we find that as the proportion of highly risk-averse members or those with more volatile loss processes increases, the overall insurance demand within the entire MIC tends to rise. Second, a higher surplus-sharing ratio reduces the effective price of the policies, thereby increasing their insurance demand. Third, by comparing results with and without insurance constraints, we find that the constraints help confine strategies within a practical range and reduce the disparity in insurance strategies across different member classes, ultimately narrowing the resulting wealth gap. 

 This article is organized as follows. In Section \ref{sec:mf:formulation}, we formulate the optimal insurance problem for members within an MIC, under both the $N$-player setting and the mean field game framework. In Section \ref{sec:eqm:strategy}, we provide the generic solution of the mean field Nash equilibrium in terms of an MF-FBSDE, whose well-posedness is discussed in Section \ref{sec:MFFBSDE:wellpose}. We then confine ourselves to quadratic rewards in Section \ref{sec:lq}, and in Section \ref{sec:lq:riccati}, we further reduce the MF-FBSDE to simpler Riccati equations when no insurance constraint is imposed.  Section \ref{sec:mf:numerical} introduces a DNN architecture to numerically compute the underlying MF-FBSDE. Based on this, we perform a numerical experiments to examine the effect of the risk composition of the MIC and the surplus-sharing mechanism on the members' optimal insurance strategies under both quadratic and non-quadratic rewards. The article is concluded in Section \ref{sec:conclusion}. 

\section{Model Formulation}
\label{sec:mf:formulation}
We consider a mutual insurance company with $H$ classes of membership. Members are assumed to be homogeneous in dynamics and parameters within each class, and heterogeneous between different classes. In this section, we first introduce the $N$-player problem with a large (but finite) number of members. We then study the mean field formulation of the problem by considering a mutual insurance company (MIC) with infinite number of members. Such a formulation is justified by the notion of $\varepsilon$-Nash equilibrium, see Theorem \ref{thm:epsilon-Nash} below. 

{\bf Notation.} We fix a decision horizon $[0,T]$, where $T>0$. Let $(\Omega,\mathcal{F},\mathbb{P})$ be a probability space with $\mathbb{E}$ being the expected value taken with respect to $\mathbb{P}$. Given an $\sigma$-algebra $\mathcal{G}\subseteq \mathcal{F}$, we denote by $L^2(\Omega,\mathcal{G},\mathbb{P})$ the collection of all square-integrable, $\mathcal{G}$-measurable random variables. For a generic filtration $\mathbb{G}:=(\mathcal{G}_t)_{t\in[0,T]}$ defined on $(\Omega,\mathcal{F},\mathbb{P})$ and a set $A\subseteq \mathbb{R}$, we denote by
    \begin{equation*}
        L^2_{\mathbb{G}}([0,T];A) := \Bigg\{ (\alpha_t)_{t\in[0,T]} : \alpha_t\in A, \text{ $\mathcal{G}_t$-measurable, and }  \mathbb{E}\left[ \int_0^T |\alpha_t|^2dt \right]<\infty  \Bigg\}. 
    \end{equation*}
For any positive integer $n$, we denote $[n]:=\{1,\dots,n\}$. For any function $f$, we use a subscript to denote the partial derivative of $f$ with respect to the corresponding variable. We denote by ${\bf I}$ the $H\times H$ identity matrix. For any $H\times H$ matrix ${\bf A}$, we define $\lambda_{\min}({\bf A})$ and $\lambda_{\max}({\bf A})$ to be the smallest and largest eigenvalue of $({\bf A}+{\bf A}^\top)/2$, respectively. Finally, for any matrix ${\bf B}$, we let $\|{\bf B}\|_2 := \sqrt{\lambda_{\max}({\bf B}^\top {\bf B})}$ be its spectral norm.

\subsection{Preliminaries and the $N$-Player Problem}

Suppose that there are $N^h$ members for each risk class $h\in[H]$. In our model, each member represents a company or organization that holds a group insurance policy provided by an MIC for employee benefits such as health, accident, or disability coverage. The losses are retained by the organization itself, which is common in practice for risks such as workers’ injuries, property and casualty losses related to company infrastructure, and disability claims. The accumulated loss process of member $i$ in Class $h$, denoted by $L^{i,h}=(L^{i,h}_t)_{t\in [0,T]}$, is given by \[
    L^{i,h}_t := \sum_{j=1}^{M^{i,h}_t} L^{i,h,j} ,
\]
where $(M^{i,h}_t)_{t \in [0,T]}$ is a Poisson process with intensity $\lambda^h > 0$ representing the number of claims up to time $t$. The claim severities $(L^{i,h,j})_{i \in [N^h],\, j \ge 1}$ are assumed to be i.i.d.\ for each fixed $h$, and are independent of the claim count processes $(M^{i,h}_t)_{i \in [N^h],\, t \in [0,T]}$.

A popular approach in the actuarial literature (see, e.g., \cite{iglehart1969diffusion,grandell1991aspects,browne1995optimal}) is to approximate $L^{i,h}_t$ by the Cram\'er--Lundberg diffusion model.  The accumulated loss process of member $i$ in class $h$, denoted by $C^{i,h}=(C^{i,h}_t)_{t\in[0,T]}$, is then approximated by
\begin{equation}
    \label{eq:loss}
    dC^{i,h}_t = \mu^h\, dt - \sigma^h\, dW^{i,h}_t,
\end{equation}
where $(W^{i,h}_t)_{t \in [0,T]}$ is a standard Brownian motion such that $\{ W^{i,h} : i \in [N^h],\, h \in [H] \}$ are independent and identically distributed, $\mu^h:= \lambda^h \E[L^{i,h,j}]$, and $\sigma^h:=\sqrt{\lambda^h \E[(L^{i,h,j})^2]}.$ Our subsequent analysis shall be based on this diffusion approximation model.  %

    Each member $i\in [N^h]$ in Class $h\in [H]$ is entitled to choose a proportion $v^{i,h}\in \mathcal{A}_{\mathbb{F}^{i,h}}(I)$ of the loss to be transferred to the MIC, where $\mathcal{A}_{\mathbb{F}^{i,h}}(I) := L^2_{\mathbb{F}^{i,h}}([0,T];I)$ is the admissible set of proportional insurance strategies in the constraint set $I$. We assume that $I\subseteq \mathbb{R}$ is a closed interval of the form $I=[a,b]$, where $a,b\in\mathbb{R}$, $b>a$, and the filtration $\mathbb{F}^{i,h}:=(\mathcal{F}^{i,h}_t)_{t\in[0,T]}$ is defined as follows:
      \begin{align*}
       \mathcal{F}^{i,h}_t &:= \sigma\left( \xi^{i,h}, W^{i,h}_s : 0\leq s\leq t  \right) \vee \hat{\mathcal{F}}_t, \\
       \hat{\mathcal{F}}_t &:= \sigma\left( \frac{\sum_{j=1}^{N^k}{v^{j,k}_s}}{N^k} : k\in[H], \ 0\leq s\leq t \right) \bigvee \sigma\left(\frac{\sum_{i=1}^{N^k} y^{i,k}_s }{N^k} : k\in[H], 0\leq s\leq t \right)\vee\mathcal{N},
    \end{align*}
where $\mathcal{N}$ is the collection of all $\mathbb{P}$-null sets, $\xi^{i,h}$, $i\in[N^h]$, $h\in[H]$, are i.i.d.~square-integrable random variables representing the initial wealth of member $i$ from Class $h$, and $(y^{i,h}_t)_{t\in[0,T]}$ is her wealth process; see \eqref{eq:y:empirical} below. Thus, each member makes her decision based on her own wealth, and the public information consisting of the average position and wealth of all other members within the MIC contained in the filtration $(\hat{\mathcal{F}}^h_t)_{t\in[0,T]}$. 

A common choice of the constraint would be $a=0$ and $b=1$, which indicates that the member is not allowed to transfer more than her actual loss or to take a short position, although our analysis is not limited to this specific case. The rate of premium she has to pay is then given by $v^{i,h}_t c^h$, where $c^h:=  \mu^h(1 + \theta^h)$ is the premium rate charged by the MIC, and $\theta^h>0$ is the safety loading for Class $h$. We remark that the insurance constraint limits the instantaneous premium rate payable in the range $[a c^h,b c^h]$. This aligns with the practical scenario where the premium rate remains relatively stable without drastic fluctuations. In addition, each member in Class $h$ is required to pay a membership fee of $e^h\geq 0$ to be able to get a share of the surplus.  
    
    Let $U=(U_t)_{t\geq 0}$ be the surplus of the MIC, which is defined as the aggregate premium income, membership fee,  less the shared loss and management costs: 
        \begin{equation}\label{sde.surplus}
            dU_t = \underbrace{\sum_{h=1}^H \sum_{j=1}^{N^h} \left[(c^h-d^h)v^{j,h}_t + e^h-d_e^h  \right]dt}_{\text{premium income and membership fee less expenses}} - \underbrace{\sum_{h=1}^H \sum_{j=1}^{N^h} v^{j,h}_tdC^{j,h}_t}_{\text{shared loss}},
        \end{equation}
    where  $d^h,d_e^h>0$ are the common proportional and fixed management fee rate, respectively. Let $\pi^h>0$ be the proportion of shares acquired by Class $h$. The surplus or loss $U$ will then be distributed according to a simple pro-rate basis, where each member from Class $h$ receives $\pi^h/\sum_{k=1}^H \pi^kN^k$ of it. A similar pro-rata sharing mechanism is popular in practice and in the literature. For instance, \cite{ALBRECHT2017180} considered a sharing mechanism where each member receives a proportion of the surplus based on the amount of insurance they purchased. Herein, the parameter $\pi^h$ can be chosen to reflect the risk exposure, safety loading, and the membership fee rate within each risk class. Since the proportion of insurance $v^{i,h}$ is bounded within a practical range $I$ and the membership fee rate does not fluctuate significantly, using a fixed parameter $\pi^h$ provides a stable proxy for the relative premium size. This approach keeps the surplus-sharing mechanism simple and avoids the need for frequent adjustments of sharing ratios, thereby reducing administrative complexity.


    In sum, the wealth process $y^{i,h}$ of member $i$ from Class $h$  is governed by the following components. First, she earns a risk-free rate $r>0$ based on her current wealth. Second, according to her insurance strategy, she needs to pay the premium, and is responsible for the retained loss that has not been transferred to the MIC. Third, in addition to the proceeds from the MIC mentioned in the last paragraph, she also receives an exogenous income of rate $\Tilde{l}^h$. Hence, the process $y^{i,h}$ is governed by the following SDE: $y_0^{i,h} = \xi^{i,h}$ and
    \begin{align}
                dy^{i,h}_t &= \left(ry^{i,h}_t +\tilde{l}^h - e^h  -\underbrace{c^{h}v^{i,h}_t}_{\text{ 
 premium paid}} \right)dt - \underbrace{(1-v^{i,h}_t)dC^{i,h}_t}_{\text{
 retained loss}} \nonumber\\
 &\quad + \underbrace{ \frac{\pi^h}{\sum_{l=1}^H \pi^l N^l }\left( \sum_{k=1}^H \sum_{j=1}^{N^k} \left[(c^k-d^k)v^{j,k}_t +e^k-d_e^k \right] dt - \sum_{k=1}^H \sum_{j=1}^{N^k} v^{j,k}_tdC^{j,k}_t\right)}_{\text{ 
 shared surplus/deficit from MIC}} \nonumber \\
                &= \left(ry^{i,h}_t+ {l}^{h} -\kappa^h v^{i,h}_t + \pi^h\sum_{k=1}^H  \omega^k  (\kappa^k-d^k)\frac{\sum_{j=1}^{N^k}v^{j,k}_t}{N^k}  \right)dt + \sigma^h(1-v^{i,h}_t)dW^{i,h}_t \nonumber\\
                &\quad + \underbrace{\pi^h\sum_{k=1}^H \sigma^k \omega^k\frac{\sum_{j=1}^{N^k}v^{j,k}_t}{N^k}dW^{j,k}_t}_{\text{idiosyncratic risk}}, 
        \label{eq:y:empirical}    
    \end{align}
    where $l^{h}:= \tilde{l}^h - \mu^h-e^h + \frac{\pi^h}{\sum_{l=1}^H \pi^lN^l} \sum_{k=1}^H N^k(e^k-d_e^k)$, $\kappa^h := \mu^h\theta^h$, and  $\omega^h := N^h/\sum_{k=1}^H \pi^kN^k$. The parameter $\omega^h$ represents the proportion of members in Class $h$ within the entire MIC, adjusted by the shares acquired by each risk class. We assume that $(\omega^h)_{h=1}^H$ is independent of the absolute population sizes $(N^h)_{h=1}^H$, meaning that even if the population sizes change, this ratio remains constant. Under this assumption, we have \[l^h = \tilde{l}^h - \mu^h -e^h + \pi^h \sum_{k=1}^H \omega^k(e^k-d_e^k).\]%
    In addition, it is clear that $\sum_{h=1}^H\pi^h \omega^h=1$.  We also assume that $\kappa^h-d>0$ for all $h\in[H]$ throughout the rest of the article. This condition ensures that the risk premium rate exceeds the expense rate, meaning that the premiums sufficiently cover expenses to sustain meaningful MIC operations and avoid immediate bankruptcy. 
    \begin{remark}
        We assume that members will inject new capital into the MIC in proportion to their shares to avoid it from bankruptcy.  This explains why the deficit in \eqref{eq:y:empirical} is also shared among members in our setting when $U_t<0$. In practice, when $U_t<0$, an MIC may respond by increasing premiums, which results in a net outflow from members’ wealth. However, because the owners of an MIC are the members themselves, the management of the company does not inject capital into the mutual; instead, they provide services and collect management fees.
    \end{remark}

 Each member $i$ from Class $h$ aims to take an insurance strategy $v^{i,h}\in \mathcal{A}_{\mathbb{F}^{i,h}}(I)$ to maximize the following objective:
  \begin{equation}
       \begin{aligned}
         &\mathcal{J}^{i,h}(v^{i,h}):= 
         \mathcal{J}^{i,h}\left(v^{i,h}; {\bf y}^{-i,h}, {\bf v}^{-i,h} \right)\\
          &\quad := \mathbb{E}\left[\int_0^T f^h\left(t,y^{i,h}_t, \frac{\sum_{j=1,j\neq i}^{N^h} y^{j,h}_t}{N^h-1} , v^{i,h}_t,\frac{\sum_{j=1, j\neq i}^{N^h}v^{j,h}_t}{N^h-1}  \right)dt + g^h\left(y^{i,h}_T, \frac{\sum_{j=1,j\neq i}^{N^h} y^{j,h}_T}{N^h-1} \right)\right],
        \end{aligned}
        \label{eq:empirical:objective}
  \end{equation}
where  
${\bf v}^{-i,h}:= (v^{j,h})_{j\in[N^h], j \neq i}$, ${\bf y}^{-i,h} = (y^{j,h})_{j\in[N^h], j\neq i}$ are the associated wealth processes under the $N$-player game; $f^h:[0,T]\times\mathbb{R}\times \mathbb{R} \times \mathbb{R} \times \mathbb{R} \to \mathbb{R}$ and $g^h : \mathbb{R}\times \mathbb{R}\to \mathbb{R}$. 
   In other words, each member within a given risk class shares the same preference, which accounts for her own wealth, her insurance strategy relative to the class average, and the average wealth of members across all classes. Assumptions on $f^h$ and $g^h$ are deferred to Section \ref{sec:ass}. 

%
In practice, several factors lead to insurance purchase behavior that depends on the coverage level \( v^{i,h} \) in a non-linear and concave manner, a feature we capture through the reward function \( f^h \) in our model. First, as shown by \cite{mossin:1968}, full coverage is generally not optimal when premiums include loadings, since diminishing marginal utility of wealth and actuarially unfair pricing produce an interior optimum. Second, regulatory frameworks often impose minimum coverage requirements such as auto third-party liability or workplace injury insurance that members or group managers must meet but are not required to exceed, especially when risks or potential losses are low. For example, rather than fully insuring depreciated equipment or property, members may opt to save on premiums and replace the item if damaged. Third, prospect theory (\cite{prospect:1979}) suggests that individuals tend to be myopic and underweight low-probability events, or exhibit loss aversion relative to reference wealth levels, which contributes to under-insurance even in situations involving severe but infrequent losses (\cite{pauly:neglect:disaster:2004}). These considerations motivate the incorporation of \( v^{i,h} \) into the reward function.

Under the setting \eqref{eq:y:empirical} and \eqref{eq:empirical:objective}, the decision problems of members within the MIC are coupled via the surplus/deficit sharing mechanism and their objective functions. Problem \ref{p:mic:empirical} below formulates the notion of optimal strategies for all members within the MIC in terms of a Nash equilibrium, where a member would be worse-off if she deviates from the equilibrium strategy.  
\begin{problem}
\label{p:mic:empirical}
    Find a Nash equilibrium strategy $(v^{i,h})_{h\in [H],i\in[N^h]}$ such that $v^{i,h} \in \mathcal{A}_{\mathbb{F}^{i,h}}(I)$, and  
        \begin{equation*}
          \mathcal{J}^{i,h}\left(v^{i,h};{\bf y}^{-i,h},{\bf v}^{-i,h} \right) \geq  \mathcal{J}^{i,h}\left(u^{i,h}; \check{{\bf y}}^{-i,h}, {\bf v}^{-i,h}\right),
        \end{equation*}
    for any $u^{i,h}\in \mathcal{A}_{\mathbb{F}^{i,h}}(I)$, and any $h\in[H]$ and $i\in [N^h]$, where $\check{{\bf y}}^{-i,h} = (\check{y}^{j,h})_{j\in [N^h], j\neq i}$, and $(\check{y}^{j,h})_{j\in[N^h]}$ are the associated wealth processes under the $N$-player game with strategies $(v^{1,h},\dots,$ $v^{i-1,h},u^{i,h},v^{i+1,h},\dots, v^{N^h,h})$.
\end{problem}

\subsection{Mean Field Game Formulation}
Due to the intricate interactions between members arising from the surplus-sharing mechanism, it is analytically challenging to obtain a Nash equilibrium strategy for Problem \ref{p:mic:empirical}. To this end, we adopt the mean field formulation of Problem \ref{p:mic:empirical}. 

We consider the case where the number of members $N^h$, $h\in[H]$, tends to infinity, and suppose that we are given a collection of exogenous and deterministic processes $(z^h)_{h\in[H]}$ and $( \bar{v}^h)_{h\in[H]}$, where $z^h=(z^h_t)_{t\in [0,T]}$ and $\bar{v}^h = (\bar{v}^h_t)_{t\in [0,T]}$. For $h\in [H]$ and $i\in[N^h]$, let $x^{i,h}:=(x^{i,h}_t)_{t\in[0,T]}$ be the wealth process of member $i$ from Class $h$, which satisfies the following mean field dynamics: 
    \begin{equation}
    \label{eq:xh}
         dx^{i,h}_t = \left(rx^{i,h}_t +l^h - \kappa^h v^{i,h}_t + \pi^h\sum_{k=1}^H \omega^k (\kappa^k-d^k){\bar{v}^k_t} \right)dt + \sigma^h(1-v^{i,h}_t)dW^{i,h}_t, \ x^h_0 = \xi^{i,h}.
    \end{equation}
Each member from Class $h$ aims to maximize the following objective: 
\begin{equation}
 \begin{aligned}
               J^{i,h}(v^{i,h})&:=  J^{i,h}\left(v^{i,h};z^h, \bar{v}^h\right) = \mathbb{E}\left[\int_0^T f^h\left(t,x^{i,h}_t, z^h_t,v^{i,h}_t,\bar{v}^h_t\right)dt + g^h\left( x^{i,h}_T,z^h_T \right)\right].
 \end{aligned}
 \label{eq:Jh}
    \end{equation}
    
    Since the number of members in each class is indefinite, we have the following observations: (i) the idiosyncratic part in \eqref{eq:y:empirical} shall vanish, and (ii) the contribution of each individual on the average terms such as $\sum_{j=1}^N v^{j,k}_t/N^k$ becomes negligible. This allows us to treat the average wealth and average insurance strategy for each of Class $h$ to be exogeneously given, which are represented by $z^h$ and $v^h$, respectively. Under this framework, the wealth and objective functions between members are essentially decoupled, which allows us to focus on the decision problem for a single representative member from each risk class. Henceforth, we shall omit the index $i$ in all the occurrence in the sequel, and simply call $x^h$ the wealth process of the representative member (or simply member below) from Class $h$. We also define the filtrations $\mathbb{F}^h:=(\mathcal{F}^h_t)_{t\in[0,T]}$ and $\mathbb{F}^{[H]}:=(\mathcal{F}^{[H]}_t)_{t\in[0,T]}$ by $\mathcal{F}^h_t := \sigma\left( \xi^h, W^h_s : 0\leq s\leq t  \right)$ and $\mathcal{F}^{[H]}_t := \bigvee_{h=1}^H \mathcal{F}^h_t$, respectively.

To achieve equilibrium, the deterministic functions $z^h$ and $v^h$ should eventually agree with the average wealth and the average strategy when optimality is achieved.  This solution approach, often known as the \textit{fixed point approach}, can be formulated in terms of the following two sub-problems.

\begin{problem} 
\label{p:mf}
    Given the deterministic functions $  (z^h)_{h\in[H]}$ and $\bar{\bf v} := (\bar{v}^h)_{h\in[H]}$, find the optimal control ${\bf v} := (v^h)_{h\in[H]}$ such that for any $h\in[H]$,
       \begin{equation*}
                    v^h= \mathop{\arg\max}_{u^h \in \mathcal{A}_{\mathbb{F}^h}(I)} J^h\left(u^h;z^h, \bar{v}^h \right).
                \end{equation*}
\end{problem}

\begin{problem} 
\label{p:fixed:point}
   Find the mean field equilibrium wealth ${\bf z} = (z^h)_{h\in[H]}$ and strategy $\bar{\bf v} = (\bar{v}^h)_{h\in[H]}$ such that for any $t\in[0,T]$,
   \begin{equation*}
               (\bar{v}^1_t,\dots,\bar{v}^H_t) = \mathbb{E}\left[ (v^1_t,\dots,v^H_t)\right] \quad \text{and} \quad (z^1_t,\dots,z^H_t)=\mathbb{E}\left[ (x^1_t,\dots,x^H_t)\right].
            \end{equation*}
\end{problem}
Since the shared surplus/deficit directly depends on the insurance strategies of the other members, an additional fixed point $ (\bar{v}^1_t,\dots,\bar{v}^H_t) = \mathbb{E}\left[ (v^1_t,\dots,v^H_t)\right]$ has to be satisfied in Problem \ref{p:fixed:point}. This formulation is called an \textit{extended mean field game} (\cite{carmona2015probabilistic,gomes2014mean}) since it includes finding the equilibrium law of the optimal control. Note also that the diffusion term in \eqref{eq:xh} is controlled. As documented in \cite{bensoussan2023degeneratemeanfieldtype}, such control in the MFG context can complicate the representation of the solution and the mathematical analysis, particularly because the control depends on the backward component of the associated BSDE  as shown in \eqref{eq:v*} below.

Theorem \ref{thm:epsilon-Nash} below establishes the $\varepsilon$-Nash equilibrium of the mean field game \eqref{eq:xh}-\eqref{eq:Jh} for the original $N$-player game \eqref{eq:y:empirical}-\eqref{eq:empirical:objective}. It says that, the optimal strategies obtained in the mean field game is very close to achieving a Nash equilibrium for the $N$-player game, where the discrepancy decays with the class sizes in the order of $\frac{1}{2}$. 

\begin{theorem}
\label{thm:epsilon-Nash}
    Let $(v^{i,h})_{h\in[H], i \in[N^h] }$, $(z^h)_{h\in[H]}$ and $(\bar{v}^h)_{h\in[H]}$ be the solution of Problems \ref{p:mf}-\ref{p:fixed:point} with wealth process and objective functions given by \eqref{eq:xh} and \eqref{eq:Jh}, respectively. Consider Problem \ref{p:mic:empirical} with class size $N^h$ for each membership class $h\in[H]$. 
    Then, under Assumption \ref{ass:concave:monotonicity:h}.A below, it holds that 
        \begin{equation*}
          \mathcal{J}^{i,h}\left(v^{i,h}; {\bf y}^{-i,h}, {\bf v}^{-i,h} \right) \geq \mathcal{J}^{i,h}\left(u^{i,h};\check{{\bf y}}^{-i,h}, {\bf v}^{-i,h} \right)  - \sum_{k=1}^H O\left( \frac{1}{\sqrt{N^k}} \right),
        \end{equation*}
    for any  $u^{i,h}\in \mathcal{A}_{\mathbb{F}^{i,h}}(I)$, where $\check{{\bf y}}^{-i,h}$ is defined as in Problem \ref{p:mic:empirical}. 
\end{theorem}

\begin{proof}
    The proof is relegated to Appendix \ref{sec:app:e-nash}. 
\end{proof}

\subsection{A Discussion of a Members' Survival Model}


\label{sec:app:survival}
In this section, we provide a discussion on  extending our model to incorporate a survival framework, allowing for the possibility that members leave the MIC involuntarily, for example, due to discontinuation of business, default, regulatory intervention, or forced lapse. Let $\tau^{i,h}$ denote the  exit time of member $i$ in Class $h$. We assume that the family of exit times $(\tau^{i,h})_{i \in [N^h],\, h \in [H]}$ is independent, and that for each $h \in [H]$, the collection $(\tau^{i,h})_{i \in [N^h]}$ is identically distributed. Moreover, each exit time $\tau^{i,h}$ is independent of the random variables associated with other members and the market variables, and is not determined by the members themselves. Note that now $N^h$ denotes the initial number of members in Class $h$. We shall assume that $\mathbb{P}(\tau^h>T)>0$ for all $h\in[H]$, where $\tau^h$ represents the common distribution of $\tau^{i,h}$, $i\in N^h$.

Under the survival mode, the surplus process of the MIC is given by 
\begin{equation}\label{sde.surplus.tau}
            dU_t = \sum_{h=1}^H \sum_{j=1}^{N^h} \left[(c^h-d^h)v^{j,h}_t + e^h-d_e^h \right] \mathds{1}_{\{\tau^{j,h} > t\}}dt-  \sum_{h=1}^H \sum_{j=1}^{N^h} v^{j,h}_t\mathds{1}_{\{\tau^{j,h} > t\}}dC^{j,h}_t,
\end{equation}
indicating that only surviving members will purchase insurance, pay the membership fee, and transfer their loss to the MIC. Furthermore, let $N^h_t:= \sum_{j=1}^{N^h}  \mathds{1}_{ \{ \tau^{j,h} > t \} }$ be the number of surviving members in Class $h$. As such, 
the wealth process $y^{i,h}$ of member $i$ from Class $h$ is given by, for $t\in[0,\tau^{i,h}\wedge T]$,
\begin{align}
                dy^{i,h}_t &=
                \left(ry^{i,h}_t +\tilde{l}^h  - c^{h}v^{i,h}_t   \right)dt - (1-v^{i,h}_t)dC^{i,h}_t  \nonumber\\
 &\quad + \frac{\pi^h}{\sum_{l=1}^H \pi^l N^l_t }\left( \sum_{k=1}^H \sum_{j=1}^{N^k} \left[(c^k-d^k)v^{j,k}_t + e^k-d_e^k \right]\mathds{1}_{\{\tau^{j,k} > t\}} dt - \sum_{k=1}^H \sum_{j=1}^{N^k} v^{j,k}_t\mathds{1}_{\{\tau^{j,k} > t\}}dC^{j,k}_t\right)  \nonumber \\
            &= \left(ry^{i,h}_t+ \tilde{l}^{h,N} -\kappa^h v^{i,h}_t + \pi^h\left( \frac{\sum_{l=1}^H \pi^lN^l }{\sum_{l=1}^H \pi^l N^l_t } \right) \sum_{k=1}^H  \omega^k  (\kappa^k-d^k)\frac{\sum_{j=1}^{N^k}v^{j,k}_t\mathds{1}_{\{\tau^{j,k} > t\}}}{N^k}  \right)dt\nonumber\\
                &\quad  + \sigma^h(1-v^{i,h}_t)dW^{i,h}_t  +\pi^h \left( \frac{\sum_{l=1}^H \pi^lN^l }{\sum_{l=1}^H \pi^l N^l_t } \right)\sum_{k=1}^H \sigma^k \omega^k\frac{\sum_{j=1}^{N^k}v^{j,k}_t\mathds{1}_{\{\tau^{j,k} > t\}}}{N^k}dW^{j,k}_t,
        \label{eq:y:empirical.tau}    
\end{align}
where $\omega^h:= N^h/\sum_{k=1}^H\pi^kN^k$, $h=1,\dots,H$, and 
\begin{equation*}
        \tilde{l}^{h,N}_t := \tilde{l}^h-\mu^h + \pi^h \left( \frac{\sum_{l=1}^H \pi^lN^l }{\sum_{l=1}^H \pi^l N^l_t } \right) \sum_{k=1}^H \omega^k \frac{N^k_t}{N^k} (e^k-d_e^k).
\end{equation*}

The objective function for the $i$-th member from Class $h$ under this involuntary exit model is given by
\begin{equation}
       \begin{aligned}
         &\mathcal{J}^{i,h}(v^{i,h}):= 
         \mathcal{J}^{i,h}\left(v^{i,h}; {\bf y}^{-i,h}, {\bf v}^{-i,h} \right)\\
          &\quad := \mathbb{E}\Bigg[\int_0^{T\wedge \tau^{i,h}} f^h\left(t,y^{i,h}_t, \bar{y}^{i,h,\tau^h}_t, \bar{v}^{i,h,\tau^h}_t \right)dt + \mathds{1}_{\{\tau^{i,h} > T\}}g^h\left(y^{i,h}_T, \bar{y}^{i,h,\tau^h}_T \right)\Bigg],
        \end{aligned}
        \label{eq:empirical:objective.tau}
\end{equation}
where  
    \begin{equation*}
         \bar{y}^{i,h,\tau^h}_t :=  \frac{\sum_{j=1,j\neq i}^{N^h}y^{j,h}_t \mathbbm{1}_{\{ \tau^{j,h}>t \} } }{N^h-1} , \  \bar{v}^{i,h,\tau^h}_t :=  \frac{\sum_{j=1,j\neq i}^{N^h}v^{j,h}_t \mathbbm{1}_{\{ \tau^{j,h}>t \} } }{N^h-1}.
    \end{equation*}

The above formulation motivates the following mean field game formulation by passing to the limit $N^h \to \infty$, $h\in[H]$. Let $(\widetilde{z}^h)_{h\in[H]}$ and $(\widetilde{v}^h)_{h\in[H]}$ be exogeneously given, deterministic functions, and denote $s^h_t:= \mathbb{P}(\tau^h>t)$.  For $i\in\mathbb{N}$ and $h\in[H]$, let $(x^{i,h}_t)_{t\in[0,T]}$ be the process that satisfies, for $t\in[0, T]$,
\begin{equation}
    \label{eq:xh:tau}
         dx^{i,h}_t = \left(rx^{i,h}_t +\widetilde{l}^h_t - \kappa^h v^{i,h}_t + \pi^h\sum_{k=1}^H \omega^k (\kappa^k-d^k){\widetilde{v}^k_t} \right)dt + \sigma^h(1-v^{i,h}_t)dW^{i,h}_t, \ x^h_0 = \xi^{i,h},
\end{equation}
where
    \begin{equation*}
        \widetilde{l}^h_t  := \tilde{l}^h - \mu^h + \frac{\pi^h}{ \sum_{l=1}^H \pi^l\omega^l s^l_t}  \sum_{k=1}^H \omega^k (e^k-d_e^k) s^k_t. 
    \end{equation*}
Note that by the strong law of large numbers, as $N^h\to\infty$ for $h\in[H]$, we have almost surely that $N^k_t/N^k \to s^k_t$ and 
    \begin{equation*}
        \frac{\sum_{l=1}^H \pi^l N^l}{ \sum_{k=1}^H \pi^k N^k_t } = \frac{1}{\sum_{k=1}^H \pi^k \left(\frac{N^k}{\sum_{l=1}^H \pi^l N^l} \right)   \frac{N^k_t}{N^k}   } = \frac{1}{\sum_{k=1}^H \pi^k \omega^k   \frac{N^k_t}{N^k}   } \to \frac{1}{ \sum_{k=1}^H \pi^k \omega^k s^k_t }.  
    \end{equation*}
In other words, $\tilde{l}^{h,N}_t \to  \widetilde{l}^h_t$ a.s.~when $N^h\to \infty$, $h\in[H]$. 

In light of the independence of $(\tau^{i,h})_{i\in\mathbb{N},h\in[H]}$, $(W^{i,h})_{i\in\mathbb{N},h\in[H]}$, and $(\xi^{i,h})_{i\in\mathbb{N},h\in[H]}$, and the fact that members' dynamically systems are decoupled under the large-population limit, we introduce the following objective under the MFG formulation: for $i\in\mathbb{N}$ and $h\in[H]$, 
\begin{equation}
 \begin{aligned}
               J^{i,h}(v^{i,h})&:=  J^{i,h}\left(v^{i,h};\widetilde{z}^h, \widetilde{v}^h\right) = \mathbb{E}\left[\int_0^{T\wedge \tau^{i,h}} f^h\left(t,x^{i,h}_t, \widetilde{z}^h_t,v^{i,h}_t,\widetilde{v}^h_t\right)dt + \mathds{1}_{\{\tau^{i,h} > T\}}g^h\left( x^{i,h}_T, \widetilde{z}^h_T \right)\right] \\
               &= \mathbb{E}\left[\int_0^{T} \tilde{f}^h\left(t,x^{i,h}_t, \widetilde{z}^h_t,v^{i,h}_t,\widetilde{v}^h_t\right)dt + \tilde{g}^h\left( x^{i,h}_T, \widetilde{z}^h_T \right)\right],
 \end{aligned}
 \label{eq:Jh.tau}
\end{equation}
where $(\widetilde{z}^h)_{h\in[H]}$ and $(\widetilde{v}^h)_{h\in[H]}$ are exogeneously given, and
    \begin{equation*}
        \tilde{f}^h(t,x,z,v,\widetilde{v}) := s^h_t f^h(t,x,z,v,\widetilde{v})  \quad \text{and} \quad \tilde{g}^h(x,z) := s^h_T g(x,z). 
    \end{equation*}

Comparing the mean field dynamics $x^{i,h}$ with $y^{i,h}$, and the mean field objective functions \eqref{eq:Jh.tau} with \eqref{eq:empirical:objective.tau}, we observe that  $\widetilde{z}^h_t$ essentially replaces the empirical average of the surviving members $\frac{\sum_{j=1}^{N^h}y^{j,h}_t\mathds{1}_{\{\tau^{j,h} > t\}}}{N^h-1}$ under the $N$-player game, while $\widetilde{v}^{h}_t$ corresponds to 
\[
\frac{\sum_{l=1}^H \pi^lN^l}{\sum_{k=1}^H \pi^k N^k_t} \frac{\sum_{j=1,j\neq i}^{N^h} v^{j,h}_t \mathds{1}_{ \{\tau^{i,h}>t \} } }{N^h-1}.
\]
These observations naturally lead to the following MFG formulation and the corresponding definition of the mean field terms:

\begin{problem} 
\label{p:mf.tau}
    Given the deterministic functions $\widetilde{\bf z} := (\widetilde{z}^h)_{h\in[H]}$ and $\widetilde{\bf v} := (\widetilde{v}^h)_{h\in[H]}$, find the optimal control ${\bf v} := (v^h)_{h\in[H]}$ such that for any $h\in[H]$,
       \begin{equation*}
                    v^h= \mathop{\arg\max}_{u^h \in \mathcal{A}_{\mathbb{F}^h}(I)} J^h\left(u^h;\widetilde{z}^h, \widetilde{v}^h \right).
                \end{equation*}
\end{problem}

\begin{problem} 
\label{p:fixed:point.tau}
   Find the mean field equilibrium wealth $\widetilde{\bf z} = (\widetilde{z}^h)_{h\in[H]}$ and strategy $\widetilde{\bf v} = (\widetilde{v}^h)_{h\in[H]}$ such that for any $t\in[0,T]$ and $h\in[H]$,
    \begin{equation*}
        \widetilde{z}^h_t = \mathbb{E}[x^h_t] s^h_t \quad \text{and} \quad \widetilde{v}^h_t = \frac{s^h_t}{\sum_{k=1}^H \pi^k \omega^k s^k_t} \mathbb{E}[v^h_t]  .
    \end{equation*}
   
\end{problem}

The survival model represents an extension that falls largely under the framework of the original formulation The main differences are that the class weight now becomes time-dependent, and the definition of the mean field terms is revised to account for involuntary exits. Moreover, the independence of the exit times $\tau^{i,h}$ allows the survival probability to be absorbed into the coefficients $f^h$ and $g^h$. Under suitable conditions on the survival probabilities $s^h_t$, the analytical results and solution methodology developed in the main formulation remain valid within this extended survival framework.

In light of the above discussions, while our framework can naturally accommodate a more general survival model with involuntary exits, we shall focus on the original formulation given by \eqref{eq:xh}-\eqref{eq:Jh} and Problems \ref{p:mf}-\ref{p:fixed:point}.

\subsection{Assumptions}
\label{sec:ass}
In the sequel, we shall impose the following assumptions on the functions $f^h$ and $g^h$.

\begin{assumption}
\label{ass:concave:monotonicity:h}
For each $h \in [H]$, the function $f^h$ is differentiable in its $x$- and $v$-arguments, and $g^h$ is differentiable in its $x$-argument. In addition, 
\begin{enumerate}[label={\bf \Alph*.}]
        \item (Lipschitz continuity) There exist $L,L^X,L^V>0$ and $L^g\geq 0$ such that, for any $h\in[H]$, $t\in[0,T]$, and any  $x_1,x_2,z_1,z_2\in\mathbb{R}$,  $v_1,v_2,\bar{v}_1,\bar{v_2} \in I$, 
            \begin{enumerate}[label=(\roman*)]
                \item $|f^h(t,x_1,z_1,v_1,\bar{v}_1) -f^h(t,x_2,z_2,v_2,\bar{v}_2)| \leq  L( 1 + |x_1|+|z_1|+|v_1|+|\bar{v}_1| + |x_2| + |z_2|   + |v_2| + |\bar{v}_2|)  \left(|x_1-x_2| + |z_1-z_2| + |v_1-v_2| + |\bar{v}_1-\bar{v}_2| \right)$,
                \item $|g^h(x_1,z_1)-g^h(x_2,z_2)| \leq L(1+|x_1|+|x_2|+|z_1|+|z_2|)(|x_1-x_2|+|z_1-z_2|) $,
                \item $|f^h_x(t,x_1,z_1,v_1,\bar{v}_1) -f^h_x(t,x_2,z_2,v_2,\bar{v}_2)| \leq  L^X (|x_1-x_2| + |z_1-z_2| + |v_1-v_2| + |\bar{v}_1-\bar{v}_2| )$,
                \item $|f^h_v(t,x_1,z_1,v_1,\bar{v}_1) -f^h_v(t,x_2,z_2,v_2,\bar{v}_2)| \leq  L^V (|x_1-x_2| + |z_1-z_2| + |v_1-v_2| + |\bar{v}_1-\bar{v}_2| )$,
                \item  $|g^h_x(x_1,z_1)-g^h_x(x_2,z_2)| \leq L^g(|x_1-x_2|+|z_1-z_2|)$;
            \end{enumerate}
        

        \item ($\alpha$-concavity)  The function $v\in I\mapsto f^h(t,x,z,v,\bar{v})$ is strictly concave for any $t\in[0,T]$,  $x,z\in\mathbb{R}$,   $\bar{v}\in I$, and $h\in[H]$. In addition, there exist  $\alpha^X_1>0$, $\alpha^X_2\geq 0$, $\alpha^V_1>0, \alpha^V_2\geq 0$, and $\alpha^g_1>0,\alpha^g_2\geq 0$  such that, for any $h\in[H]$, 
        $t\in[0,T]$,  $z_1,z_2 \in \mathbb{R},\bar{v}_1,\bar{v}_2\in I$, and any $(X^1_t)_{t\in[0,T]}$,   $(X^2_t)_{t\in[0,T]},  (V^1_t)_{t\in[0,T]}$, $(V^2_t)_{t\in[0,T]} \in L^2_{\mathbb{F}^h}([0,T];\mathbb{R}) $, 
            \begin{enumerate}[label=(\roman*)]
         
          \item $\mathbb{E}\left[\left(f_x^h(t,X^1_t,z_1,V^1_t,\bar{v}_1)-f_x^h(t,X^2_t,z_2,V^2_t,\bar{v}_2)\right)(X^1_t-X^2_t)\right] \leq-\alpha^X_1\mathbb{E}[|X^1_t-X^2_t|^2] + \alpha^X_2|z_1-z_2|^2 $,

            \item $\mathbb{E}\left[\left(f_v^h(t,X^1_t,z_1,V^1_t,\bar{v}_1)-f_v^h(t,X^2_t,z_2,V^2_t,\bar{v}_2)\right)(V^1_t-V^2_t)\right] \leq-\alpha^V_1\mathbb{E}[|V^1_t-V^2_t|^2] + \alpha^V_2|\bar{v}_1-\bar{v}_2|^2$,

            \item 
            
            $\mathbb{E}[ (g^h_x(X^1_T,z_1)-g^h_x(X^2_T,z_2))(X^1_T-X^2_T)] \leq -\alpha^g_1\mathbb{E}[|X^1_T-X^2_T|^2]+\alpha^g_2|z_1-z_2|^2$.

            
            \end{enumerate}
        
    \end{enumerate}
    
\end{assumption}

\begin{assumption}
\label{ass:contraction}
   For each $h\in[H]$, for any $t\in[0,T]$ and $x,z\in\mathbb{R}$, $\bar{v}\in I$, the function $f^h_v(t,x,z,\cdot,\bar{v})$ admits a unique inverse $(f^h_v)^{-1}(\cdot;t,x,z,\bar{v})$. In addition,  there exists $\rho^h \in (0,1)$ such that, for any $t\in[0,T]$ and $u,\bar{v}_1,\bar{v}_2,v\in I$,  $x,z\in \mathbb{R}$ 
   \begin{equation*}
       \left| (f^h_v)^{-1}\left(u;t,x,z,\bar{v}_1\right) -  (f^h_v)^{-1}\left(u;t,x,z,\bar{v}_2\right) \right| \leq \rho^h\left| \bar{v}_1-\bar{v}_2 \right|.
   \end{equation*}
\end{assumption}

Assumption \ref{ass:concave:monotonicity:h} encompasses the standard Lipschitz and concavity condition that warrants the unique existence of the FBSDEs characterizing the optimal equilibrium solution; see Section \ref{sec:MFFBSDE:wellpose}. These conditions are readily satisfied by quadratic rewards (Section \ref{sec:lq}), and more generally, by revised utility functions; see also Section \ref{sec:mixed:reward} for an example.  Assumption \ref{ass:contraction} ensures the solvability of the mean field fixed point defined in Problem \ref{p:fixed:point} below.  

\begin{remark}
\label{remark:concavity}
    By the mean value theorem, Assumption \ref{ass:concave:monotonicity:h} implies that, for any  $z\in \mathbb{R}$, $\bar{v}\in I$, and any $(X^1_t)_{t\in[0,T]}$,   $(X^2_t)_{t\in[0,T]}$, $(V^1_t)_{t\in[0,T]}$, $(V^2_t)_{t\in[0,T]} \in L^2_{\mathbb{F}^h}([0,T];\mathbb{R})$, 
        \begin{equation*}
        \begin{aligned}
             \mathbb{E}\left[f^h(t,X^1_t,z,V^1_t,\bar{v}) - f^h(t,X^2_t,z,V^2_t,\bar{v})\right] &\leq \mathbb{E}\bigg[f^h_x(t,X^2_t,z,V^2_t,\bar{v})(X^1_t-X^2_t) \\
             &\quad + f^h_v(t,X^2_t,z,V^2_t,\bar{v})(V^1_t-V^2_t)  \bigg], \\
             \mathbb{E}\left[g^h(X^1_T,z) - g^h(X^2_T,z)\right] &\leq \mathbb{E}\left[g^h_x(X^2_T,z)(X^1_T-X^2_T)\right]. 
        \end{aligned}
        \end{equation*}
\end{remark}


\section{Optimal Mean Field Insurance Strategy}
\label{sec:eqm:strategy}
In this section, we construct the optimal insurance strategy of the representative member of each risk class $h\in[H]$ under the mean field formulation (Problems \ref{p:mf} and \ref{p:fixed:point}). The proofs of statements in this section are relegated to Appendix \ref{sec:app:sec:eqm:strategy}. 
   
 By Assumption \ref{ass:concave:monotonicity:h}, one can verify that $J^h(\cdot)$ is concave and coercive, which guarantees the unique existence of optimal control of Problem \ref{p:mf}. The precise statement is formulated below. 

\begin{lemma}
\label{lem:cocercive}
    Suppose that the mean field terms $(z^h)_{h=1}^H$ and $(\bar{v}^h)_{h=1}^H$ are exogeneously given.     Under Assumption \ref{ass:concave:monotonicity:h}, for each $h\in[H]$, the mapping $v^h \in L^2_{\mathbb{F}^h}([0,T];\mathbb{R}) \mapsto J^h(v^h)$ is continuous, strictly concave, and coercive. The last property means that $J^h(v^h)\to-\infty$ as $\mathbb{E}[\int_0^T|v^h_t|^2dt]\to \infty$. 
\end{lemma}
\begin{proof}
    The proof is relegated to Appendix \ref{sec:app:pf:lem:cocercive}. 
\end{proof}

  By Lemma \ref{lem:cocercive}, if the constraint set $I$ is unbounded, the unique existence of a global maximizer of $J^h(\cdot)$ in $\mathcal{A}_{\mathbb{F}^h}(I)$ given the mean field terms $(z^h)_{h=1}^H$ and $(\bar{v}^h)_{h=1}^H$ is a consequence of Theorem 7.2.12 of \cite{drabek2007methods}. On the other hand,  if $a,b\in \mathbb{R}$, by the Banach–Alaoglu theorem,
 the set $\mathcal{A}_{\mathbb{F}^h}(I)\subset L^2_{\mathbb{F}^h}([0,T];\mathbb{R})$ is weakly compact. Hence, the unique existence of a global maximizer of $J^h(\cdot)$ in $\mathcal{A}_{\mathbb{F}^h}(I)$ is an immediate consequence of the extreme value theorem (see Theorem 7.2.4 of \cite{drabek2007methods}).

The following statement characterizes the optimal insurance strategy of Problem \ref{p:mf} using the stochastic maximum principle. 

\begin{theorem}
\label{thm:sol:p2}
  Under Assumption \ref{ass:concave:monotonicity:h}, and  given the deterministic functions $(z^h)_{h=1}^H$ and $(\bar{v}^h)_{h=1}^H$, the optimal insurance strategy for the representative member in Class $h$, $h\in[H]$, is given by 
        \begin{equation}
            \label{eq:v*}
            v^h_t = \textup{Proj}_I\left[ \left(f^h_v\right)^{-1}\left(-\left(\kappa^h p^h_t + \sigma^h \eta^h_t\right) ; t,x^h_t,z^h_t,\bar{v}^h_t\right) \right],
        \end{equation}
where 
$(x^h,p^h,\eta^h)\in L^2_{\mathbb{F}^h}([0,T]; \mathbb{R}^{3})$ is the solution of the following FBSDE:    
    \begin{equation}
    \label{eq:FBSDE:P1}
        \begin{dcases}
            dx^h_t = \left(rx^h_t + l - \kappa^h v^h_t + \pi^h\sum_{j=1}^H\omega^j (\kappa^j-d^j)\bar{v}^j_t \right)dt + \sigma^h(1-v^h_t)dW^h_t, \\ 
            -dp^h_t = \left(rp^h_t - f^h_x(t,x^h_t,z^h_t,v^h_t,\bar{v}^h_t) \right)dt - \eta^h_t dW^h_t, \\
            x^h_0 = \xi^h, \\ 
            p^h_T = -g_x(x^h_T,z^h_T).
        \end{dcases}
    \end{equation}
\end{theorem}    
\begin{proof}
    The proof is relegated to Appendix \ref{sec:app:pf:thm:sol:p2}. 
\end{proof}

Theorem \ref{thm:sol:p2} presents the solution of Problem \ref{p:mf} given the mean field terms $(z^h)_{h=1}^H$ and $(v^h)_{h=1}^H$. By taking expectations on \eqref{eq:FBSDE:P1}, we see that the solution of the mean field game is characterized by, for each $h\in[H]$,  
    \begin{equation}
    \label{eq:MFFBSDE:component}
    \left\{
        \begin{aligned}
             dx^h_t &= \left(rx^h_t + l - \kappa^h v^h_t + \pi^h\sum_{j=1}^H\omega^j(\kappa^j-d^j)\bar{v}^j_t\right)dt + \sigma^h(1-v^h_t)dW^h_t, \\ 
            -dp^h_t &= \left(rp^h_t - f^h_x(t,x^h_t,z^h_t,v^h_t,\bar{v}^h_t) \right)dt - \eta^h_t dW^h_t, \\
            dz^h_t &= \left(rz^h_t + l - \kappa^h \bar{v}^h_t + \pi\sum_{j=1}^H \omega^j(\kappa^j-d^j)\bar{v}^j_t\right) dt, \\
            x^h_0 &= \xi^h, \\ 
            z^h_0 &= \mathbb{E}[\xi^h],\\ 
            p^h_T &= -g_x(x^h_T,z^h_T),
        \end{aligned}\right. 
    \end{equation}
where for $t\in[0,T]$, 
    \begin{align*}
             v^h_t &= \text{Proj}_I\left[ \left(f^h_v\right)^{-1}\left(-\left(\kappa^h p^h_t + \sigma^h \eta^h_t\right) ;t, x^h_t,z^h_t,\bar{v}^h_t \right) \right],   \\
           \bar{v}^h_t &= \mathbb{E}\left[ \text{Proj}_I\left[ \left(f^h_v\right)^{-1}\left(-\left(\kappa^h p^h_t + \sigma^h \eta^h_t\right) ; t,x^h_t,z^h_t,\bar{v}^h_t\right) \right] \right]. 
    \end{align*}

Notice that by Assumption \ref{ass:contraction}, for any $u,\bar{v}_1,\bar{v}_2\in I$, $x,z\in \mathbb{R}$,
    \begin{align*}
       &\ \left|\text{Proj}_I (f_v^h)^{-1}(u;t,x,z,\bar{v}_1) -\text{Proj}_I (f_v^h)^{-1}(u;t,x,z,\bar{v}_2) \right| \\ \leq&\  \left|(f_v^h)^{-1}(u;t,x,z,\bar{v}_1)  -  (f_v^h)^{-1}(u;t,x,z,\bar{v}_2) \right| \leq \rho^h|\bar{v}_1-\bar{v}_2|. 
    \end{align*}
Since $\rho^h<1$, and the mapping 
    \begin{equation*}
        \bar{v}^h  \mapsto  \mathbb{E}\left[ \text{Proj}_I\left[ \left(f^h_v\right)^{-1}\left(-\left(\kappa^h p^h_t + \sigma^h \eta^h_t\right) ; t,x^h_t,z^h_t,\bar{v}^h_t\right) \right] \right]
    \end{equation*}
is clearly invariant in $I$, by the Banach fixed point theorem, there exists a unique fixed point $\bar{v}^h$ that solves the last equation of \eqref{eq:MFFBSDE:component}, provided that the MF-FBSDE is solvable. 
    
Collecting all $H$ representative members, we obtain the following system of MF-FBSDE: 
  \begin{equation}
    \label{eq:MFBSDE:general}
        \left\{\begin{aligned}
            d{\bf x}_t &= \left(r{\bf x}_t + {\bf l} - {\bf K}{\bf v}_t + {\bf \Pi} \bar{{\bf v}}_t \right)dt + {\bf \Sigma}\left({\bf I} - \text{diag}({\bf v}_t)\right)d{\bf W}_t, \\
            -d{\bf p}_t &= \left(r{\bf p}_t - \partial_{\bf x} {\bf F}(t,{\bf x}_t,{\bf z}_t,{\bf v}_t, \bar{{\bf v}}_t)\right)dt  - \text{diag}(\boldsymbol{\eta}_t) d{\bf W}_t, \\
            {\bf x}_0 &= (\xi^1,\dots,\xi^h)^\top, \\
            {\bf p}_T &= -\partial_{\bf x} {\bf G}({\bf x}_T,{\bf z}_T), 
        \end{aligned}\right.
    \end{equation}
where for $t\in[0,T]$, 
    \begin{align*}
    {\bf z}_t &= \mathbb{E}[{\bf x}_t], \\ 
          \bar{{\bf v}}_t &= \mathbb{E}\left[\text{Proj}_{I^H}\left[ \left(\partial_{\bf v} {\bf F}\right)^{-1}\left(-\left({\bf K}{\bf p}_t + {\bf \Sigma} \boldsymbol{\eta}_t\right) ;t, {\bf x}_t,{\bf z}_t,\bar{{\bf v}}_t\right) \right]\right], \\
            {\bf v}_t &=  \text{Proj}_{I^H}\left[ \left(\partial_{\bf v} {\bf F}\right)^{-1}\left(-\left({\bf K}{\bf p}_t + {\bf \Sigma} \boldsymbol{\eta}_t\right) ;t, {\bf x}_t,{\bf z}_t,\bar{{\bf v}}_t\right) \right],
    \end{align*}
and the vectors and matrices in \eqref{eq:MFBSDE:general} are defined as follows:
\begin{align*}
          &  {\bf x}_t = \begin{pmatrix}
                x^1_t\\
                \vdots \\
                x^H_t
            \end{pmatrix},{\bf z}_t = \begin{pmatrix}
                z^1_t\\
                \vdots \\
                z^H_t
            \end{pmatrix}, {\bf p}_t = \begin{pmatrix}
                p^1_t\\ \vdots\\ p^H_t  
            \end{pmatrix}, \boldsymbol{\eta}_t =  \begin{pmatrix}
                \eta^1_t \\
                \vdots \\ \eta^H_t
            \end{pmatrix} , {\bf v}_t = \begin{pmatrix}
                v^1_t \\
                \vdots \\ v^H_t
            \end{pmatrix},  {\bf \bar{v}}_t = \begin{pmatrix}
                \bar{v}^1_t \\
                \vdots \\ \bar{v}^H_t
            \end{pmatrix}, 
             {\bf W}_t = \begin{pmatrix}
                W^1_t\\
                \vdots\\
                W^H_t
            \end{pmatrix}, \\
            & {\bf l} = \begin{pmatrix}
                l^1 \\ \vdots \\ l^H
            \end{pmatrix},  \boldsymbol{\Pi} =    \begin{pmatrix}
                \pi^1\omega^1(\kappa^1-d^1) & \cdots & \pi^1\omega^H(\kappa^H-d^H) \\
                \vdots &\vdots&\vdots \\
                   \pi^H  \omega^1(\kappa^1-d^1) & \cdots & \pi^H\omega^H(\kappa^H-d^H) 
            \end{pmatrix},  \  {\bf F} = \begin{pmatrix}
                f^1 \\ \vdots \\ f^H 
            \end{pmatrix}, \ {\bf G} = \begin{pmatrix}
                g^1 \\ \vdots \\ g^H
            \end{pmatrix},
        \end{align*}
${\bf K} = \text{diag}( (\kappa^h)_{h=1}^H)$, $ {\bf \Sigma} = \text{diag}( (\sigma^h)_{h=1}^H)$, $\partial_{\bf x} = \text{diag}((\partial_{x^h})_{h=1}^H)$, $\partial_{\bf v} = \text{diag}((\partial_{v^h})_{h=1}^H)$, and $(\partial_{\bf v}{\bf F})^{-1}=( (f_v^1)^{-1},\dots,(f_v^H)^{-1})^\top$.  
    
\section{Well-posedness of the MF-FBSDE \eqref{eq:MFBSDE:general}}
\label{sec:MFFBSDE:wellpose}
In this section, we  establish the global existence and uniqueness of the MF-FBSDE   \eqref{eq:MFBSDE:general}, which therefore warrants the solvability of Problems \ref{p:mf}-\ref{p:fixed:point}. In the sequel, the term solution always refers to a triple $({\bf x},{\bf p},\boldsymbol{\eta})$ that satisfies \eqref{eq:MFBSDE:general} and lies in $L^2_{\mathbb{F}^{[H]}}([0,T];\mathbb{R}^{3H})$. The proofs of statements in this section are relegated to Appendix \ref{sec:app:pf:wellpose}. 

In the technical perspective, canonical results in the literature concerning the global existence of MF-FBSDE cannot be directly applied herein due to two major aspects. First, as a result of the insurance constraint, the coefficients of the MF-FBSDE fail to satisfy the standard monotonicity property. Second, the forward equations depend directly on the mean field insurance strategies of the representative members from the other risk classes under the extended mean field game framework. Our approach thus involves adaptations of the well-known continuation approach (see e.g.~\cite{bensoussan2017linear}) by utilizing the properties of a projection map.

\subsection{Assumptions for Well-posedness of MF-FBSDE}
Before proceeding to the main results and proofs of this section, we introduce the following additional assumptions.  
\begin{assumption}  \hfill 
    \label{ass:fg:sep}
    \begin{enumerate}[label=(\alph*)]  
        \item (Separability) For any $h\in [H]$,  $f^h$ is separable in the following sense:
    \begin{equation*}
        f^h(t,x,z,v,\bar{v}) = f^{X,h}(t,x,z) + f^{V,h}(t,v,\bar{v}),
    \end{equation*}
where $f^{X,h}:[0,T]\times \mathbb{R}\times \mathbb{R}\to \mathbb{R},f^{V,h}:[0,T]\times \mathbb{R}\times\mathbb{R}\to \mathbb{R}$. 

    \item (Lipschitzity of $f^{X,h}_x$) For any $h\in [H]$, $t\in[0,T]$ and $x_1,x_2,z_1,z_2\in \mathbb{R}$,  
        \begin{equation*}
            |f^{X,h}_x(t,x_1,z_1)-f^{X,h}_x(t,x_2,z_2)|\leq L^X(|x_1-x_2|+|z_1-z_2|), 
        \end{equation*}
    where $L^X>0$ is the constant in Assumption \ref{ass:concave:monotonicity:h}. 
   \item (Convexity, and Lipschitzity of $f^{V,h}_v$) There exist  $\alpha^V,L^V_1,L^V_2>0$ with $\alpha^V > L^V_2$ such that, for any $h\in[H]$, $t\in[0,T]$, $h\in[H]$, and  $v,v_1,v_2,\bar{v},\bar{v}_1,\bar{v}_2\in \mathbb{R}$, 
    \begin{align*}
       & (v_1-v_2)(f^{V,h}_v(t,v_1,\bar{v}) - f^{V,h}_v(t,v_2,\bar{v})  ) \leq -\alpha^V(v_1-v_2)^2,  \\
       & \left|f^{V,h}_v(t,v_1,\bar{v})  - f^{V,h}_v(t,v_2,\bar{v})  \right| \leq L^V_1|v_1-v_2|, \\ 
       &  |f^{V,h}_v(t,v,\bar{v}_1) -f^{V,h}_v(t,v,\bar{v}_2) |\leq L^V_2|\bar{v}_1-\bar{v}_2|.
    \end{align*}
     \end{enumerate}

\end{assumption}


To introduce the next assumption, we define the matrix ${\bf M}$ by 
    \begin{equation}
        \label{eq:M}
            {\bf M} :=  {\bf \Pi}\left( {\bf \Pi} -{\bf K} \right)^{-1}. 
        \end{equation}
The following result shows that ${\bf \Pi}-{\bf K}$ is invertible, whence ${\bf M}$ is well-defined.

\begin{lemma}\label{lem:invertible}
	The matrix ${\bf \Pi}-{\bf K}$ is invertible. 
\end{lemma}
\begin{proof}
	Define
\begin{equation}
\label{eq:pi:v}
    \boldsymbol{\pi} := 
    \begin{pmatrix} 
        \pi^1 & \cdots & \pi^h 
    \end{pmatrix}^{\!\top}, 
    \ 
    \boldsymbol{\upsilon} := 
    \begin{pmatrix} 
        \omega^1(\kappa^1- d^h) & \cdots & \omega^h(\kappa^h- d^h)
    \end{pmatrix}^{\!\top}.
\end{equation}
Note that ${\bf \Pi}-{\bf K} = \boldsymbol{\pi} \boldsymbol{\upsilon}^{\top} - {\bf K}$, ${\bf K}$ is invertible and $\boldsymbol{\upsilon}^{\top}{\bf K}^{-1}\boldsymbol{\pi} = \sum_{h=1}^H \pi^h\omega^h\frac{\kappa^h-d}{\kappa^h} < \sum_{h=1}^H \pi^h\omega^h = 1$. By the Sherman–Morrison-Woodbury formula (see Section 2.1.3 on page 50 of \cite{golub2013matrix}), ${\bf \Pi}-{\bf K}$ is invertible and
    \begin{equation}
    \label{eq:Pi-K:inv}
        ({\bf \Pi} - {\bf K})^{-1} = -{\bf K}^{-1} - \frac{{\bf K}^{-1}\boldsymbol{\pi} \boldsymbol{\upsilon}^{\top}{\bf K}^{-1}}{1-\boldsymbol{\upsilon}^{\top}{\bf K}^{-1}\boldsymbol{\pi} }. 
    \end{equation}

\end{proof}

\begin{assumption}
\label{ass:M} \hfill 
    \begin{enumerate}[label=(\alph*)]
     

    \item    $\lambda_{\min}({\bf I}-{\bf M})>0$;
        \item  There exist $\alpha_{\bf M},\alpha^{\bf G}_{\bf M}  >0$  such that, for any ${\bf x}^i, {\bf v}^i \in L^2_{\mathbb{F}^{[H]}}([0,T];\mathbb{R}^H)$, $i=1,2$,  

        \begin{equation}
        \label{eq:concavity:U}
            \begin{aligned}
             & \ \  \    \mathbb{E}\bigg[  \bigg\langle {\bf x}^1_t - {\bf x}^2_t - {\bf M}\mathbb{E}[{\bf x}^1_t-{\bf x}^2_t], \partial_{\bf x}{\bf F}({\bf x}^1_t,\mathbb{E}[{\bf x}^1_t],{\bf v}^1_t,\mathbb{E}[{\bf v}^1_t]) \\
             &\quad - \partial_{\bf x}{\bf F}({\bf x}^2_t,\mathbb{E}[{\bf x}^2_t],{\bf v}^2_t,\mathbb{E}[{\bf v}^2_t])    \bigg\rangle \bigg]\leq -\alpha_{{\bf M}}\mathbb{E}\left[ \|{\bf x}^1_t-{\bf x}^2_t \|^2 \right], \\
             & \ \ \  \mathbb{E}\left[  \left\langle {\bf x}^1_T - {\bf x}^2_T - {\bf M}\mathbb{E}[{\bf x}^1_T-{\bf x}^2_T], \partial_{\bf x}{\bf G}({\bf x}^1_T,\mathbb{E}[{\bf x}^1_T]) - \partial_{\bf x}{\bf G}({\bf x}^2_T,\mathbb{E}[{\bf x}^2_T])    \right\rangle \right]\\
             &\leq -\alpha_{\bf M}^{\bf G}  \mathbb{E}\left[ \|{\bf x}^1_T-{\bf x}^2_T \|^2 \right]. 
            \end{aligned}
        \end{equation}
    \end{enumerate}

\end{assumption}

\begin{remark}
    Assumption \ref{ass:fg:sep} implies Assumption \ref{ass:concave:monotonicity:h}.A.(iii)-(iv), and B.(ii). 
\end{remark}

The first condition of Assumption \ref{ass:fg:sep} requires the running rewards to be separable into two parts, namely, a state-dependent and a control-dependent component. This aligns with popular choices of rewards where a separate term is included to penalize extreme actions. The third condition requires $f^{V,h}_v$ to be Lipschitz in the $\bar{v}$-variable, where the Lipschitz constant $L^V_2$ shall be smaller than the concavity constant $\alpha^V$ with respect to the $v$-variable. This reflects a smaller sensitivity of the representative member's preference with respect to the mean field term, which thus captures a small mean field effect in practical MFGs. 

Assumption \ref{ass:M} loosely requires that the matrix ${\bf M}$ has a moderate impact. In Proposition \ref{pp:I-M>0} below, we provide equivalent formulations of Assumption \ref{ass:M}(a), along with a sufficient condition on the model parameters that implies it.  On the other hand,  Assumption \ref{ass:M}(b) can be fulfilled by reward functions satisfying a slightly stronger monotonicity condition; see e.g.~Section \ref{sec:lq} below. Alternatively, if the concavity constants $\alpha^X_1,\alpha^X_2,\alpha^g_1,\alpha^g_2$ in Assumption \ref{ass:concave:monotonicity:h}.B satisfy $\alpha^X_1>\alpha^X_2 \geq 0$ and $\alpha^g_1>\alpha^g_2\geq 0$,  then \eqref{eq:concavity:U} holds if 
\begin{equation}
        \label{eq:alpha:M}
           \alpha^X - 2L^X \|{\bf M}\|_2 >0, \ \alpha^g-2L^g\|{\bf M}\|_2>0,
        \end{equation}
    where $\alpha^X:=\alpha^X_1-\alpha^X_2$ and $\alpha^g:= \alpha^g_1-\alpha^g_2$. To see this, for any ${\bf x}^i, {\bf v}^i \in L^2_{\mathbb{F}^{[H]}}([0,T];\mathbb{R}^H)$, $i=1,2$, using Assumptions \ref{ass:concave:monotonicity:h}, \ref{ass:fg:sep}, and Jensen's inequality, 
        \begin{align*}
           & \ \ \ \  \mathbb{E}\bigg[  \bigg\langle {\bf x}^1_t - {\bf x}^2_t - {\bf M}\mathbb{E}[{\bf x}^1_t-{\bf x}^2_t], \partial_{\bf x}{\bf F}({\bf x}^1_t,\mathbb{E}[{\bf x}^1_t],{\bf v}^1_t,\mathbb{E}[{\bf v}^1_t])- \partial_{\bf x}{\bf F}({\bf x}^2_t,\mathbb{E}[{\bf x}^2_t],{\bf v}^2_t,\mathbb{E}[{\bf v}^2_t])    \bigg\rangle \bigg] \\
           &\leq -\alpha^X \mathbb{E}\left[\left|{\bf x}^1_t-{\bf x}^2_t \right|^2\right] + L^X\|{\bf M}\|_2\mathbb{E}\left[ \left|{\bf x}^1_t-{\bf x}^2_t \right|\left( \left|{\bf x}^1_t-{\bf x}^2_t \right| +  \left|\mathbb{E}[{\bf x}^1_t-{\bf x}^2_t] \right| \right)  \right] \\
           &\leq -(\alpha^X - 2L^X\|{\bf M}\|_2)\mathbb{E}\left[\left|{\bf x}^1_t-{\bf x}^2_t \right|^2\right],
        \end{align*}
so that we can pick $\alpha_{\bf M} = \alpha^X - 2L^X\|{\bf M}\|_2>0$. The second inequality in \eqref{eq:concavity:U}  can be shown in the same manner. Following this idea, if ${\bf F}$ and ${\bf G}$ are independent of the argument ${\bf z}$, the condition can be further relaxed to requiring  
    \begin{equation}
        \label{eq:alpha:M:2}
        \alpha^X_1 - L^X\|{\bf M}\|_2 ,\  \alpha^g_1 - L^g\|{\bf M}\|_2 >0. 
    \end{equation}

\begin{proposition}\label{pp:I-M>0}
	The following conditions are equivalent: 
		\begin{enumerate}
                \item $\lambda_{\min}({\bf I}-{\bf M})>0$.
			\item  $\lambda_{\max}({\bf \Pi}{\bf K}^{-1}) < 1$;
                \item  $\sum_{h=1}^H \frac{\pi^h\omega^h(\kappa^h-d^h)}{\kappa^h} + \sqrt{\left(\sum_{h=1}^H (\pi^h)^2\right)\left( \sum_{h=1}^H \left(\frac{\omega^h(\kappa^h-d^h)}{\kappa^h}\right)^2\right)} < 2$. 
		\end{enumerate}
   In addition, the above conditions hold provided that 
    \begin{enumerate}
    \setcounter{enumi}{3}
        \item $ \sup_{h\in[H]}\left\{\frac{\pi^h}{\omega^h}\right\} < \inf_{h\in[H]}\left\{\frac{\pi^h}{\omega^h} \frac{\kappa^h}{\kappa^h-d}\right\}$.
    \end{enumerate}
\end{proposition}
\begin{proof}
	The proof is relegated to Appendix \ref{app:lem:I-M>0}.
\end{proof}

 Proposition \ref{pp:I-M>0} indicates that the condition $\lambda_{\min}({\bf I}-{\bf M})>0$ is met if and only if the effect of the surplus distribution on members' wealth, as captured by the magnitude of ${\bf \Pi}$, remains sufficiently moderate relative to the premium rate and safety loading represented by ${\bf K}$. This again echoes the small mean field requirement. In particular, the condition is satisfied when the ratios $(\frac{\pi^h}{\omega^h})_{h=1}^H$ do not deviate significantly between classes.

 \subsection{Uniqueness of Solution}
We begin by establishing the uniqueness of solutions to the MF-FBSDE \eqref{eq:MFBSDE:general}. The proof relies on the properties of the projection map (see Lemmas \ref{lem:convex}–\ref{lem:proj}), which enable us to derive a weaker form of monotonicity. Combined with Assumption \ref{ass:M}, this allows us to bypass the stronger monotonicity conditions commonly assumed in the literature, which no longer hold herein due to the non-expansive nature of the projection map.

\begin{theorem}
    \label{thm:unique:general}
    Under Assumptions \ref{ass:concave:monotonicity:h}, \ref{ass:contraction}, \ref{ass:fg:sep}, and \ref{ass:M},
    the MF-FBSDE \eqref{eq:MFBSDE:general} admits at most one adapted solution. 
\end{theorem}

\begin{proof}
    Consider two solutions $({\bf x}^i,{\bf p}^i,\boldsymbol{\eta}^i)$, $i=1,2$, and let $(\tilde{{\bf x}}, \tilde{{\bf p}}, \tilde{\boldsymbol{\eta}}) :=({\bf x}^1-{\bf x}^2,{\bf p}^1-{\bf p}^2, \boldsymbol{\eta}^1-\boldsymbol{\eta}^2)$. By applying It\^o's lemma to $\langle \tilde{{\bf x}}_t,\tilde{{\bf p}}_t\rangle$, we have 
    \begin{align}
    \label{eq:unique:1}
     &\ \ \ \    - \mathbb{E}\left[\langle \tilde{{\bf x}}_T, \partial_{\bf x} {\bf G}({\bf x}^1_T,{\bf z}^1_T) - \partial_{\bf x} {\bf G}({\bf x}^2_T,{\bf z}^2_T) \rangle\right]\nonumber \\
      &= \mathbb{E}\Bigg[\int_0^T \Bigg(\bigg\langle {\bf r}\tilde{\bf x}_t - {\bf K}\tilde{\bf v}_t +{\bf \Pi} \mathbb{E}[\tilde{\bf v}_t]  , \Tilde{{\bf p}}_t\bigg\rangle  - \langle \tilde{\boldsymbol{\eta}}_t, {\bf \Sigma}\tilde{{\bf v}}_t \rangle \nonumber  \\
      &\qquad - \left\langle \tilde{{\bf x}}_t, r\tilde{{\bf p}}_t -\left( \partial_{\bf x} {\bf F}(t,{\bf x}^1_t,{\bf z}^1_t,{\bf v}^1_t, \bar{{\bf v}}^1_t)- \partial_{\bf x} {\bf F}(t,{\bf x}^2_t,{\bf z}^2_t,{\bf v}^2_t, \bar{{\bf v}}^2_t) \right) \right\rangle \Bigg)dt \Bigg]  \nonumber\\
       &=  \mathbb{E}\Bigg[\int_0^T \Bigg( \langle \tilde{{\bf x}}_t, \partial_{\bf x} {\bf F}(t,{\bf x}^1_t,{\bf z}^1_t,{\bf v}^1_t, \bar{{\bf v}}^1_t)- \partial_{\bf x} {\bf F}(t,{\bf x}^2_t,{\bf z}^2_t,{\bf v}^2_t,\bar{{\bf v}}^2_t) \rangle- \langle \tilde{{\bf v}}_t, {\bf K}\tilde{{\bf p}}_t + {\bf \Sigma}\Tilde{\boldsymbol{\eta}}_t \rangle \nonumber \\
       &\qquad + \langle   \tilde{{\bf p}}_t, {\bf \Pi} \mathbb{E}[\tilde{\bf v}_t] \rangle   \Bigg)dt \Bigg],    \end{align}
where $\bar{\bf v}^i_t := \mathbb{E}[{\bf v}^i_t]$, $i=1,2$. 

To proceed, we establish the following weaker form of monotonicity: 
    \begin{equation}
    \label{eq:bad:1}
        \mathbb{E}\left[\int_0^T\langle \tilde{{\bf v}}_t, {\bf K}\tilde{{\bf p}}_t + {\bf \Sigma}\Tilde{\boldsymbol{\eta}}_t \rangle dt\right] \geq 0. 
    \end{equation}
Indeed, for each $h\in[H]$, 
 \begin{align*}
        & \ \ \ \  (v^{h,1}_t-v^{h,2}_t)\left(\kappa^h(p^{h,1}_t-p^{h,2}_t) + \sigma^h(\eta^{h,1}_t-\eta^{h,2}_t) \right)\\  
        &= (v^{h,1}_t-v^{h,2}_t)\left[ f^{V,h}_v\left(t,\hat{v}^{h,2}_t,\bar{v}^{h,2}_t\right) -  f^{V,h}_v\left(t,\hat{v}^{h,1}_t,\bar{v}^{h,1}_t\right) \right],
    \end{align*}
where for $i=1,2$, 
    \begin{equation*}
        \hat{v}^{h,i}_t := (f^{V,h}_v)^{-1}(-\kappa^h p^{h,i}_t - \sigma^h\eta^{h,i}_t;t, \bar{v}^{h,i}_t). 
    \end{equation*}
Hence, 
\begin{align}
\label{eq:bad:1:1}
         & \ \ \ \   (v^{h,1}_t-v^{h,2}_t)\left(\kappa^h(p^{h,1}_t-p^{h,2}_t) + \sigma^h(\eta^{h,1}_t-\eta^{h,2}_t) \right) \nonumber  \\
        &= -(v^{h,1}_t-v^{h,2}_t)\left[ f^{V,h}_v\left(t,v^{h,1}_t, \bar{v}^{h,1}_t\right) - f^{V,h}_v\left(t,v^{h,2}_t,\bar{v}^{h,1}_t\right) \right]  \nonumber \\
        &\quad -  (v^{h,1}_t-v^{h,2}_t)\Bigg[ \left(f^{V,h}_v\left(t,\hat{v}^{h,1}_t, \bar{v}^{h,1}_t\right) - f^{V,h}_v\left(t,v^{h,1}_t, \bar{v}^{h,1}_t\right)\right) \nonumber \\
        &\qquad - \left(f^{V,h}_v\left(t,\hat{v}^{h,2}_t, \bar{v}^{h,1}_t\right) - f^{V,h}_v\left(t,v^{h,2}_t, \bar{v}^{h,1}_t\right)\right) \Bigg] \nonumber \\
        &\quad - (v^{h,1}_t-v^{h,2}_t)\left[ f^{V,h}_v\left(t,\hat{v}^{h,2}_t, \bar{v}^{h,1}_t\right) -  f^{V,h}_v\left(t,\hat{v}^{h,2}_t,\bar{v}^{h,2}_t\right) \right].
    \end{align}
By Assumption \ref{ass:fg:sep}, we have 
    \begin{align}
    \label{eq:bad:1:2}
       -(v^{h,1}_t-v^{h,2}_t)\left( f^{V,h}_v\left(t,v^{h,1}_t, \bar{v}^{h,1}_t\right) - f^{V,h}_v\left(t,v^{h,2}_t,\bar{v}^{h,1}_t\right) \right)    \geq \alpha^V|v^{h,1}_t-v^{h,2}_t|^2. 
    \end{align}
On the other hand, by noticing that $f^{V,h}_v(t,\cdot,\bar{v})$ is non-increasing and $v^{h,i}_t = \text{Proj}_I[ \hat{v}^{h,i}_t ]$, using Lemma \ref{lem:proj}, we arrive at 
    \begin{align}
    \label{eq:bad:1:3}
      &\  -  (v^{h,1}_t-v^{h,2}_t)\Bigg[ \left(f^{V,h}_v\left(t,\hat{v}^{h,1}_t, \bar{v}^{h,1}_t\right) - f^{V,h}_v\left(t,v^{h,1}_t, \bar{v}^{h,1}_t\right)\right) \nonumber \\
        &\qquad - \left(f^{V,h}_v\left(t,\hat{v}^{h,2}_t, \bar{v}^{h,1}_t\right) - f^{V,h}_v\left(t,v^{h,2}_t, \bar{v}^{h,1}_t\right)\right) \Bigg] \geq 0. 
    \end{align}
Next, by Assumption \ref{ass:fg:sep}, 
    \begin{align}
    \label{eq:bad:1:4}
        -(v^{h,1}_t-v^{h,2}_t)\left[ f^{V,h}_v\left(t,\hat{v}^{h,2}_t, \bar{v}^{h,1}_t\right) -  f^{V,h}_v\left(t,\hat{v}^{h,2}_t,\bar{v}^{h,2}_t\right) \right] &\geq - L^V_2 |v^{h,1}_t-v^{h,2}_t||\bar{v}^{h,1}_t-\bar{v}^{h,2}_t|. 
    \end{align}
Hence, by combining \eqref{eq:bad:1:1}-\eqref{eq:bad:1:4} using Jensen's inequality, we have 
    \begin{align*}
     & \ \ \ \  \mathbb{E}\left[\int_0^T (v^{h,1}_t-v^{h,2}_t)\left(\kappa^h(p^{h,1}_t-p^{h,2}_t) + \sigma^h(\eta^{h,1}_t-\eta^{h,2}_t) \right)dt\right] \\
     &\geq  \int_0^T \left(\alpha^V\mathbb{E}\left[|v^{h,1}_t-v^{h,2}_t|^2\right]  -L^V_2 \mathbb{E}\left[|v^{h,1}_t-v^{h,2}_t||\bar{v}^{h,1}_t-\bar{v}^{h,2}_t|\right] \right) dt \\
     &\geq \int_0^T (\alpha^V-L^V_2)\mathbb{E}\left[|v^{h,1}_t-v^{h,2}_t|^2\right] dt \geq 0,
    \end{align*}
and thus \eqref{eq:bad:1} follows.

Next, by differentiating $\langle {\bf M} \tilde{{\bf z}}_t, \mathbb{E}[\tilde{{\bf p}}_t]\rangle$ with respect to $t$, we have 
    \begin{align}
    \label{eq:bad:2}
        \int_0^T  \langle  \mathbb{E}[\tilde{{\bf p}}_t], {\bf \Pi} \mathbb{E}[\tilde{\bf v}_t] \rangle dt &= - \langle {\bf M}\tilde{{\bf z}}_T,  \mathbb{E}\left[ \partial_{\bf x}{\bf G}({\bf x}^1_T,{\bf z}^1_T) - \partial_{\bf x}{\bf G}({\bf x}^2_T,{\bf z}^2_T)  \right]  \rangle\nonumber  \\
        &\hspace{-1cm} - \int_0^T \left\langle {\bf M}\tilde{{\bf z}}_t, \mathbb{E}\left[\partial_{\bf x} {\bf F}(t,{\bf x}^1_t,{\bf z}^1_t,{\bf v}^1_t, \bar{{\bf v}}^1_t)- \partial_{\bf x} {\bf F}(t,{\bf x}^2_t,{\bf z}^2_t,{\bf v}^2_t,\bar{{\bf v}}^2_t) \right] \right\rangle dt. 
    \end{align}

 By combining \eqref{eq:unique:1}, \eqref{eq:bad:1}, \eqref{eq:bad:2}, and using Assumptions \ref{ass:concave:monotonicity:h}, \ref{ass:fg:sep}, and \ref{ass:M}, we have  
    \begin{align}
        \label{eq:proof:matrix:concavity}
       0&\geq   -\mathbb{E}\left[\langle \tilde{{\bf x}}_T, \partial_{\bf x} {\bf G}({\bf x}^1_T,{\bf z}^1_T) - \partial_{\bf x} {\bf G}({\bf x}^2_T,{\bf z}^2_T) \rangle\right] +   \langle {\bf M}\tilde{{\bf z}}_T,  \mathbb{E}\left[ \partial_{\bf x}{\bf G}({\bf x}^1_T,{\bf z}^1_T) - \partial_{\bf x}{\bf G}({\bf x}^2_T,{\bf z}^2_T)  \right]\rangle\nonumber \\
       &\quad  - \mathbb{E}\left[\int_0^T  \langle \tilde{{\bf x}}_t, \partial_{\bf x} {\bf F}(t,{\bf x}^1_t,{\bf z}^1_t,{\bf v}^1_t, \bar{{\bf v}}^1_t)- \partial_{\bf x} {\bf F}(t,{\bf x}^2_t,{\bf z}^2_t,{\bf v}^2_t,\bar{{\bf v}}^2_t) \rangle dt\right]\nonumber \\
       &\quad +  \int_0^T \left\langle {\bf M}\tilde{{\bf z}}_t, \mathbb{E}\left[\partial_{\bf x} {\bf F}(t,{\bf x}^1_t,{\bf z}^1_t,{\bf v}^1_t, \bar{{\bf v}}^1_t)- \partial_{\bf x} {\bf F}(t,{\bf x}^2_t,{\bf z}^2_t,{\bf v}^2_t,\bar{{\bf v}}^2_t) \right] \right\rangle dt \nonumber \\
       &= -\mathbb{E}\left[\langle \tilde{{\bf x}}_T-{\bf M}\tilde{{\bf z}}_T, \partial_{\bf x} {\bf G}({\bf x}^1_T,{\bf z}^1_T) - \partial_{\bf x} {\bf G}({\bf x}^2_T,{\bf z}^2_T) \rangle\right] \nonumber\\ 
       &\quad - \mathbb{E}\left[\int_0^T  \langle \tilde{{\bf x}}_t-{\bf M}\tilde{{\bf z}}_t, \partial_{\bf x} {\bf F}(t,{\bf x}^1_t,{\bf z}^1_t,{\bf v}^1_t, \bar{{\bf v}}^1_t)- \partial_{\bf x} {\bf F}(t,{\bf x}^2_t,{\bf z}^2_t,{\bf v}^2_t,\bar{{\bf v}}^2_t) \rangle dt\right] \nonumber \\
       &\geq   \alpha_{\bf M}^{{\bf G}}\mathbb{E}\left[|\tilde{\bf x}_T|^2\right] +  \alpha_{\bf M} \mathbb{E}\left[\int_0^T |\tilde{{\bf x}}_t|^2 dt\right].  
    \end{align}
By standard a priori estimates of (F)BSDEs (see e.g.~\eqref{eq:exist:general:proof:8} with $\delta=0$ below) and Gr\"onwall's inequality, we conclude that ${\bf p}^1\equiv {\bf p}^2$ and $\boldsymbol{\eta}^1\equiv \boldsymbol{\eta}^2$. 

 
\end{proof}

\subsection{Global Existence of Solution}
\label{sec:exist}
We proceed to prove the global existence of solution of the MF-FBSDE \eqref{eq:MFBSDE:general} by the continuation approach. To this end, we consider the following MF-FBSDE parameterized by $\mu\in[0,1]$: 
  \begin{equation}
     \label{eq:MFFBSDE:general:parametrized}
    \left\{
        \begin{aligned}
            d\hat{{\bf x}}_t &= \left[ -(1-\mu)\hide{\beta_2} \hat{{\bf p}}_t + \mu\left( r\hat{{\bf x}}_t + {\bf l} - {\bf K}\hat{{\bf v}}_t + {\bf \Pi} \mathbb{E}[\hat{\bf v}_t] \right) + \boldsymbol{\phi}_t  \right]dt \\
            &\quad + \left[ -(1-\mu)\hide{\beta_2}\text{diag}(\hat{\boldsymbol{\eta}}_t)  + \mu{\bf \Sigma}({\bf I}-\text{diag}(\hat{{\bf v}}_t)) +\boldsymbol{\psi}_t \right]d{\bf W}_t, \\
            -d\hat{{\bf p}}_t &= \left[ (1-\mu)\hide{\beta_1} \hat{{\bf x}}_t + \mu\left( r\hat{{\bf p}}_t - \partial_{\bf x}{\bf F}(t,\hat{\bf x}_t,\hat{\bf z}_t,\hat{\bf v}_t,\mathbb{E}[\hat{{\bf v}}_t])\right) + \boldsymbol{\xi}_t \right]dt - \text{diag}(\hat{\boldsymbol{\eta}}_t)d{\bf W}_t,  \\
            \hat{\bf x}_0 &= {\bf x}_0, \\
            \hat{{\bf p}}_T &= -\mu\partial_{\bf x}{\bf G}(\hat{{\bf x}}_T,\hat{{\bf z}}_T) + (1-\mu) \hat{{\bf x}}_T + \boldsymbol{\zeta}_T, 
      \end{aligned}\right.
    \end{equation}
where
    \begin{align*}
      \hat{{\bf z}}_t &= \mathbb{E}[\hat{{\bf x}}_t] , \\
        \hat{{\bf v}}_t &=  \text{Proj}_{I^H}\left[ \left(\partial_{\bf v} {\bf F}\right)^{-1}\left(-\left({\bf K}\hat{\bf p}_t + {\bf \Sigma} \hat{\boldsymbol{\eta}}_t\right) ;t, \hat{\bf x}_t,\hat{\bf z}_t,\mathbb{E}[\hat{\bf v}_t]\right) \right], \\
            \mathbb{E}[\hat{\bf v}_t] &= \mathbb{E}\left[\text{Proj}_{I^H}\left[ \left(\partial_{\bf v} {\bf F}\right)^{-1}\left(-\left({\bf K}\hat{\bf p}_t + {\bf \Sigma} \hat{\boldsymbol{\eta}}_t\right) ;t, \hat{\bf x}_t,\hat{\bf z}_t,\mathbb{E}[\hat{\bf v}_t]\right) \right]\right], 
    \end{align*}
$\boldsymbol{\phi},\boldsymbol{\xi} \in L^2_{\mathbb{F}^{[H]}}([0,T];\mathbb{R}^H)$, $\boldsymbol{\psi} \in L^2_{\mathbb{F}^{[H]}}([0,T];\mathbb{R}^{H}\times \mathbb{R}^H)$, and $\boldsymbol{\zeta}_T\in L^2(\Omega,\mathcal{F}_T,\mathbb{P})$. 

It is clear that  \eqref{eq:MFFBSDE:general:parametrized} admits a solution when $\mu=0$; see Lemma 2.5 of \cite{peng1999fully}. The following lemma establishes a contraction property such that, if the system admits a solution for some $\mu_0\in[0,1)$, then it also admits a solution for any $\mu \in [\mu_0,\mu_0+\delta]$ for some $\delta>0$ independent of $\mu_0$. Using this property recursively, we can extend the existence of a solution to $\mu=1$, thus proving the solvability of \eqref{eq:MFBSDE:general}.  

\begin{lemma}
\label{lem:exist:general}
   Assume that there exists a constant $\mu_0\in[0,1)$ such that, for any $\boldsymbol{\phi},\boldsymbol{\xi} \in L^2_{\mathbb{F}^{[H]}}([0,T];\mathbb{R}^H)$, $\boldsymbol{\psi} \in L^2_{\mathbb{F}^{[H]}}([0,T];\mathbb{R}^{H}\times \mathbb{R}^H)$, $\boldsymbol{\zeta}_T\in L^2(\Omega,\mathcal{F}_T,\mathbb{P})$, the MF-FBSDE \eqref{eq:MFFBSDE:general:parametrized} admits a solution. Then, under Assumptions \ref{ass:concave:monotonicity:h}, \ref{ass:contraction}, \ref{ass:fg:sep}  and \ref{ass:M}, there exists a $\delta_0  \in (0,1)$  which only depends on $\hide{\beta_1,}\hide{\beta_2}T$, and independent of $\mu_0$, such that for any $\mu\in [\mu_0, \mu_0 + \delta_0]$, the MF-FBSDE \eqref{eq:MFFBSDE:general:parametrized} admits a solution for any $\boldsymbol{\phi},\boldsymbol{\xi},\boldsymbol{\psi}$ and $\boldsymbol{\zeta}_T$. 

\end{lemma}

\begin{proof}
    The proof is relegated to Appendix \ref{sec:app:pf:lem:exist}.
\end{proof}

As an immediate consequence of Theorem \ref{thm:unique:general} and Lemma \ref{lem:exist:general}, we state the main result of this section. 

\begin{theorem}
\label{thm:exist:unique:MFFBSDE:general}
    Under Assumptions \ref{ass:concave:monotonicity:h}, \ref{ass:contraction}, \ref{ass:fg:sep}  and \ref{ass:M}, the MF-FBSDE \eqref{eq:MFBSDE:general} admits a unique solution for any $T>0$. 
\end{theorem}

\section{Quadratic Rewards}
\label{sec:lq}
In this section, we consider a particular class of reward functions that are quadratic in representative members' wealth and strategies, which can be interpreted as simultaneously maximizing wealth while penalizing deviations from a given target. This specification is analytically tractable and has been extensively applied in economics (\cite{HANSEN19807,hansen1995discounted,hansen2013recursive}), and in the actuarial context (\cite{NGWIRA2007283,HUANG2010208,DELONG2019196}). To be exact, for $h\in[H]$, we let
    \begin{equation}
    \label{eq:quadratic:f:g}
        \begin{aligned}
      f^h(t,x,{\bf z},v,\bar{v})&=  f^h(t,x,z^h,v,\bar{v}) := -\frac{Q^h_t}{2}\left(x-S^h_tz^h\right)^2 - \frac{P^h_t}{2}\left(v-R^h_t\bar{v}\right)^2,  \\
     g^h(x,{\bf z})&=   g^h(x,z^h) := \gamma^hx - \frac{Q^h_T}{2}\left(x-S^h_Tz^h\right)^2,\\
        \end{aligned}
    \end{equation}
where $\gamma^h>0$, and $Q^h_\cdot,P^h_\cdot,S^h_\cdot$ and $R^h_\cdot$ are bounded deterministic functions with $\inf_{t\in[0,T]} Q_t>0$ and $\inf_{t\in[0,T]} P^h_t>0$. In other words, each member aims to maximize her own wealth while taking into account the fluctuations from the average wealth and strategies of other members from the same class, which mirrors a mean-variance objective. To facilitate the subsequent analysis, we define the following $\mathbb{R}^H\times \mathbb{R}^H$-valued functions: 
    \begin{align}
    \label{eq:PQR:lq}
    &{\bf P}_t := \text{diag}\left((P^h_t)_{h=1}^H\right), \     {\bf Q}_t := \text{diag}\left((Q^h_t)_{h=1}^H\right), \notag \\
    &{\bf R}_t := \text{diag}\left((R^h_t)_{h=1}^H\right), \ {\bf S}_t := \text{diag}\left((S^h_t)_{h=1}^H\right),
    \end{align} 
    and a $\mathbb{R}^H$ column vector $\boldsymbol{\gamma}=(\gamma^1,\dots,\gamma^H)^\top$. 


\subsection{Equilibrium Solution}\label{sec:lq:assumption}
Before stating the equilibrium solution under the quadratic reward functions \eqref{eq:quadratic:f:g}, we introduce the following assumption.

\begin{assumption}
\label{ass:lq}
\hfill 
     \begin{enumerate}[label=(\alph*)]
            \item $\sup_{t\in[0,T]} |S^h_t|<1$ for all $h\in[H]$;
            \item $\sup_{t\in[0,T]} |R^h_t| < 1$ for all  $h\in[H]$;
            \item   $ \lambda_{\min}({\bf I} - {\bf M}) > 0 $.  
            \item $\inf_{t\in[0,T]}  \lambda_{\min}(({\bf I}-{\bf M}^\top){\bf Q}_t({\bf I}-{\bf S}_t)) >0$.

        \end{enumerate}
\end{assumption}
Under the quadratic rewards \eqref{eq:quadratic:f:g}, it is clear that Assumptions \ref{ass:concave:monotonicity:h}, \ref{ass:contraction}, and \ref{ass:fg:sep}  are fulfilled given  Assumption \ref{ass:lq}(a)-(b). In addition,  Assumption \ref{ass:M} is a consequence of (c) and (d) in Assumption \ref{ass:lq}. To see this, consider for any ${\bf x}^i, {\bf v}^i \in L^2_{\mathbb{F}^H}([0,T];\mathbb{R}^H)$, $i=1,2$, and any $t\in[0,T]$, 
    \begin{align}\label{ineq.x-MEx}
      \nonumber & \     \mathbb{E}\left[  \bigg\langle {\bf x}^1_t - {\bf x}^2_t - {\bf M}\mathbb{E}[{\bf x}^1_t-{\bf x}^2_t], \partial_{\bf x}{\bf F}(t,{\bf x}^1_t,\mathbb{E}[{\bf x}^1_t],{\bf v}^1_t,\mathbb{E}[{\bf v}^1_t])- \partial_{\bf x}{\bf F}(t,{\bf x}^2_t,\mathbb{E}[{\bf x}^2_t],{\bf v}^2_t,\mathbb{E}[{\bf v}^2_t])    \bigg\rangle \right] \\
      \nonumber =&\ \mathbb{E}\left[\left\langle \tilde{{\bf x}}_t - {\bf M}\tilde{{\bf z}}_t,  -{\bf Q}_t \left(\tilde{{\bf x}}_t - {\bf S}_t \tilde{{\bf z}}_t \right) \right\rangle \right] \\
      \nonumber =&\ -\mathbb{E}\left[\langle \tilde{{\bf x}}_t , {\bf Q}_t \tilde{{\bf x}}_t\rangle\right] + \left\langle\tilde{{\bf z}}_t, \left({\bf Q}_t{\bf S}_t + {\bf M}^\top{\bf Q}_t({\bf I}-{\bf S}_t)\right) \tilde{{\bf z}}_t  \right\rangle \\
      \nonumber 
      \leq&\  -  \min\left\{\lambda_{\min}({\bf Q}_t), \lambda_{\min}(({\bf I}-{\bf M}^\top){\bf Q}_t({\bf I}-{\bf S}_t))  \right\} \mathbb{E}\left[|\tilde{{\bf x}}_t|^2\right], 
    \end{align}
where $\tilde{{\bf x}}_t := {\bf x}^1_t-{\bf x}^2_t$ and $\tilde{{\bf z}}_t := {\bf z}^1_t-{\bf z}^2_t$, and the last line follows from Lemma \ref{lem:EX:Z:inequality}. By Assumption \ref{ass:lq}(d), we can take $$\alpha_{\bf M} := \inf_{t\in[0,T]}\left\{\lambda_{\min}({\bf Q}_t), \lambda_{\min}(({\bf I}-{\bf M}^\top){\bf Q}_t({\bf I}-{\bf S}_t))\right\}>0.$$ 
Likewise, one can show that the same constant $\alpha_{{\bf M}}^g = \alpha_{{\bf M}}$ can be used to satisfy \eqref{eq:concavity:U}, thereby fulfilling Assumption \ref{ass:M}.  Although Assumption \ref{ass:lq}(c) and (d) are not equivalent, they both share the same key feature: the matrix ${\bf M}$ has only a moderate impact.    

By   Theorems  \ref{thm:sol:p2} and \ref{thm:exist:unique:MFFBSDE:general}, the solution of Problems \ref{p:mf}-\ref{p:fixed:point} under the quadratic rewards \eqref{eq:quadratic:f:g} is immediately characterized by the following. 

\begin{corollary}
\label{cor:lq:optimal}
 Under quadratic reward \eqref{eq:quadratic:f:g}, the MF-FBSDE \eqref{eq:MFBSDE:general} can be written as 
              \begin{equation}
         \label{eq:FBSDE:lq:insurance:constrained}
         \left\{
            \begin{aligned}
            d{\bf x}_t &= \left(r{\bf x}_t + {\bf l} - {\bf K}{\bf v}_t + {\bf \Pi} \bar{{\bf v}}_t \right)dt + {\bf \Sigma}\left({\bf I} - \text{diag}({\bf v}_t)\right)d{\bf W}_t, \\
            -d{\bf p}_t &= \left(r{\bf p}_t + {\bf Q}_t({\bf x}_t-{\bf S}_t{\bf z}_t)\right)dt  - \text{diag}(\boldsymbol{\eta}_t) d{\bf W}_t, \\
            {\bf x}_0 &= (\xi^1,\dots,\xi^H)^\top, \\
            {\bf p}_T &= {\bf Q}_T({\bf x}_T-{\bf S}_T{\bf z}_T) - \boldsymbol{\gamma}, 
        \end{aligned}\right.
        \end{equation}
    where 
        \begin{align*}
           {\bf z}_t = \mathbb{E}[{\bf x}_t],\      {\bf v}_t =  \text{Proj}_{I^H}\left[ {\bf P}^{-1}_t  \left({\bf K}{\bf p}_t + {\bf \Sigma} \boldsymbol{\eta}_t\right) + {\bf R}_t \bar{{\bf v}}_t\right],\ 
            \bar{{\bf v}}_t = \mathbb{E}\left[ {\bf v}_t \right], 
        \end{align*}
       and ${\bf P}^{-1}_t$ is the inverse matrix of ${\bf P}_t$. In addition, Equation \eqref{eq:FBSDE:lq:insurance:constrained} is uniquely solvable under Assumption  \ref{ass:lq}. 
    \end{corollary}

\subsection{Equilibrium without Insurance Constraints}
\label{sec:lq:riccati}
When no insurance constraint is imposed, i.e., $I=\mathbb{R}$, the MF-FBSDE \eqref{eq:FBSDE:lq:insurance:constrained} admits a closed form solution, which can be represented in terms of the solutions of certain Riccati equations. In this case, the mean field term $(\bar{v}^h_t)_{t\in[0,T]}$ and the optimal strategy $(v^h_t)_{t\in[0,T]}$ can be reduced to the following: 
    \begin{equation}
    \label{eq:v:ansatz}
        v^h_t = \frac{\kappa^hp^h_t+\sigma^h\eta^h_t}{P^h_t} + R^h_t \bar{v}^h_t \quad \text{and} \quad \bar{v}^h_t = \frac{\kappa^h\bar{p}^h_t + \sigma^h\mathbb{E}[\eta^h_t]}{P^h_t(1-R^h_t)}.
    \end{equation}
    
    Let ${\bf \Gamma}_\cdot=\text{diag}((\Gamma^h_\cdot)_{h=1}^H):[0,T]\to \mathbb{R}^H$ be the solution of the following Riccati equation:
    \begin{equation}
    \label{eq:Gamma:mf}
    \begin{dcases}
        \frac{d\Gamma^h_t}{dt}  -\frac{ (\kappa^h)^2(\Gamma^h_t)^2}{P^h_t +(\sigma^h)^2\Gamma^h_t} +2r\Gamma^h_t + Q^h_t = 0,\\
        \Gamma^h_T = Q^h_T. 
    \end{dcases}
    \end{equation}
Equation \eqref{eq:Gamma:mf} enables us to characterize the system \eqref{eq:FBSDE:lq:insurance:constrained} and the optimal strategy by the deterministic functions $\bar{{\bf p}}=(\bar{p}^h)_{h\in[H]}$ and ${\bf z}=(z^h)_{h\in[H]}$ in an affine relationship. Indeed, using the \textit{ansatz} and It\^o's lemma, it is straightforward to verify that ${\bf p}_t = {\bf \Gamma}_t({\bf x}_t-{\bf z}_t)+ \bar{\bf p}_t$, where $(\bar{\bf p},{\bf z})$ satisfies the following FBODE: 
\begin{equation}
\label{eq:FBODE:unconstraint}
    \begin{dcases}
    d{\bf z}_t = \left(r{\bf z}_t + {\bf l} +\left({\bf \Pi} -{\bf K}\right)\left({\bf A}_t\bar{{\bf p}}_t +{\bf b}_t\right)   \right)dt, \\
    -d\bar{{\bf p}}_t = \left(r\bar{{\bf p}}_t + {\bf Q}_t({\bf I}-{\bf S}_t){\bf z}_t \right)dt, \\
    {\bf z}_0 = (\mathbb{E}[\xi^1],\dots,\mathbb{E}[\xi^H]),\\
    \bar{{\bf p}}_T = {\bf Q}_T({\bf I}-{\bf S}_T){\bf z}_T - \boldsymbol{\gamma},
    \end{dcases}
\end{equation}
and
     \begin{align*}
     & {\bf A}_t = {\bf K}\left({\bf \Sigma}^2\boldsymbol{\Gamma}_t + {\bf P}_t({\bf I}-{\bf R}_t) \right)^{-1}, \ {\bf b}_t = {\bf \Sigma}^2 \left({\bf \Sigma}^2\boldsymbol{\Gamma}_t + {\bf P}_t({\bf I}-{\bf R}_t) \right)^{-1}\text{vec}(\boldsymbol{\Gamma}_t), \\
       & {\bf C}_t = {\bf K}\left({\bf \Sigma}^2\boldsymbol{\Gamma}_t + {\bf P}_t \right)^{-1}, \ {\bf D}_t =  {\bf P}_t{\bf R}_t\left({\bf \Sigma}^2\boldsymbol{\Gamma}_t + {\bf P}_t  \right)^{-1}. 
    \end{align*}

The discussion of the well-posedness of \eqref{eq:FBODE:unconstraint} is relegated to Appendix \ref{sec:well-posed:unconstraint}. Indeed, the FBODE \eqref{eq:FBODE:unconstraint} can further be reduced by considering the following \textit{ansatz}:
    \begin{equation}
        \label{eq:fbode:ansatz}
        \bar{{\bf p}}_t = {\bf \Xi}_t {\bf z}_t + \boldsymbol{\zeta}_t,
    \end{equation}
where ${\bf \Xi}_\cdot : [0,T] \to \mathbb{R}^{H}\times \mathbb{R}^H$ and $\boldsymbol{\zeta}_\cdot : [0,T] \to \mathbb{R}^H$  satisfy the following equation:
    \begin{equation}
    \label{eq:FBODE:riccati}
        \begin{dcases}
            \frac{d{\bf \Xi}_t}{dt} + 2r{\bf \Xi}_t + {\bf \Xi}_t\left({\bf \Pi} -{\bf K}\right){\bf A}_t{\bf \Xi}_t + {\bf Q}_t({\bf I}-{\bf S}_t) = 0, \\
            \frac{d\boldsymbol{\zeta}_t}{dt} + \left(r{\bf I} + {\bf \Xi}_t\left({\bf \Pi} -{\bf K}\right){\bf A}_t\right)\boldsymbol{\zeta}_t + {\bf \Xi}_t\left({\bf l} + \left({\bf \Pi} -{\bf K}\right){\bf b}_t\right)=0, \\
            {\bf \Xi}_T = {\bf Q}_T({\bf I}-{\bf S}_T), \\
            \boldsymbol{\zeta}_T = -\boldsymbol{\gamma}. 
        \end{dcases}
    \end{equation}
Hence, the well-posedness of \eqref{eq:FBODE:unconstraint} can be implied by that of \eqref{eq:FBODE:riccati}, and the complete solution of the MFG can be characterized by \eqref{eq:Gamma:mf} and \eqref{eq:FBODE:riccati}. The following summarizes the findings in this section. 
\begin{theorem}
\label{thm:riccati}
    Suppose that the system \eqref{eq:FBODE:riccati} admits a unique solution. Then, the optimal insurance strategy $({\bf v}_t)_{t\in[0,T]}$ is given by 
        \begin{equation*}
          {\bf v}_t = {\bf C}_t\left({\bf \Gamma}_t({\bf x}_t-{\bf z}_t) + \bar{{\bf p}}_t \right) + {\bf D}_t\bar{{\bf v}}_t + {\bf e}_t,
        \end{equation*}
    where ${\bf e}_t = {\bf \Sigma}^2\left({\bf \Sigma}^2\boldsymbol{\Gamma}_t + {\bf P}_t  \right)^{-1} \boldsymbol{\Gamma}_t$,   $\bar{{\bf v}}_t = {\bf A}_t\bar{{\bf p}}_t + {\bf b}_t$, $\bar{\bf p}_t$ is given by \eqref{eq:fbode:ansatz}, and $({\bf x}_t)_{t\in[0,T]}$, $({\bf z}_t)_{t\in[0,T]}$ are the solution of the following SDE and ODE, respectively: 
        \begin{align*}
       & \left\{ \begin{aligned}
            d{\bf x}_t &= \left(r{\bf x}_t + {\bf l} - {\bf K}{\bf v}_t +  {\bf \Pi}\bar{{\bf v}}_t   \right)dt +{\bf \Sigma}({\bf I}-\textup{diag}({\bf v}_t))  d{\bf W}_t,\\
              {\bf x}_0 &= (\xi^h)_{h=1}^H, 
              \end{aligned} \right.\\
           &   \left\{\begin{aligned}
            d{\bf z}_t &=  \left(r{\bf z}_t + {\bf l} + \left({\bf \Pi} -{\bf K}\right)\bar{{\bf v}}_t \right)dt, \\
             {\bf z}_0 &= (\mathbb{E}[\xi^h])_{h=1}^H.
        \end{aligned}\right.
        \end{align*}
\end{theorem}
\begin{proof}
    The proof is relegated to Appendix \ref{sec:app:pf:thm:riccati}. 
\end{proof}

\section{Numerical Experiments}
\label{sec:mf:numerical}
In this section, we perform comprehensive numerical experiments to examine the equilibrium insurance strategies and the resulting wealth of representative members. All computations are performed using an NVIDIA RTX A5500 GPU.\footnote{The implementation code is publicly available on GitHub at: \href{https://github.com/WenyuanLi-HKU-SAAS/Mean-Field-Analysis-of-Mutual-Insurance-Market.git}{https://github.com/WenyuanLi-HKU-SAAS/Mean-Field-Analysis-of-Mutual-Insurance-Market.git.}} Supplementary tables in this section (Tables \ref{app:sec:tables}-\ref{tab:v:values}) are provided in Appendix \ref{sec:app:Supp:tables}.

In the first part of the experiment, we consider an MIC with two membership classes ($H=2$), and the members exhibit quadratic rewards as described in \eqref{eq:quadratic:f:g}. The following parameters are chosen as the baseline scenario:
    \begin{align}
        & r = 0.03,\   \tilde{l}^1 - \mu^1 = \tilde{l}^2 - \mu^2 = 0.02,\ P^1_\cdot  = P^2_\cdot = Q^1_\cdot = Q^2_\cdot \equiv 1, \ R^1_\cdot = R^2_\cdot \equiv 0.1,\notag\\ 
        &  S^1_\cdot  = S^2_\cdot \equiv 0.6,\  \kappa^1 = \kappa^2 = 0.5,\   e^1 = e^2 = 0.01,\ d^1_e=d^2_e=0.1e^1=0.001,  \notag\\ 
        & \omega^1=\omega^2=0.5,\ \sigma^1=\sigma^2 = 0.3,\ \gamma^1=\gamma^2 = 1, \ \xi^1 = \xi^2=2, \ d = 0.05,\  T = 1. \label{LQ_baseline_scenario}
    \end{align}
 The sharing proportion  $\pi^h$ of the surplus, and the fixed management fee rate $d_e^h$, $h=1,2$, are taken to be proportional to the membership fee as follows: 
    \begin{equation}
    \label{eq:prop:e:pi}
        \pi^h := \frac{e^h}{\sum_{k=1}^2 e^k \omega^k }, \ d_e^h = 0.1e^h.
    \end{equation}
Under the baseline scenario, we have $\pi^1=1=\pi^2$ and $d_e^1=d_e^2 = 0.001$. On the other hand, the net income rates are
\begin{align*}
     l^1 = \tilde{l}^1 - \mu^1 -e^1 + \pi^1 \sum_{k=1}^2 \omega^k(e^k-d^k_e)=0.019 = l^2. 
\end{align*}%
Furthermore, for $h=1,2$, we set the cases listed in Table \ref{tab:parameter} to study the effect of the volatility $\sigma^h$, the safety loading $\kappa^h$, the risk aversion $\gamma^h$,  the membership fee $e^h$, the net income rate before sharing $\tilde{l}^h - \mu^h$, and the relative class size $\omega^h$.  

\begin{table}[!h]
    \centering    
  \caption{Parameters used across different test cases } 
    \smallskip
    \begin{tabular}{c c} \hline 
           Case & Parameters  \\ \hline 
       1(a) & $\sigma^1=0.1$, $\sigma^2 = 0.3$ \\
       1(b)  & $ \omega^1=0.8$, $\omega^2 = 0.2$; $\sigma^1 = 0.1$, $\sigma^2 = 0.3$\\
       1(c) & $\omega^1=0.2$, $\omega^2 = 0.8$; $\sigma^1 = 0.1$, $\sigma^2 = 0.3$\\
        \hline
        2(a) & $\gamma^1=1.0$, $\gamma^2 = 1.6$;\\ 
        2(b) & $\omega^1=0.8$, $\omega^2 = 0.2$; $\gamma^1=1.0$, $\gamma^2 = 1.6$\\ 
       2(c) & $\omega^1=0.2$, $\omega^2 = 0.8$; $\gamma^1=1.0$, $\gamma^2 = 1.6$\\ 
        \hline
       3(a) & $\kappa^1 = 0.1$, $\kappa^2=0.5$, $\gamma^1=\gamma^2 = 1.6$\\
       3(b) & $\kappa^1 = 0.1$, $\kappa^2=0.5$, $\gamma^1=\gamma^2 = 1.0$\\
        \hline
      4(a) & $\tilde{l}^1 - \mu^1=0.02$, $\tilde{l}^2 - \mu^2=0.1$\\
       4(b) & $e^1=0.1$, $e^2 = 0.01$; $\tilde{l}^1 - \mu^1=0.02$, $\tilde{l}^2 - \mu^2=0.1$\\
      4(c) & $e^1=0.01$, $e^2 = 0.1$; $\tilde{l}^1 - \mu^1=0.02$, $\tilde{l}^2 - \mu^2=0.1$\\
        \hline

    \end{tabular}
    \label{tab:parameter}
\end{table}


For all cases, we consider two scenarios: with and without an insurance constraint. In the former case, we impose an insurance constraint $I=[0,1]$, i.e., members are prohibited from taking a ``short position", nor transferring an amount more than their actual losses to the MIC. We remark that all the combinations of parameters above satisfy Assumption \ref{ass:lq}. 

To obtain the optimal insurance strategies and the equilibrium wealth under insurance constraint, we solve the FBSDE \eqref{eq:FBSDE:lq:insurance:constrained} numerically using a deep BSDE approach, and the details are elaborated in the next subsection.

\subsection{Neural Network Architectures}
\label{sec:NN}
The deep BSDE approach presented in this subsection is an adaptation of the methods introduced in \cite{germain2022numerical, carmona2022convergence, han2024learning}. The central idea is to approach the backward equation $({\bf p}_t)_{t\in[0,T]}$ in a forward manner, treating the initial value ${\bf p}_0$ as a trainable component by a neural network. The system is then simulated forward in time, solving both the forward equation for $({\bf x}_t)_{t\in[0,T]}$ and the backward equation for $({\bf p}_t)_{t\in[0,T]}$ using Monte Carlo methods. The loss function for the neural network is defined as the deviation between the simulated value ${\bf p}_T$ at the terminal time, and the prescribed terminal condition of the original backward equation.  To accommodate the extended mean field game framework, a penalty term is introduced for the mean field term  $(\bar{{\bf v}}_t)_{t\in[0,T]}$ to ensure the additional fixed point condition is satisfied. The complete architecture is described as follows.

We begin by building six neural networks for $\bar{v}^1,\bar{v}^2,\eta^1,\eta^2,p^1_0,p^2_0$: for $t\in[0,T]$,  
\begin{align*} 
    &\bar{v}^1_t = \mathcal{NN}^{\phi_1}_1(t), ~\bar{v}^2_t = \mathcal{NN}^{\phi_2}_2(t), \\
    &\eta^1_t = \mathcal{NN}^{\phi_3}_3(t,x^1_t,z^1_t,p^1_t), 
    ~\eta^2_t = \mathcal{NN}^{\phi_4}_4(t,x^2_t,z^2_t,p^2_t),\\
    &~p^1_0 = \mathcal{NN}^{\phi_5}_5(x^1_0), ~p^2_0 = \mathcal{NN}^{\phi_6}_6(x^2_0),
\end{align*}
where $\phi_i$ are the weights and biases of neural network $\mathcal{NN}_i$. The optimal strategies are then computed by 
    \begin{equation*}
        v^h_t = \text{Proj}_I\left[(f^h_v)^{-1}(-(\kappa^h p^h_{t} + \sigma^h\eta^h_{t});t,\bar{v}^h_{t})  \right], \ h=1,2.
    \end{equation*}

\begin{figure}[htbp]
	\centering
	\begin{tikzpicture}[x=1.5cm, y=1.3cm, >=stealth]
		
		\tikzset{%
			every neuron/.style={
				circle,
				draw,
				minimum size=1.0cm
			},
			neuron missing/.style={
				draw=none, 
				scale=3.5,
				text height=0.222cm,
				execute at begin node=\color{black}$\vdots$
			},
		}
		\foreach \m [count=\y] in {1,2,3,4}
		\node [every neuron/.try, neuron \m/.try, fill = green!95!black!10] (input-\m) at (0,4.2-\y){};
		
		\foreach \m [count=\y] in {1,missing,missing,2}
		\node [every neuron/.try, neuron \m/.try] (hidden1-\m) at (2,5.0-\y*1.3) {};
		
		\node [every neuron/.try, neuron 1/.try, fill = blue!95!black!10] (hidden1-1) at (2,5.0-1*1.3) {};
		\node [every neuron/.try, neuron 2/.try, fill = blue!95!black!10] (hidden1-2) at (2,5.0-4*1.3) {};
		
		\foreach \m [count=\y] in {1,missing,missing,2}
		\node [every neuron/.try, neuron \m/.try ] (hidden2-\m) at (4,5.0-\y*1.3) {};
		
		\node [every neuron/.try, neuron 1/.try, fill = blue!95!black!10] (hidden2-1) at (4,5.0-1*1.3) {};
		\node [every neuron/.try, neuron 2/.try, fill = blue!95!black!10] (hidden2-2) at (4,5.0-4*1.3) {};
		
		\foreach \m [count=\y] in {1}
		\node [every neuron/.try, neuron \m/.try, fill = red!95!black!10] (output-\m) at (6,4.2-\y*2.4) {};
		
		\node at (input-1) {$t$};
		\node at (input-2) {$x^1_t$};
		\node at (input-3) {$z^1_t$};
		\node at (input-4) {$p^1_t$};
		
		\foreach \l [count=\i] in {1}
		\node at (hidden1-\i) {$H^{(1)}_{\l}$};
		\node at (hidden1-2) {$H^{(1)}_{32}$};
		
		\foreach \l [count=\i] in {1}
		\node at (hidden2-\i) {$H^{(2)}_{\l}$};
		\node at (hidden2-2) {$H^{(2)}_{32}$};
		
		\node at (output-1) {$\eta^1_t$};
		
		\foreach \i in {1,...,4}
		\foreach \j in {1,...,2}
		\draw [->] (input-\i) -- (hidden1-\j);
		
		\foreach \i in {1,...,2}
		\foreach \j in {1,...,2}
		\draw [->] (hidden1-\i) -- (hidden2-\j);
		
		\foreach \i in {1,...,2}
		\foreach \j in {1}
		\draw [->] (hidden2-\i) -- (output-\j);
		
		\foreach \l [count=\x from 0] in {Input, Hidden 1, Hidden 2, Ouput}
		\node [align=center, above] at (\x*2,4.2) {\l};
		
	\end{tikzpicture}
	\caption{Neural network for $\eta^1_t$ with a ``$4-32-32-1$'' structure. 
	}\label{eta1_NN3}
\end{figure}
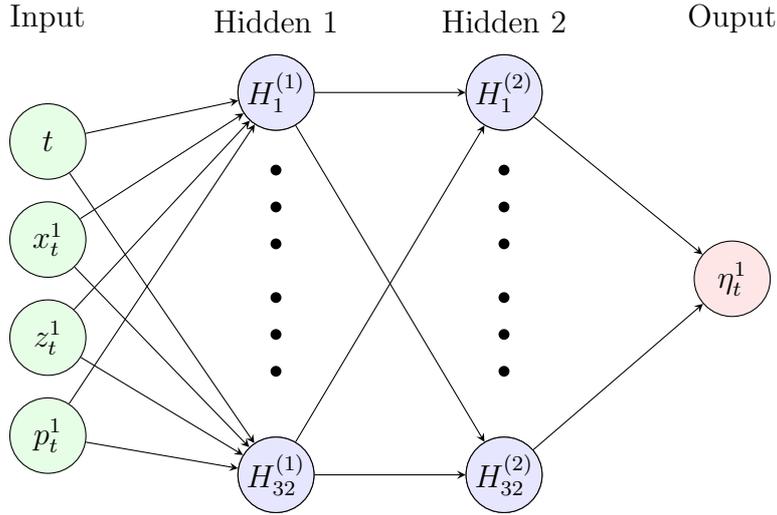

Each neural network $\mathcal{NN}_i^{\phi_i}$ above is chosen to have two hidden layers, and each layer consists of 32 hidden nodes. The  \textit{Rectified Linear Unit} (ReLU) function and the identity function are chosen as the activation function in the hidden layer and the output layer, respectively.   Figure \ref{eta1_NN3} shows the structure of the neural network for $(\eta^1_t)_{t\in[0,T]}$. For cases with constraint ${\bf v}_t \in [0,1]^H$,   $\bar{v}^h_t$ are defined by projecting the output of the neural network  $\mathcal{NN}_h$ to $I=[0,1]$:
\begin{align*}
    &  \bar{v}^1_t = \text{Proj}_{[0,1]}[\mathcal{NN}^{\phi_1}_1(t)], ~\bar{v}^2_t = \text{Proj}_{[0,1]}[\mathcal{NN}^{\phi_2}_2(t)]. 
\end{align*}

To simulate the SDEs using the Euler-Maruyama method, we discretize $[0, T]$ into a partition $\mathcal{T} = \{t_i: i\Delta t, i=0,1, ..., M\}$, where $\Delta t = T/M$.  Then, we can formulate the loss function and the simulation scheme as follows:
\begin{align}
    &\min \limits_{\phi_1,\phi_2,\phi_3,\phi_4,\phi_5,\phi_6} \sum \limits_{h=1}^2 \mathbb{E}\left[(p^h_T + \gamma^h - Q^h_T(x^h_T - S^h_T z^h_T))^2\right] + \frac{\lambda}{M} \sum \limits_{i=0}^{M-1} \sum \limits_{h=1}^2 (\mathbb{E}[v^h_{t_i}]-\bar{v}^h_{t_i})^2,\label{NN_MF_BSDE}\\
    &\text{s.t.}~
    x^h_{t_{i+1}} = x^h_{t_i} + \left(rx^h_{t_i}+l^h-\kappa^hv^h_{t_i}+\pi^h \sum \limits_{j=1}^2 \omega^j  (\kappa^j-d^j)\bar{v}^j_{t_i}\right)\Delta t + \sigma^h(1-v^h_{t_i})\Delta W^h_{t_i},\notag\\
    &~~~~~p^h_{t_{i+1}} = p^h_{t_i} - [rp^h_{t_i} - f^{h,X}_x(t,x^h_{t_i},z^h_{t_i}) ] \Delta t + \eta^h_{t_i} \Delta W^h_{t_i}, \notag\\
    &~~~~~z^h_{t_{i+1}} = z^h_{t_i} + \left(rz^h_{t_i}+l^h-\kappa^h\bar{v}^h_{t_i}+\pi^h \sum \limits_{j=1}^2 \omega^j  (\kappa^j-d^j)\bar{v}^j_{t_i}\right)\Delta t,\notag\\
    &~~~~~x^h_0 = \xi^h, ~z^h_0 = \mathbb{E}[\xi^h], ~p^h_T = -\gamma^h + Q^h_T(x^h_T-S^h_Tz^h_T),\notag\\
    &~~~~~\bar{v}^1_t = \text{Proj}_I[\mathcal{NN}^{\phi_1}_1(t_i)], ~\bar{v}^2_t = \text{Proj}_I[\mathcal{NN}^{\phi_2}_2(t_i)],\notag\\
    &~~~~~v^1_{t_i} = \text{Proj}_I\left[ (f^1_v)^{-1}(-(\kappa^1 p^1_{t_i} + \sigma^1\eta^1_{t_i});t,\bar{v}^1_{t_i})  \right], v^2_{t_i} = \text{Proj}_I\left[(f^2_v)^{-1}(-(\kappa^2 p^2_{t_i} + \sigma^2\eta^2_{t_i});t,\bar{v}^2_{t_i}) \right],\notag\\
    &~~~~~\eta^1_{t_i} = \mathcal{NN}^{\phi_3}_3(t_i,x^1_{t_i},z^1_{t_i},p^1_{t_i}), 
    ~\eta^2_{t_i} = \mathcal{NN}^{\phi_4}_4(t,x^2_{t_i},z^2_{t_i},p^2_{t_i}),\notag\\
    &~~~~~p^1_0 = \mathcal{NN}^{\phi_5}_5(x^1_0), ~p^2_0 = \mathcal{NN}^{\phi_6}_6(x^2_0),\notag
\end{align}
where $\Delta W^h_{t_i} = W^h_{t_{i+1}}-W^h_{t_i}$, $\lambda>0$ is the penalty parameter, and the expectations are computed by the average of the simulated paths. In other words, the loss function is the sum of expected squared loss of the terminal condition of the backward equations, and a penalty term for the difference between $\mathbb{E}[v^h_t]$ and $\bar{v}^h_t$.

In this study, we perform Monte-Carlo simulations for the system \eqref{NN_MF_BSDE} with $10,000$ sample paths and $M=100$ time steps. For each case, we train the neural network $1,000$ times. An Adam optimizer is applied to minimize the objective \eqref{NN_MF_BSDE}. The learning rate is set as $5 \times 10^{-4}$. To demonstrate the accuracy of the algorithm, under quadratic rewards, we use the non-constrained case as a benchmark and compute the relative error between the neural network approach and the ordinary differential equation (ODE) benchmark (Theorem \ref{thm:riccati}), which is defined as
\begin{align*}
    \frac{1}{4M} \sum \limits_{i=0}^{M-1} \sum \limits_{h=1}^2 \left| \frac{\mathcal{NN}^{\phi_h}_h(t_i)-\bar{v}^{h,ODE}_{t_i}}{\max \limits_{j\in\{0,\dots,M-1\}}\left| \bar{v}^{h,ODE}_{t_j}\right|  }\right| + \frac{1}{4M} \sum \limits_{i=1}^{M} \sum \limits_{h=1}^2 \left|\frac{z^{h,NN}_{t_i}-z^{h,ODE}_{t_i}}{\max \limits_{j\in\{0,\dots,M-1\}} \left|z^{h,ODE}_{t_j}\right| }\right|,
\end{align*}
where the superscripts ``ODE'' and ``NN"  indicate that the values are generated by the ODE benchmark from Theorem \ref{thm:riccati}, and by the neural network approach, respectively. The factor of 4 in the error definition accounts for averaging over the four functions ($\bar{v}^1_t, \bar{v}^2_t, z^1_t, z^2_t$). Table \ref{app:sec:tables} in Appendix \ref{sec:app:Supp:tables} presents the relative errors under different choices of the penalty coefficient $\lambda$.  Based on the result, we choose $\lambda = 10$ for Cases 1, 4(b) and 4(c), and $\lambda = 1$ for other cases to minimize the training errors.  In practice, we recommend choosing $\lambda$ from 1 to 10 to obtain the smallest computation errors.

\begin{figure}[htbp]
    \centering
	\includegraphics[width=3.0in,height=2.7in]{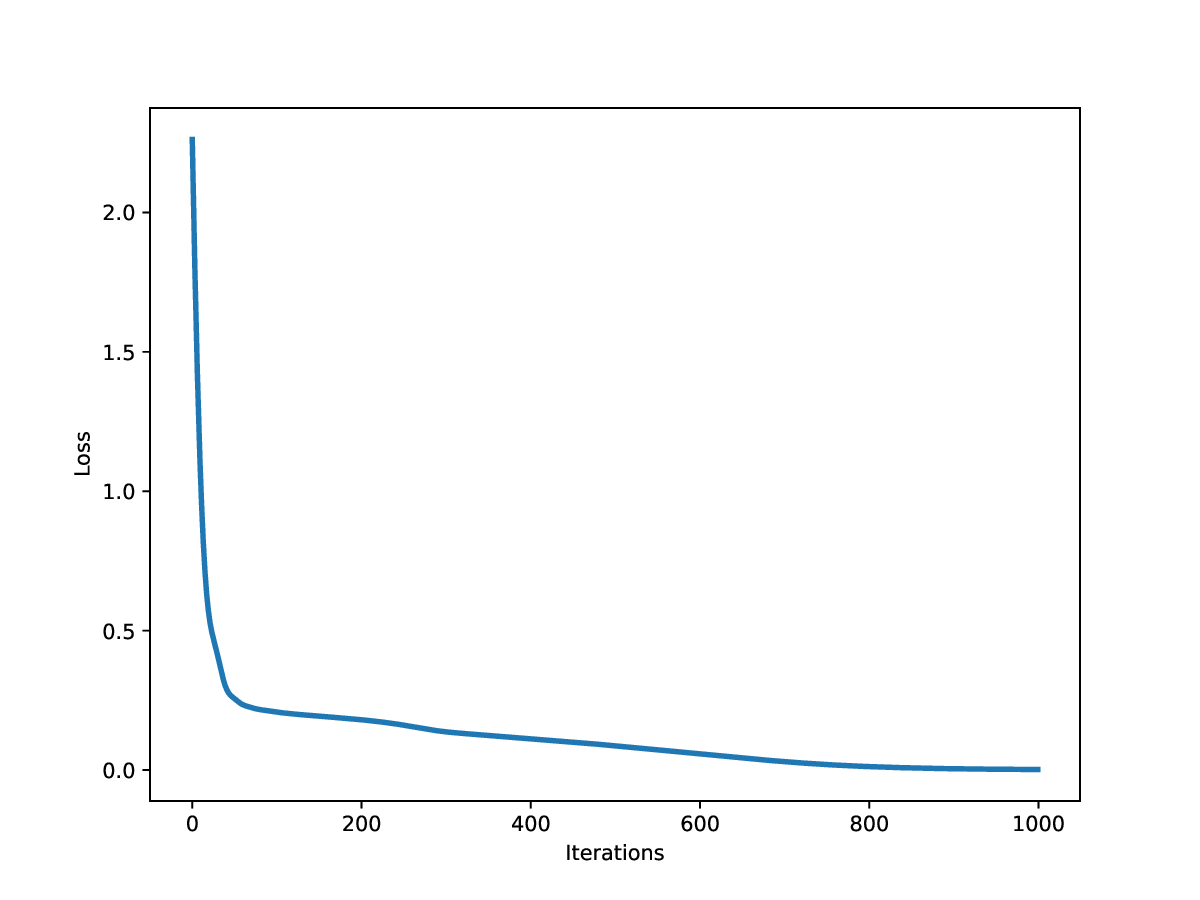}
    \caption{Loss curve for Case 1(a) with insurance constraints.}\label{fig_loss_curve}
\end{figure}

Figure \ref{fig_loss_curve} shows the loss curve for Case 1(a) with insurance constraint, illustrating that the loss function \eqref{NN_MF_BSDE} decays rapidly to zero with the number of training iterations. The numerical values of the loss functions for all cases considered are provided in  Tables \ref{loss_penalty_table1} (without constraint) and \ref{loss_penalty_table2} (with constraint) in Appendix \ref{sec:app:Supp:tables}. From the tables, we observe that both components of the training error, corresponding to the two summands in \eqref{NN_MF_BSDE}, are small, on the order of $10^{-3}$. This demonstrates the accuracy of the proposed algorithm in satisfying the BSDE’s terminal condition and approximating the mean field term $\bar{v}^h$.   

\subsection{Equilibrium Wealth and Strategies}
Figures \ref{fig_case1:figures}-\ref{fig_case4c:figures} present the results for Cases 1-4, respectively. The equilibrium strategies $\bar{{\bf v}}_t=(\bar{v}^1_t,\bar{v}^2_t)$, and the equilibrium wealth ${\bf z}_t=(z^1_t,z^2_t)$ are displayed in the left and right panels, respectively. In each figure, we  distinguish the curves without constraint by solid line, and those with constraint by dashed line. The curve for Class 1 and Class 2 are plotted in blue and yellow respectively.  
 Table \ref{tab:v:values} supplements the figures by providing the numerical values of the equilibrium strategies of all cases for $t=0$ and near the end of the planning horizon.     


\subsubsection{The impact of $\sigma^h$}
\begin{figure}[!h]
    \centering
    \subfigure[$\bar{v}^h$ for Case 1(a)]
    {
        \includegraphics[width=3.0in,height=2.7in]{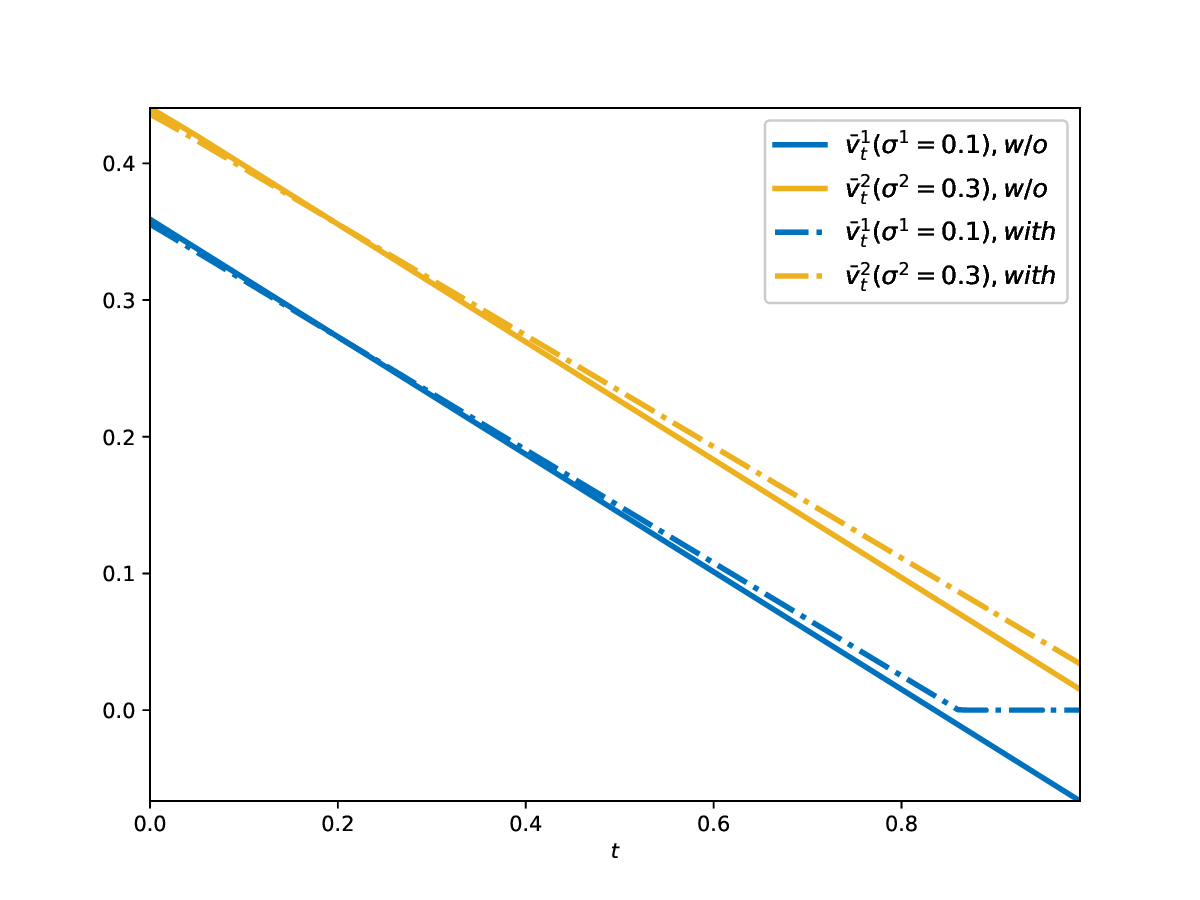}
        \label{fig_case1:figure1}
    }
    \hspace{-0.3in}
    \subfigure[$z^h$ for Case 1(a)]
    {
	\includegraphics[width=3.0in,height=2.7in]{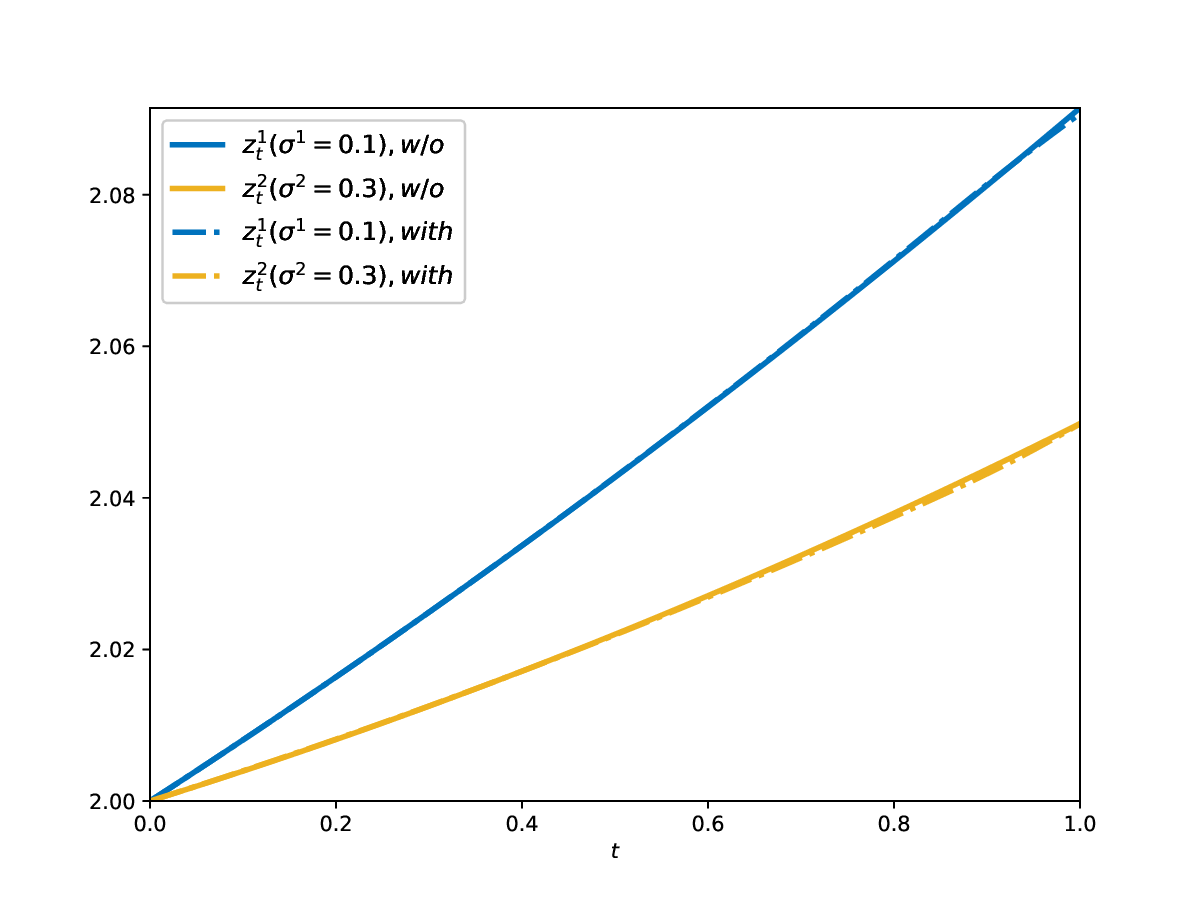}
        \label{fig_case1:figure2}
    }\\
    \caption{The equilibrium insurance strategies and wealth for representative members under Cases 1(a).}
    \label{fig_case1:figures}
\end{figure}

Figure \ref{fig_case1:figures} depicts the effect of the volatility of the loss process on the equilibrium strategies.  Figure \ref{fig_case1:figure1}  shows that the equilibrium strategy increases with volatility, with the representative member from Class 2 ($\sigma^2=30\%$) purchasing more insurance than her counterpart in Class 1 ($\sigma^1=10\%$). Intuitively, when there is greater uncertainty about the severity of the loss, members tend to purchase more insurance to transfer the uncertainty to the insurance company. Consequently, with a higher insurance purchase and, therefore, higher premium payments, the equilibrium wealth of the representative member in Class 2 tends to be smaller than that of one in Class 1; see Figure \ref{fig_case1:figure2}.

Examining the effect of the insurance constraint, we see that the constraint becomes binding for Class 1 near the end of the planning horizon. This restriction truncates the insurance strategy, leading members from Class 1 to ultimately forgo purchasing insurance. On the other hand, the constraint remains non-binding for Class 2, so their strategies are largely unchanged by the constraint. Nevertheless, small deviations do appear, stemming from the indirect influence of Class 1’s binding constraint through the sharing mechanism; see also Table \ref{tab:v:values}. Owing to the small deviations of the equilibrium strategies under the two scenarios, the effect of the insurance constraint on the equilibrium wealth is relatively small.

The impact of relative class sizes on the equilibrium strategies is illustrated in Table \ref{tab:v:values}. While the changes are not dramatic, we observe that members from both classes tend to reduce their insurance positions when the proportion of more risky members is smaller (Case 1(b), $\omega^1=0.8, \sigma^1=10\%$, $\omega^2 = 0.2, \sigma^2=30\%$), and increase their positions when the proportion of more risky members is higher (Case 1(c), $\omega^1=0.2, \sigma^1=10\%$, $\omega^2 = 0.8, \sigma^2=30\%$). Compared to Case 1(a),  when insurance constraint is imposed, the initial equilibrium strategy $\bar{v}^1_0$ for Class 1 has been reduced by {\color{black}$2.20\%$} in Case 1(b), and increased by {\color{black}$1.94\%$} in Case 1(c). This can be explained by changes in the aggregate risk of the mutual as the composition of member riskiness varies. For instance, in Case 1(c), the greater presence of high-risk members incentivizes all members to take on larger insurance positions.



\subsubsection{The impact of $\gamma^h$}
\begin{figure}[!h]
    \centering
    \subfigure[$\bar{v}^h$ for Case 2(a)]
    {
        \includegraphics[width=3.0in,height=2.7in]{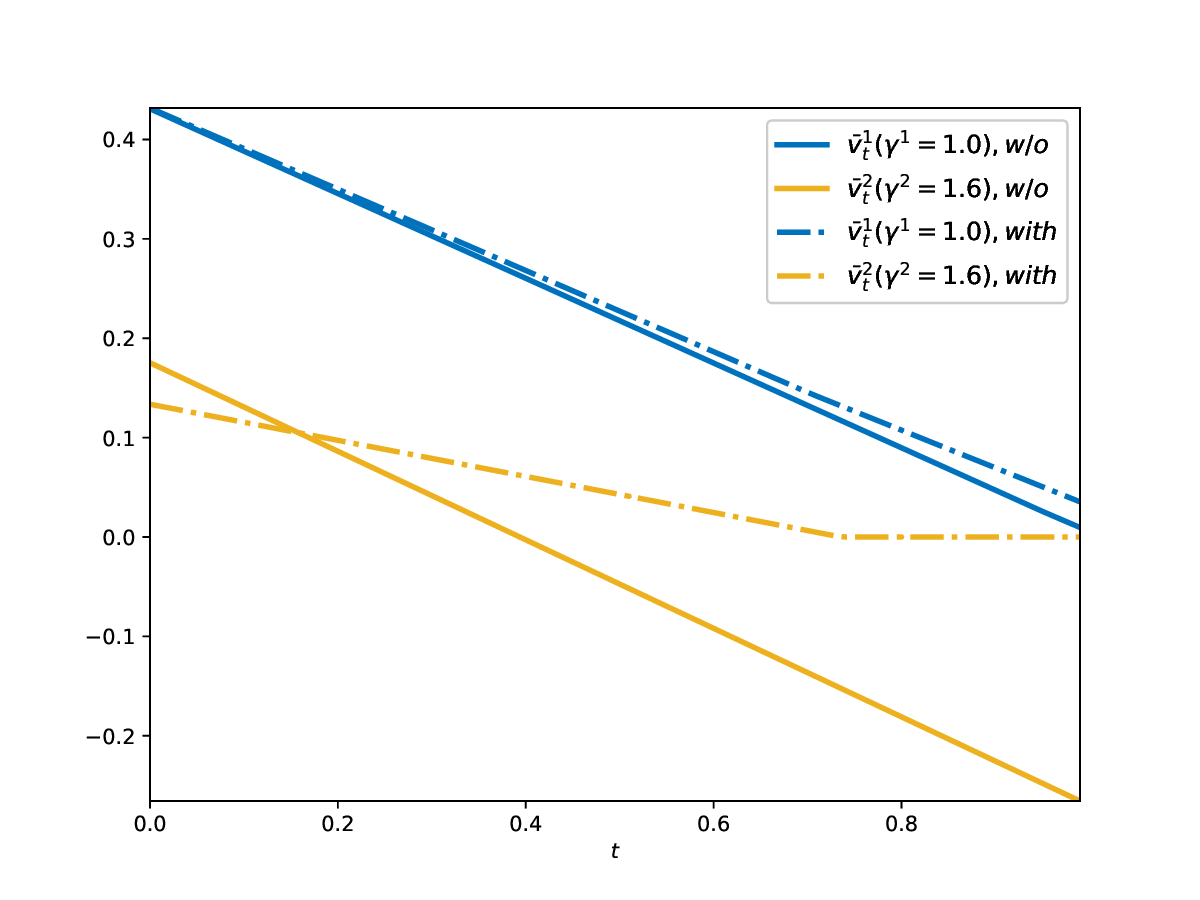}
        \label{fig_case2:figure1}
    }
    \hspace{-0.3in}
    \subfigure[$z^h$ for Case 2(a)]
    {
	\includegraphics[width=3.0in,height=2.7in]{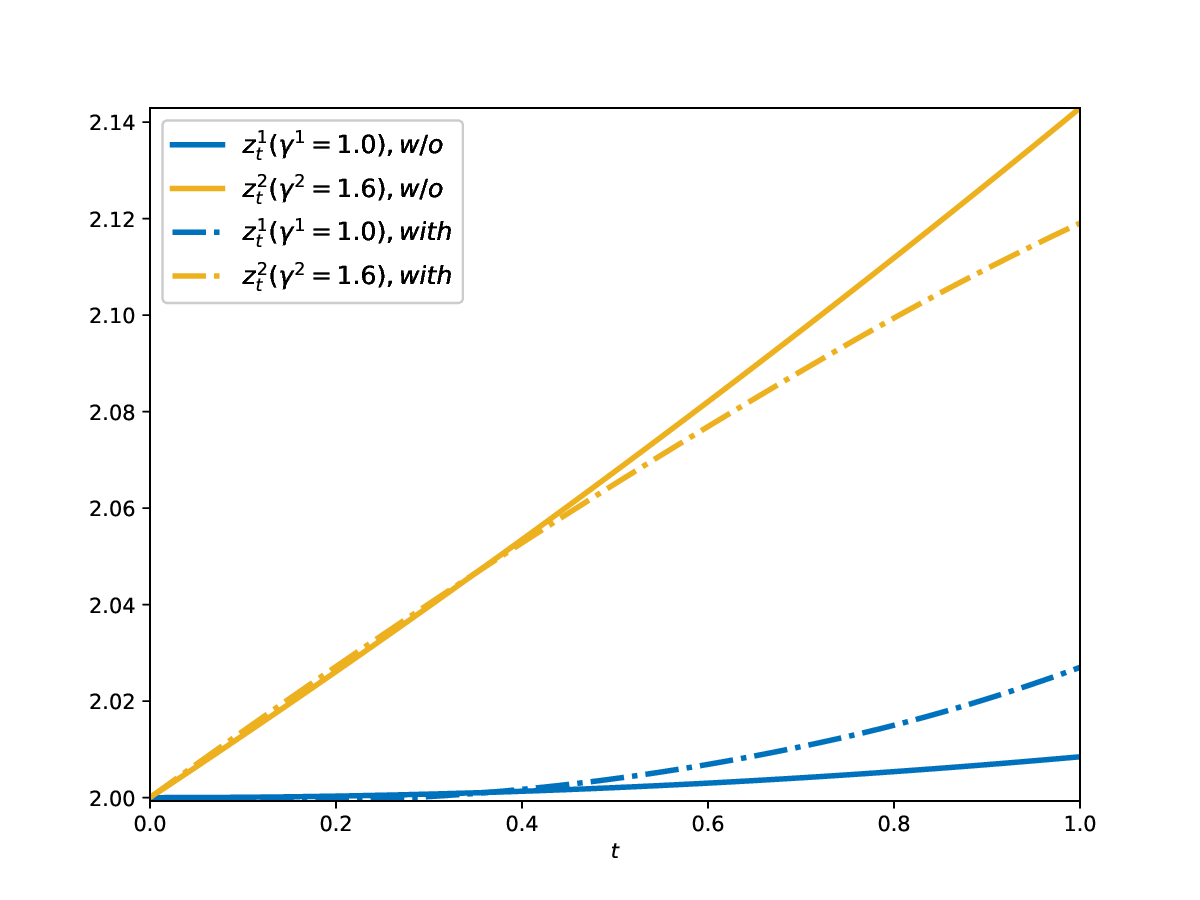}
        \label{fig_case2:figure2}
    }
  \caption{The equilibrium insurance strategies and wealth for representative members under Case 2(a).}
    \label{fig_case2:figures}
\end{figure}

The effect of the parameter $\gamma^h$, $h=1,2$, is depicted in Figure \ref{fig_case2:figures}. This parameter can serve as a measure of the risk aversion of the member. Specifically, when $\gamma^h$ is high (resp.~small), the member is more (resp.~less) concerned about her absolute terminal wealth relative to its fluctuation, indicating that the member is less (resp.~more) risk-averse.  Clearly, members who are more risk-averse tend to purchase more insurance to transfer the risk to the MIC (see Class $1$ ($\gamma^1=1$) in Figure \ref{fig_case2:figure1}). This results in a lower equilibrium wealth as opposed to Class 2 ($\gamma^2=1.6$), since (i) members of Class $1$ are less aware of the dollar amount of their terminal wealth, and (ii) more premiums are paid due to higher insurance demand. Specifically, Figure \ref{fig_case2:figure1} shows that the risk-averse member from Class 1 purchases insurance, in contrast to the short position taken by the risk-seeking member from Class 2 when no insurance constraint is imposed. Due to the difference in the risk-aversion and the insurance strategies, the representative member from Class 2 has a higher equilibrium wealth than that from Class 1.

Imposing the insurance constraint has a prominent effect on the equilibrium strategy of Class 2, since it restricts members from taking a short position. Consequently, it reduces the difference in the insurance strategies and the wealth gap between the two classes. Despite the constraint is unbinding for Class 1, the drastic change in the insurance strategy of Class 2 under the constraint induces an increase in the insurance strategy of Class 1.

The impact of the membership class composition can be assessed by comparing Cases 2(b) and 2(c) with Case 2(a) in Table \ref{tab:v:values}. When the proportion of risk-averse members is high (Case 2(b), $\omega^1=0.8$), members in both classes tend to purchase more insurance. Compared to Case 2(a), when the insurance constraint is imposed, the initial equilibrium strategy has increased by {\color{black}2.03\%} for Class 1, and {\color{black}2.73\%} for Class 2. The reasons are twofold. First, the overall risk awareness of the mutual has increased, driven by the larger share of risk-averse members. Second, the higher premium income contributed by Class 1 members leads to greater shared surplus, from which the more risk-seeking Class 2 members also benefit. This enhanced surplus distribution boosts their ability to afford more coverage. Conversely, the insurance strategies for members from both classes decrease when there is a smaller proportion of risk-averse members (Case 2(c), $\omega^1=0.2$).

\subsubsection{The impact of $\kappa^h$}
Figures \ref{fig_case3a:figures}-\ref{fig_case3b:figures} manifest the scenario under different $\kappa^h$, which is proportional to the safety loading $\theta^h$  and the rate of loss $\mu^h$. In the study, members in Class 2 ($\kappa^2=0.5)$ are charged with a higher cost of insurance   than their Class 1 ($\kappa^1=0.1$) counterparts, which can be due to higher rate of loss and safety loading of the policy.

\begin{figure}[!h]
    \centering
    \subfigure[$\bar{v}^h$ for Case 3(a)]{
        \includegraphics[width=3.0in,height=2.7in]{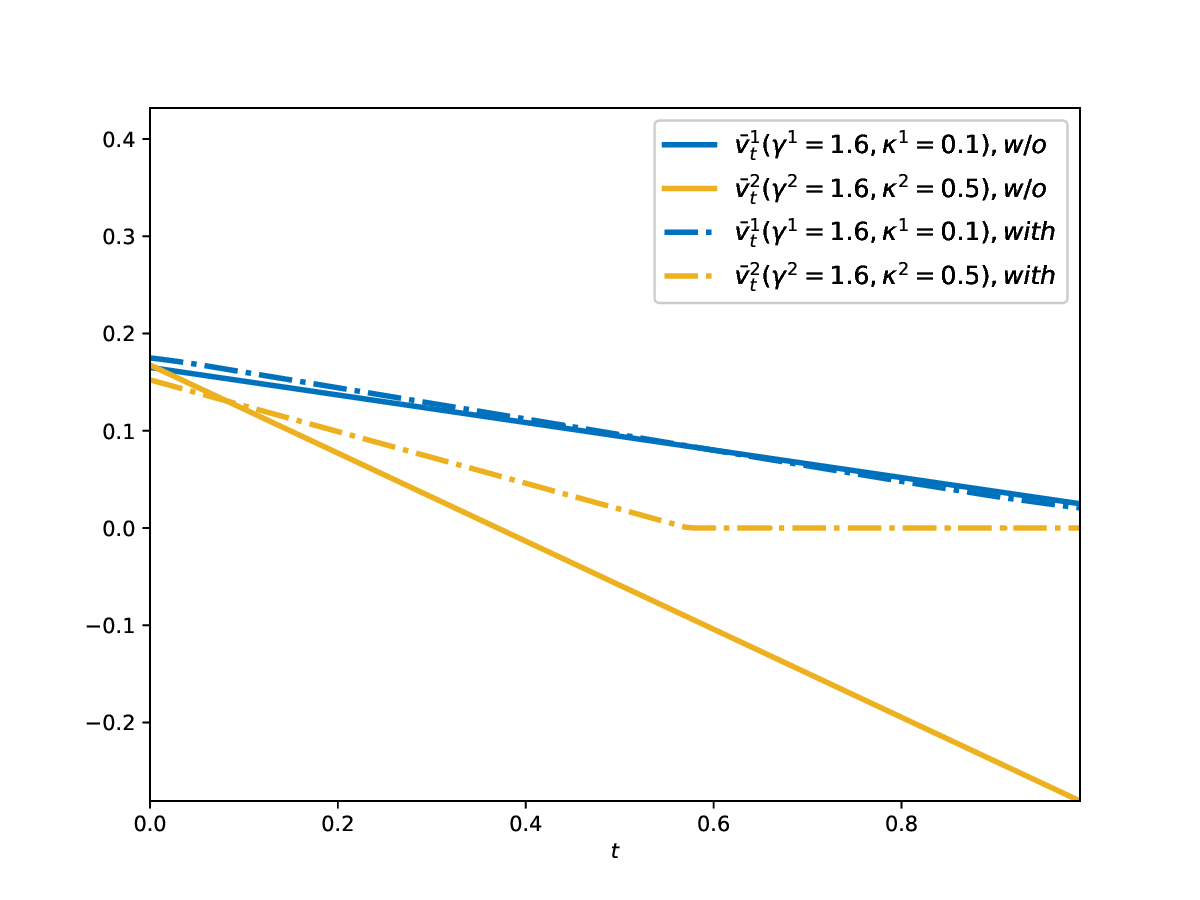}
        \label{fig_case3:figure1}
    }
    \hspace{-0.3in}
    \subfigure[$z^h$ for Case 3(a)]{
        \includegraphics[width=3.0in,height=2.7in]{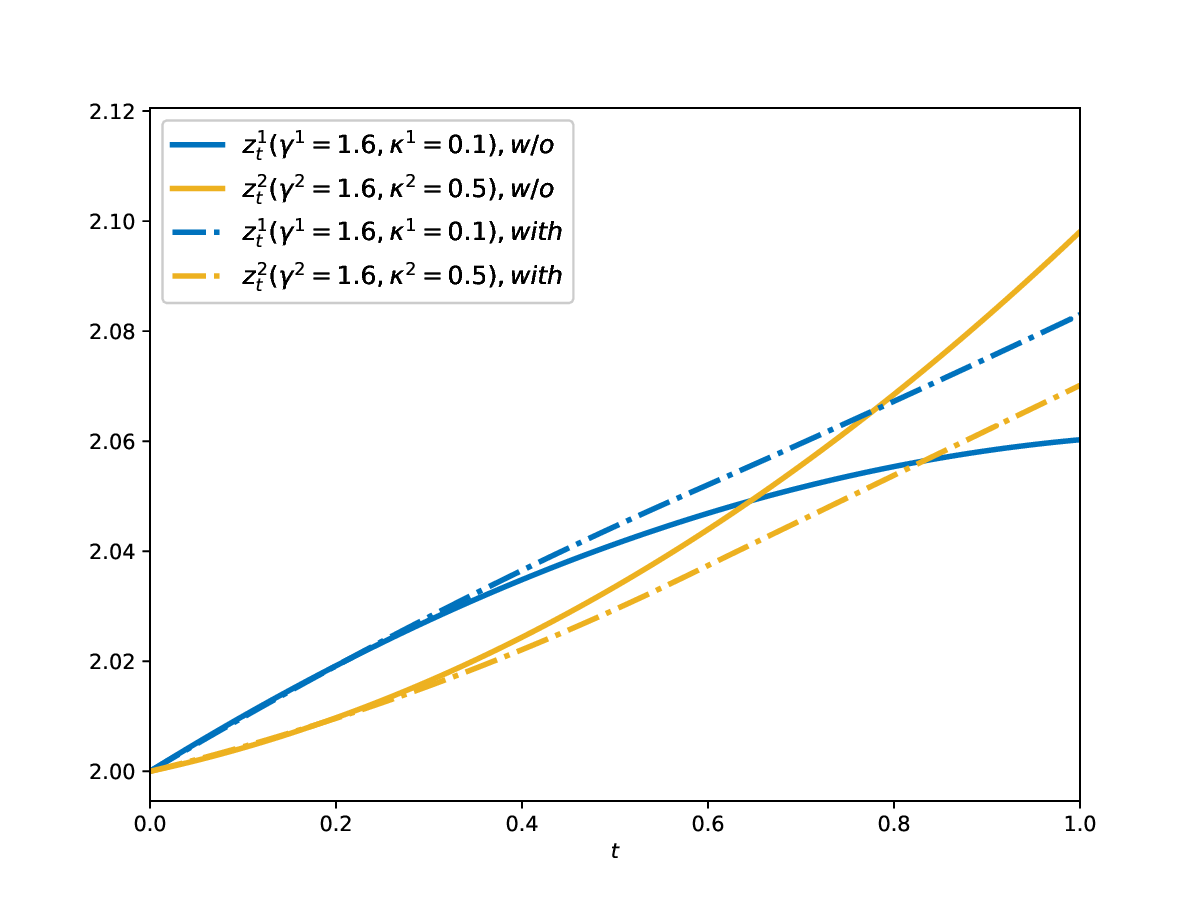}
        \label{fig_case3:figure2}
    }
    \caption{Equilibrium insurance strategies and wealth for representative members under Case 3(a).}
    \label{fig_case3a:figures}
\end{figure}

Figures \ref{fig_case3a:figures}–\ref{fig_case3b:figures} reveal that variations in $\kappa^h$ substantially influence how the equilibrium insurance strategy evolves over time. With a higher $\kappa^h$ (Class 2 in Cases 3(a)-(b)), the equilibrium strategy tends to decay at a faster rate.  Consequently, the relative size of the equilibrium wealth changes over time and is influenced by other parameters such as  $\gamma^h$. When $\gamma^h=1.6$ (Case 3(a)), indicating a relatively low level of risk aversion, a higher $\kappa^h$ (see Class 2 in Figure \ref{fig_case3:figure1}) leads to a reduction in the insurance purchases of members. The reason is straightforward: members are less inclined to buy overpriced insurance. Consequently, this lowers the equilibrium wealth for members in Class 2. The higher premium rate also lowers the equilibrium wealth for members in Class 2 compared to Class 1 when insurance constraint is imposed. In addition, for Class 2, the insurance constraint binds for roughly half of the planning horizon. In the unconstrained case, $\kappa^h$ decays rapidly over time, whereas under the constraint, this decline forces $\bar{v}^2_t=0$. Consequently, the equilibrium insurance strategies differ substantially between the constrained and unconstrained settings for Class 2.


In contrast, when the level of risk aversion is relatively high (Case 3(b), $\gamma^h = 1$), a higher $\kappa^h$ does not necessarily lower the initial insurance demand. As shown in Figure~\ref{fig_case3:figure3}, members in Class 2 purchase more insurance than their Class 1 counterparts until the end of the planning horizon. This can be explained as follows. Given that the net income remains unchanged, an increase in $\kappa^h$ may arise from both a higher $\mu^h$ and $\tilde{l}^h$. In this case, members face a greater expected loss intensity, which encourages them to purchase more insurance despite the higher premium cost. Moreover, when risk aversion is high, members in Class~2 place greater emphasis on mitigating wealth volatility within their class, resulting in higher insurance demand even at the expense of slower wealth accumulation. This pattern is reflected in Figure~\ref{fig_case3:figure4}, where their wealth growth is slower than their Class 1 counterpart. Owing to this and the higher time sensitivity of the insurance strategy under higher $\kappa^h$, their focus gradually shifts toward maximizing terminal wealth, leading to reduced insurance purchases when approaching $T$.


\begin{figure}[htbp]
    \centering
    \subfigure[$\bar{v}^h$ for Case 3(b)]{
        \includegraphics[width=3.0in,height=2.7in]{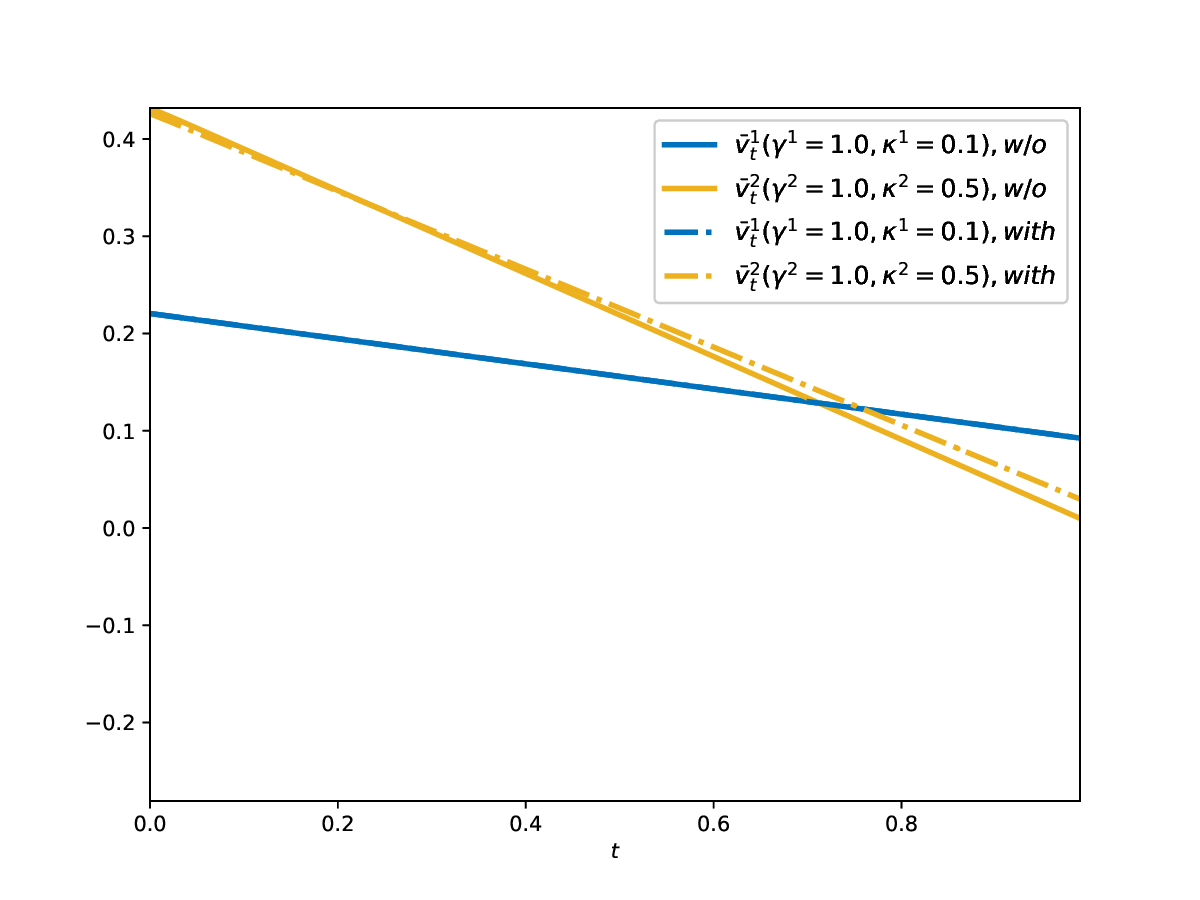}
        \label{fig_case3:figure3}
    }
    \hspace{-0.3in}
    \subfigure[$z^h$ for Case 3(b)]{
        \includegraphics[width=3.0in,height=2.7in]{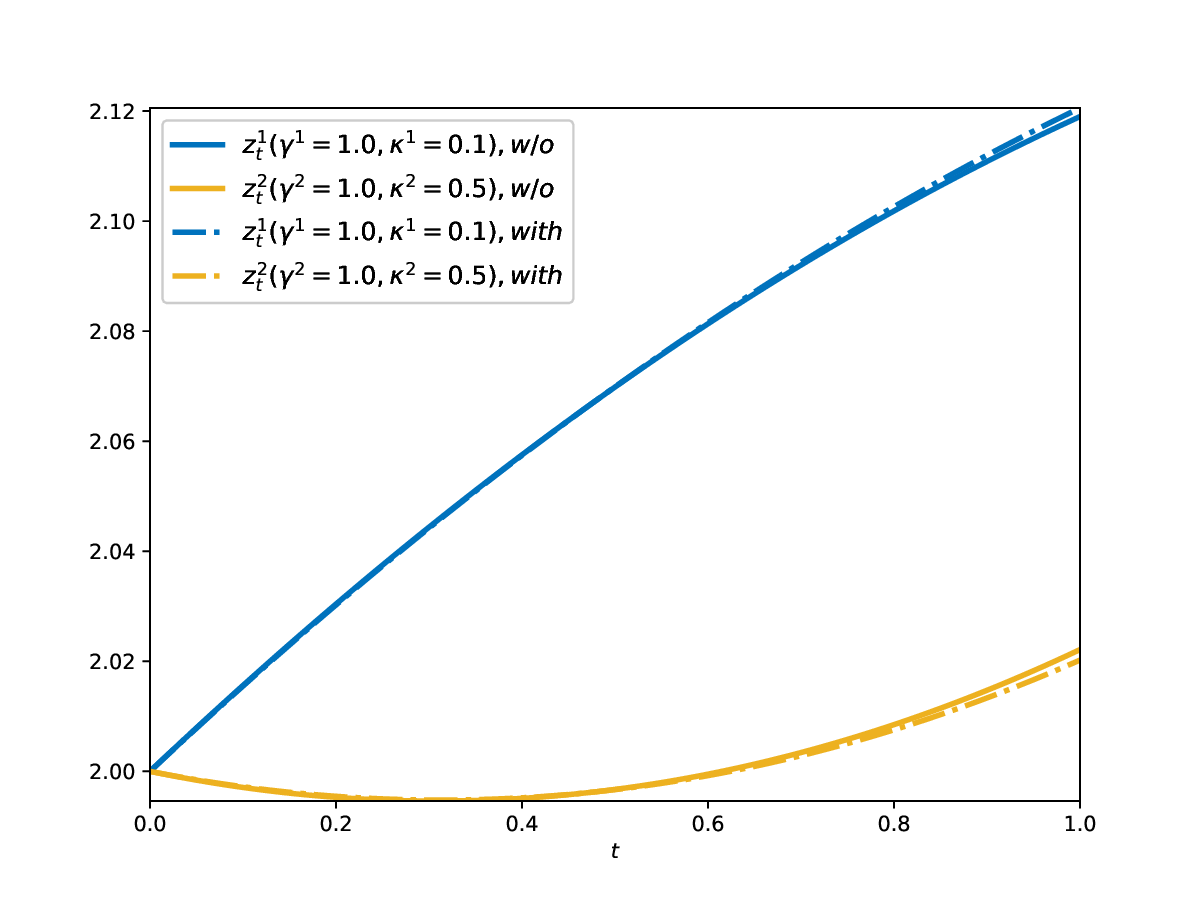}
        \label{fig_case3:figure4}
    }
    \caption{(continued) Equilibrium insurance strategies and wealth for representative members under Case 3(b).}
    \label{fig_case3b:figures}
\end{figure}

 

\subsubsection{The impact of $\tilde{l}^h-\mu^h$}
\begin{figure}[htbp]
   \centering
   \subfigure[$\bar{v}^h$ for Case 4(a)]{
       \includegraphics[width=3.0in,height=2.7in]{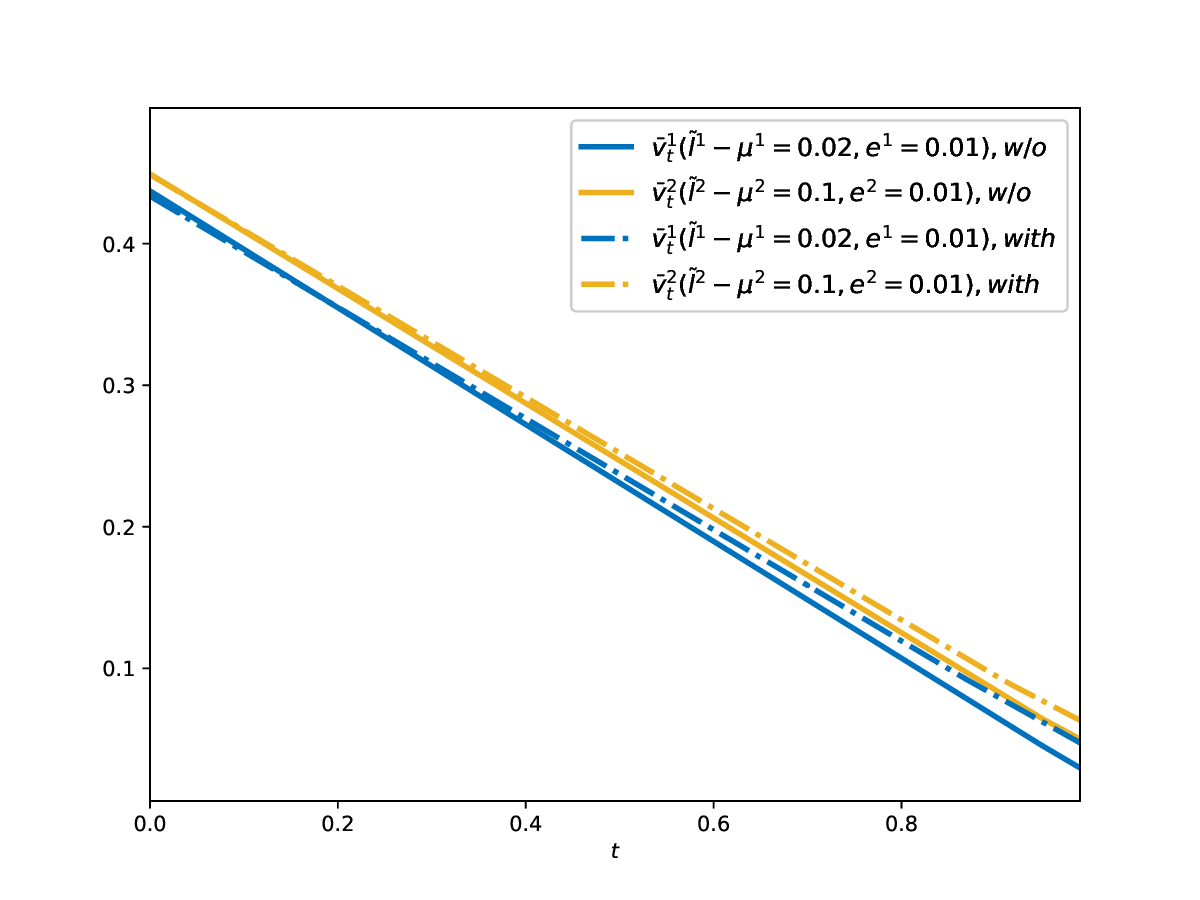}
       \label{fig_case4a:vbar}
   }
   \hspace{-0.3in}
   \subfigure[$z^h$ for Case 4(a)]{
       \includegraphics[width=3.0in,height=2.7in]{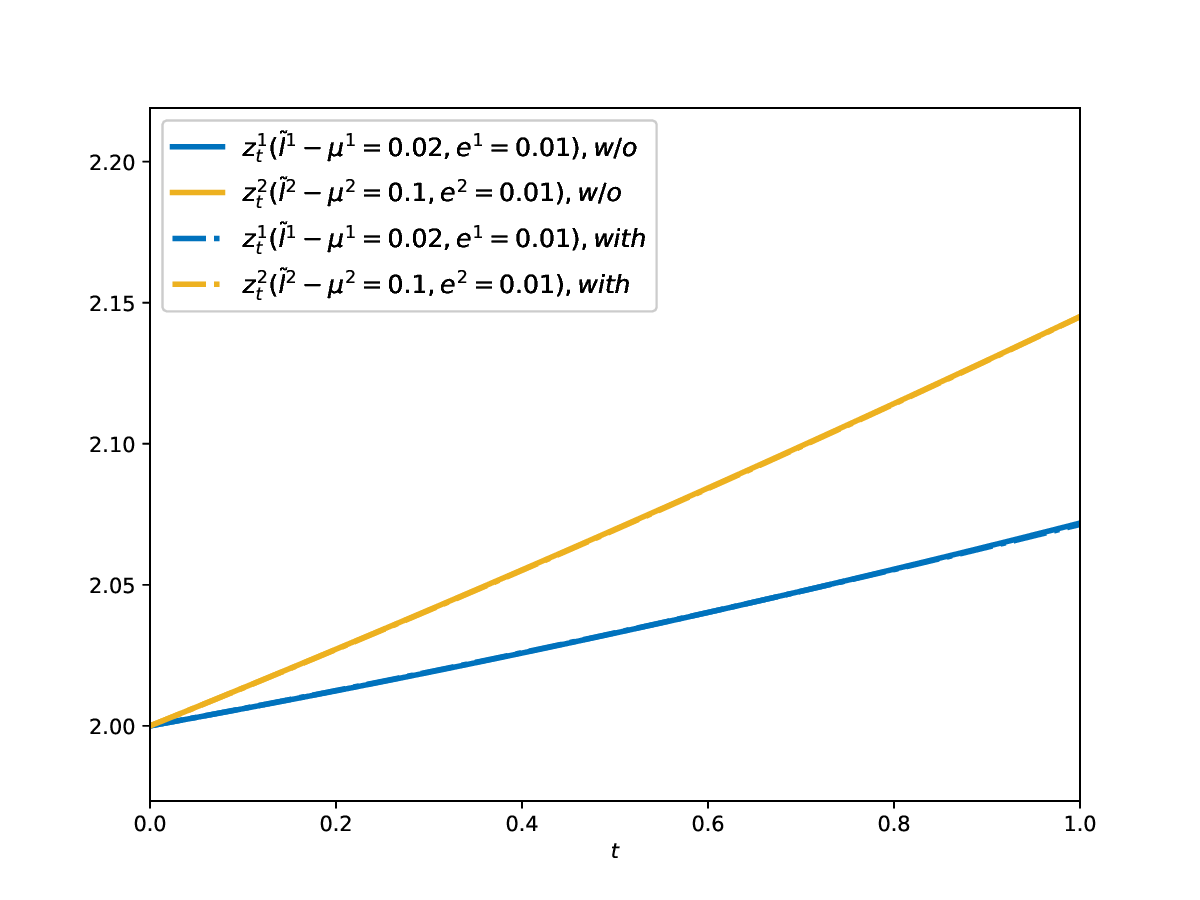}
       \label{fig_case4a:z}
   }
   \caption{Equilibrium insurance strategies and wealth for representative members under Case 4(a).}
   \label{fig_case4a:figures}
\end{figure}

Figure~\ref{fig_case4a:figures} illustrates the effect of the sharing-independent net income rate $\tilde{l}^h - \mu^h$, that is, the net income prior to any surplus or deficit transfers under the MIC, in Case~4(a). It is clear that the representative member from Class 2 $(\tilde{l}^2-\mu^2=0.1)$, who earns a higher net income rate than her Class 1 ($\tilde{l}^1-\mu^1=0.02$) counterpart, tends to purchase more insurance due to the higher purchasing power. This high income rate also offsets the higher premium rate, leading to a higher equilibrium wealth for the member in Class 2. 

\subsubsection{Impact of $e^h$}
 The effect of the membership fee rate $e^h$ can be examined by comparing Cases 4(b)-4(c) with Case 4(a). In particular, under the proportional relation \eqref{eq:prop:e:pi}, a change in the membership fee rate would also alter the sharing proportion $\pi^h$ and the management fee rate $d_e^h$. 

\begin{figure}[!h]
   \centering
   \subfigure[$\bar{v}^h$ for Case 4(b)]{
       \includegraphics[width=3.0in,height=2.7in]{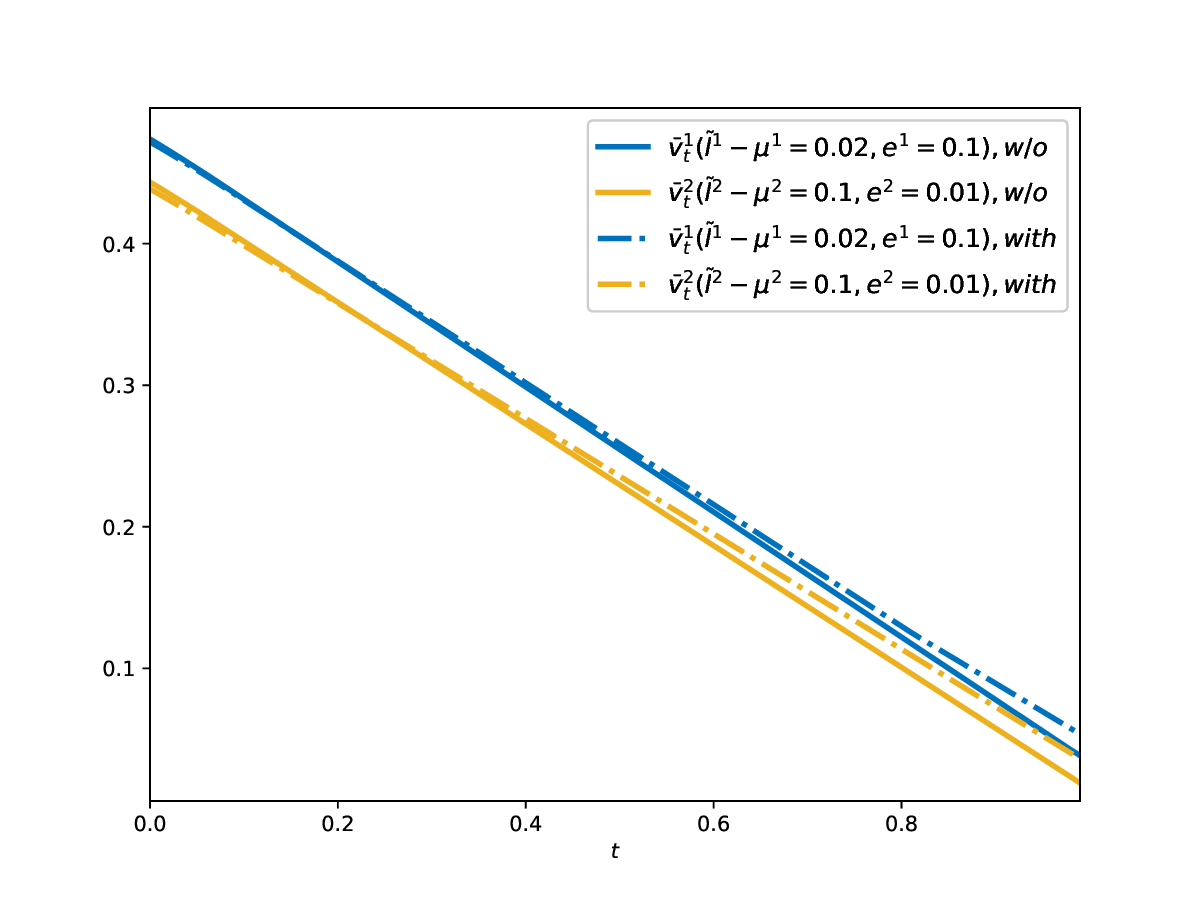}
       \label{fig_case4b:vbar}
   }
   \hspace{-0.3in}
   \subfigure[$z^h$ for Case 4(b)]{
       \includegraphics[width=3.0in,height=2.7in]{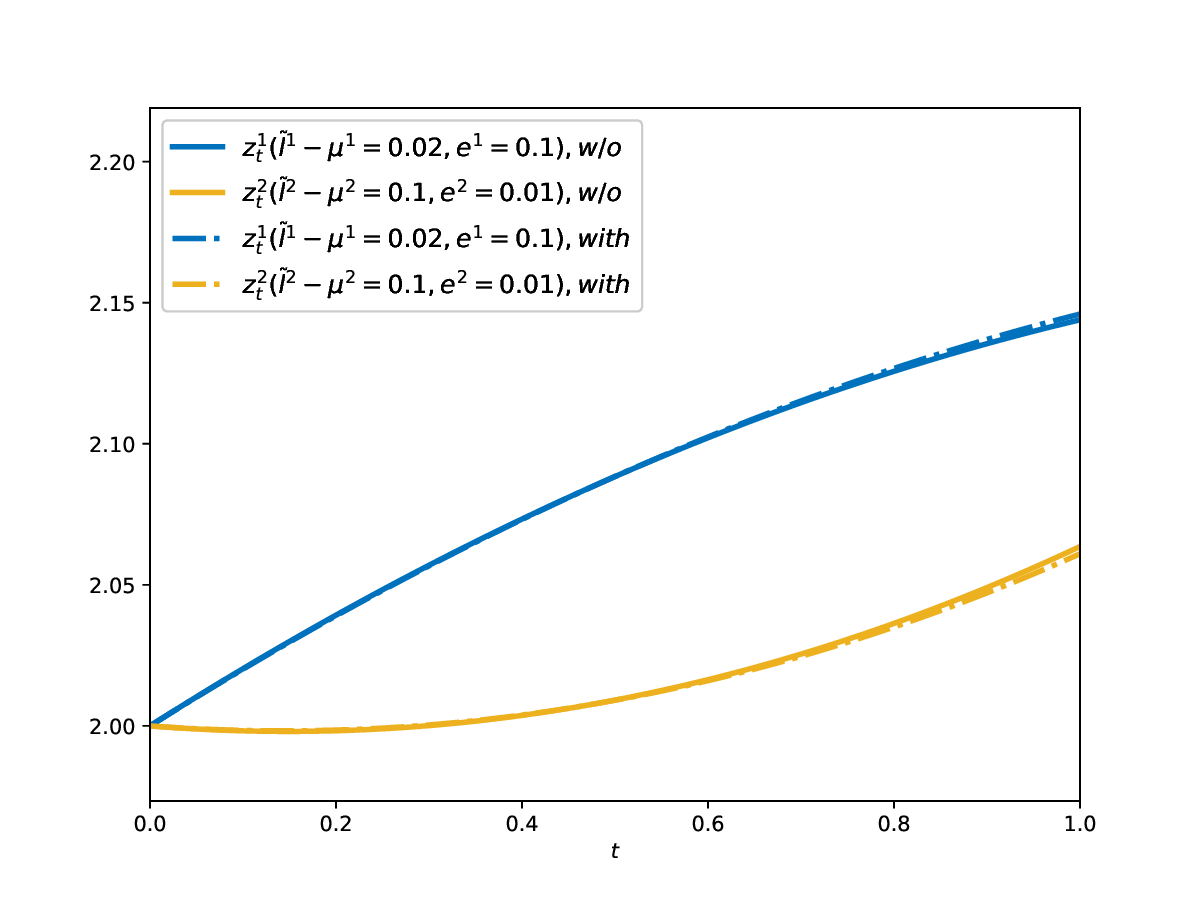}
       \label{fig_case4b:z}
   }
   \caption{Equilibrium insurance strategies and wealth for representative members under Case 4(b).}
   \label{fig_case4b:figures}
\end{figure}

In Case 4(b) (Figure \ref{fig_case4b:figures}), the increased sharing proportion {\color{black} $\pi^1=1.8182$} for Class 1 compensate the income advantage of Class 2. As a result, the wealth of Class 1 exceeds that of Class 2, accompanied by a higher insurance position.

\begin{figure}[!h]
   \centering
   \subfigure[$\bar{v}^h$ for Case 4(c)]{
       \includegraphics[width=3.0in,height=2.7in]{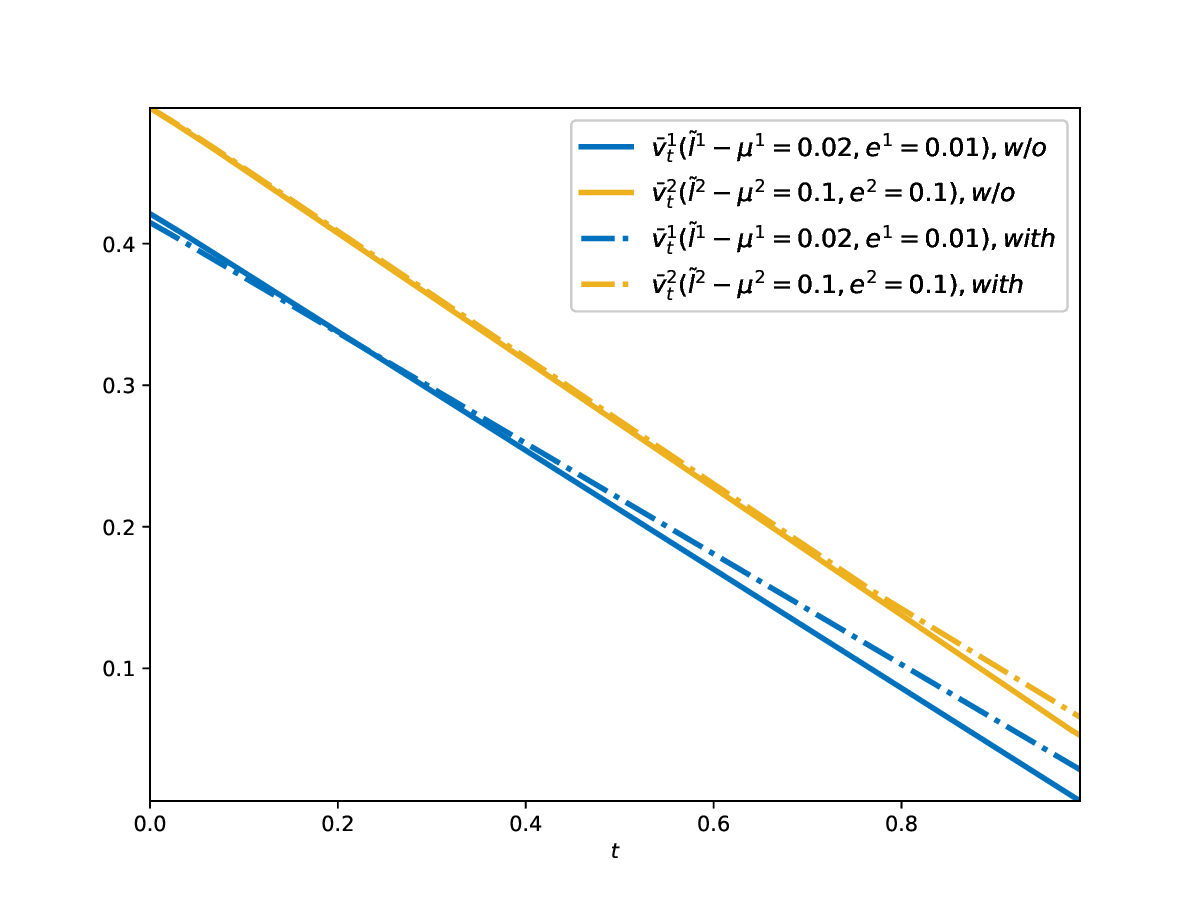}
       \label{fig_case4c:vbar}
   }
   \hspace{-0.3in}
   \subfigure[$z^h$ for Case 4(c)]{
       \includegraphics[width=3.0in,height=2.7in]{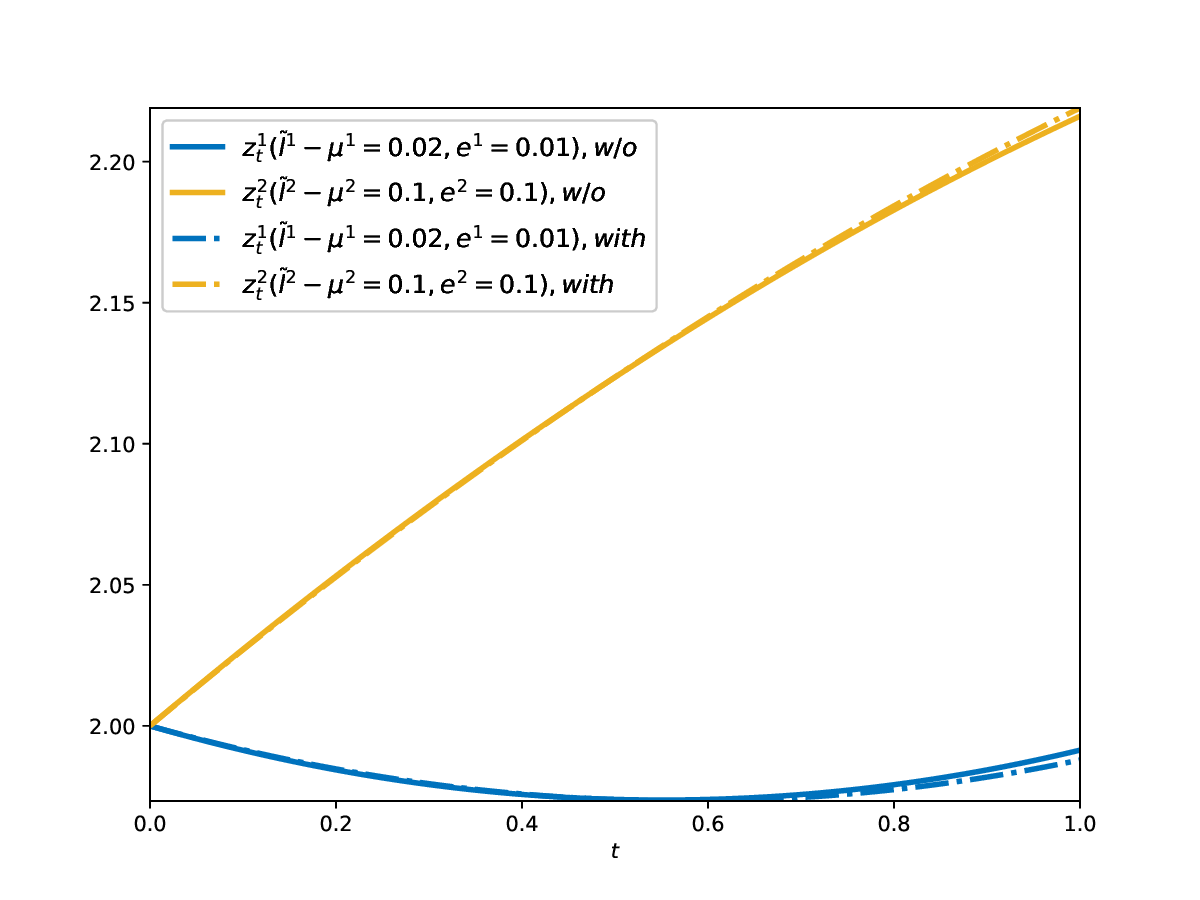}
       \label{fig_case4c:z}
   }
   \caption{Equilibrium insurance strategies and wealth for representative members under Case 4(c).}
   \label{fig_case4c:figures}
\end{figure}

In Case 4(c) (Figure~\ref{fig_case4c:figures}), with $e^1 = 0.01$ and $e^2 = 0.1$, members from Class 1 face a reduction in the shared surplus, leading to lower equilibrium wealth and a reduced insurance strategy compared to Case 4(a). In contrast, members in Class 2 receive both higher incomes and surplus from the MIC then her counterpart in Class 1, leading to an even higher insurance strategy and equilibrium wealth compared to Case 4(a).

\subsubsection{Behavior with respect to time}
Lastly, in all cases, we observe that the equilibrium insurance strategies decrease with time. The reasons are twofold. 
First, the value of the protection provided by insurance often declines with time, as the window for significant losses to occur in the future has shortened. Consequently, the uncertainty of future losses decreases, leading to a lower demand for coverage. Second, as the length of the planning horizon shortens, members would prioritize maximizing terminal wealth over long-term risk management, further contributing to the reduction in insurance strategies. The time-decaying nature of indemnity functions is also documented in the actuarial literature, see e.g.~\cite{zeng2011optimal}, \cite{li2012optimal}, and \cite{yi2013robust}. 


\subsection{General Mixture of Reward Functions}
\label{sec:mixed:reward}
The second study is based on an alternative class of reward functions. Specifically, we define
\begin{equation*}
        f^h(t,x,z,v,\bar{v}) := \begin{dcases}
            \frac{\gamma^h}{1-\gamma^h}\left(\frac{a^hx}{\gamma^h} + b^h \right)^{1-\gamma^h} - \frac{\gamma^h(b^h)^{1-\gamma^h}}{1-\gamma^h} - \frac{Q^h}{2}(x-B^h)^2 - \frac{P^h}{2}(v-R^h\bar{v})^2, \\
            \qquad\qquad\qquad\qquad\qquad\qquad\qquad \text{if }x\geq 0;\\
          a^h(b^h)^{-\gamma^h}x - \frac{Q^h}{2}(x-B^h)^2 - \frac{P^h}{2}(v-R^h\bar{v})^2, \quad \text{if }x <0,
        \end{dcases}
    \end{equation*}
and 
    \begin{equation*}
        g^h(x,z) := \begin{dcases}
            \frac{\gamma^h}{1-\gamma^h}\left(\frac{a^hx}{\gamma^h} + b^h \right)^{1-\gamma^h} - \frac{\gamma^h(b^h)^{1-\gamma^h}}{1-\gamma^h} - \frac{Q^h}{2}(x-B^h)^2, &\text{if }x\geq 0;\\
         a^h(b^h)^{-\gamma^h}x - \frac{Q^h}{2}(x-B^h)^2, & \text{if }x <0. 
        \end{dcases}
    \end{equation*}

These reward functions combine a hyperbolic absolute risk aversion (HARA) utility with a penalty relative to a specified benchmark. The parameter  $\gamma^h>0, \gamma^h \neq 1$ represents the degree of relative risk aversion, while $a^h > 0$ scales the utility function and governs its curvature. The  parameter $b^h > 0$ both shifts wealth to ensure positivity of the argument and governs how rapidly absolute risk aversion declines as wealth increases. Finally, $B^h > 0$ specifies a benchmark wealth level, penalizing deviations from the desired target. It is clear that the above choice of functions verifies  Assumptions \ref{ass:concave:monotonicity:h}, \ref{ass:contraction} and \ref{ass:fg:sep}, with $\alpha^X_1 =  \alpha^g_1 = Q^h$, $\alpha^X_2 = \alpha^g_2 = 0$, $L^X=L^g = Q^h + \frac{(a^h)^2}{(b^h)^{1+\gamma^h}} $. Hence, by \eqref{eq:alpha:M:2}, Assumption \ref{ass:M} would be fulfilled provided that 
    \begin{equation}
    \label{eq:cond:mixture:M}
     Q^h > \|{\bf M}\|_2\left(Q^h + \frac{(a^h)^2}{(b^h)^{1+\gamma^h}} \right).
    \end{equation}

In this experiment, we consider $H=2$  and choose the same parameters as in base scenario \eqref{LQ_baseline_scenario}, except   
    \begin{align*}
       \gamma^1 = 0.5, \gamma^2 = 3.0,\ a^1=a^2=1.0, \ b^1 = b^2 = 5.0,  \  B^1 = B^2 = 2.5, \ \kappa^1=\kappa^2=0.08,
    \end{align*}
so that \eqref{eq:cond:mixture:M} is fulfilled. We also define $\pi^h$ (and thus $l^h$) using the same formula as in \eqref{eq:prop:e:pi}. We refer to this study as Case 5.

 The last row of Table \ref{loss_penalty_table2} presents the training errors corresponding to the selected parameters and the reward functions in Case 5 using the training scheme \eqref{NN_MF_BSDE}. The results show that the training errors remain comparable, and even improved, to the quadratic case, while the training time increases modestly due to the more complex derivatives.

\begin{figure}[htbp]
    \centering
    \subfigure[$\bar{v}^h$ for Case 5]
    {
        \includegraphics[width=3.0in,height=2.7in]{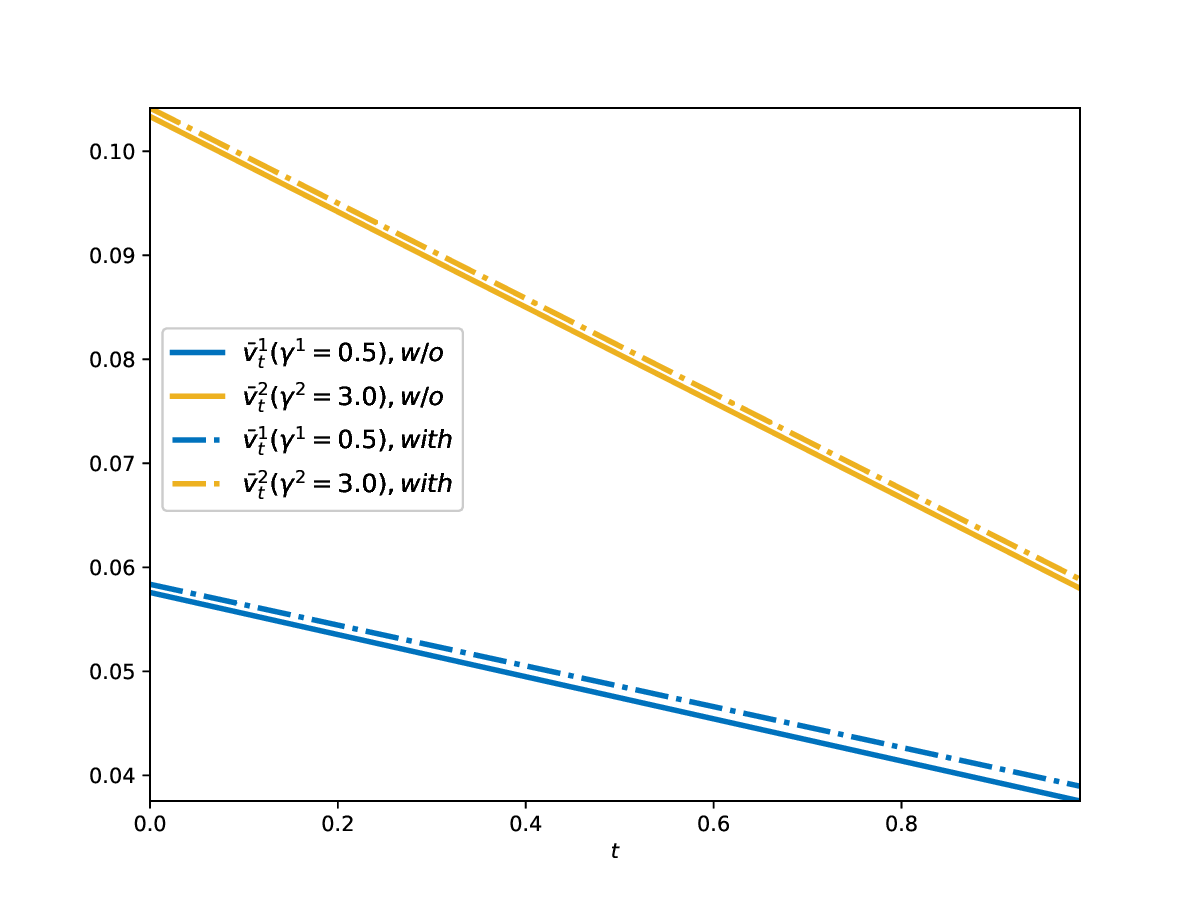}
        \label{fig_case5:figure1}
    }
    \hspace{-0.3in}
    \subfigure[$z^h$ for Case 5]
    {
	\includegraphics[width=3.0in,height=2.7in]{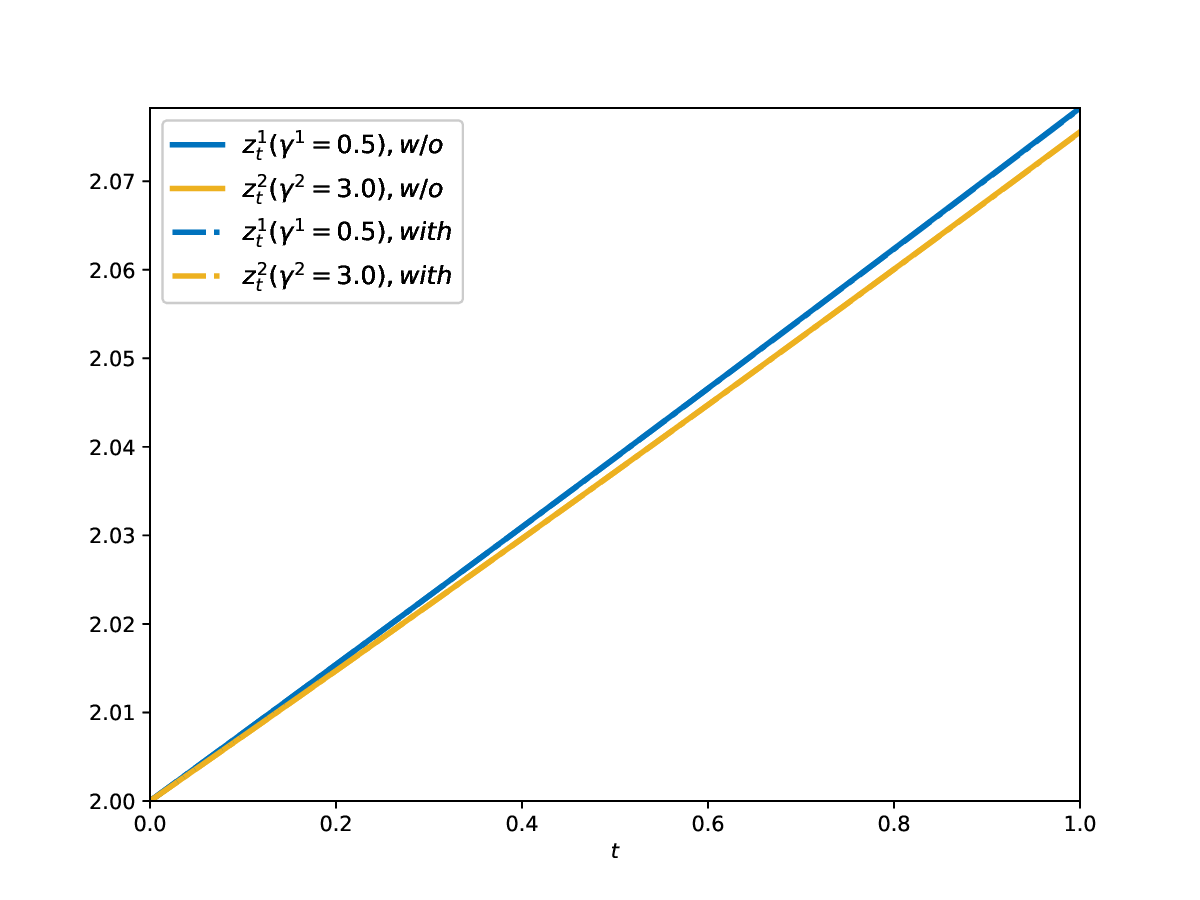}
        \label{fig_case5:figure2}
    }
  \caption{The equilibrium insurance strategies and wealth for representative members under Cases 5.}
    \label{fig_case5:figures}
\end{figure}

Figure~\ref{fig_case5:figures} illustrates the mean field equilibrium insurance strategies and the corresponding wealth levels of members in the two classes under different choices of the risk aversion parameter~$\gamma^h$, where a higher value indicates greater risk aversion. As expected, members in Class~2, with a higher risk aversion parameter $\gamma^2 = 3$, tend to purchase more insurance coverage than their counterparts in Class~1 ($\gamma^1 = 0.5$). Consequently, Class~1 members attain slightly higher equilibrium wealth due to lower premium payments. Notably, the imposed constraints are non-binding in this case, resulting in identical outcomes for the constrained and unconstrained settings.

\FloatBarrier

\section{Concluding Remarks}
\label{sec:conclusion}
In this article, we formulated a dynamic optimal insurance problem for a mutual insurance company within an extended mean field game framework. The optimal insurance strategies are characterized by a system of mean field forward-backward stochastic differential equations (MF-FBSDEs), where the global existence and uniqueness of solutions were established using the continuation method. To numerically solve the MF-FBSDEs and determine the optimal strategies, we proposed a deep BSDE approach.

This work opens several avenues for future research. First, incorporating a jump-diffusion setting could better capture the stochastic behavior of claim arrivals. Second, relaxing the separability of the objective function, as assumed in Assumption \ref{ass:fg:sep}, would broaden the model's applicability. Additionally, analogizing the mutual insurance company sharing mechanisms with those in decentralized insurance, this extended mean field game framework could be adapted to model optimal decision-making in decentralized insurance pools in future studies.

\section*{Acknowledgments}
Bohan Li is supported by the National Natural Science Foundation of China under grant No.12501661.  Wenyuan Li gratefully acknowledges a start-up grant from the University of Hong Kong. Kenneth Ng acknowledges the financial support from the Univeristy of Illinois Urbana-Champaign, the Chinese University of Hong Kong, and the start-up fund from the Ohio State University. Phillip Yam acknowledges the financial supports from HKGRF-14301321 with the project title ``General Theory for Infinite Dimensional Stochastic Control: Mean Field and Some Classical Problem'', HKGRF-14300123 with the project title ``Well-posedness of Some Poisson-driven Mean Field Learning Models and their Applications'', and  HKGRF-14300025 with the project title ``A Generic Theory for Stochastic Control against Fractional Brownian Motions''. The work described in this article was also supported by a grant from the Germany/Hong Kong Joint Research Scheme sponsored by the Research Grants Council of Hong Kong and the German Academic Exchange Service of Germany (Reference No. G-CUHK411/23). He also thanks The University of Texas at Dallas for the kind invitation to be a Visiting Professor in Naveen Jindal School of Management.


\bibliographystyle{apacite}
\bibliography{ref}

\begin{appendices}

\section{Auxiliary Lemmas}

In this section, we provide some auxiliary lemmas that are useful in constructing the optimal insurance strategies and establishing the well-posedness of the MF-FBSDE \eqref{eq:MFBSDE:general}. The first lemma is an elementary result in convex analysis. 

\begin{lemma}
    \label{lem:convex}
    Let $A\subset \mathbb{R}^n$ be a non-empty, closed, convex set. For any $x\in \mathbb{R}^n$, there exists an $x^*\in A$ such that $|x-x^*| = \min_{x'\in A}|x-x'|$. In addition, $x^*$ is characterized by the following inequality:
        \begin{equation*}
            \langle y-x^*, x-x^*\rangle \leq 0
        \end{equation*}
    for any $y\in A$. 
\end{lemma}

The next result is used in establishing the unique existence of solution of \eqref{eq:MFBSDE:general}, which can be verified by a straightforward manner. 
    \begin{lemma}
    \label{lem:proj}
        Let $a,b,u,l\in \mathbb{R}$ with $u<l$. Denote by $a_p = \text{Proj}_{[l,u]}(a)$ and $b_p = \text{Proj}_{[l,u]}(b)$. For any non-decreasing function $\varphi : \mathbb{R}\to\mathbb{R}$, it holds that 
            \begin{equation*}
                (a_p-b_p)(\varphi(a)-\varphi(b)) \geq (a_p-b_p)(\varphi(a_p)-\varphi(b_p)). 
            \end{equation*}
    \end{lemma}

In the next lemma, we demonstrate a simple inequality used in establishing the existence of solution of the MF-FBSDE under the general setup in  Theorem \ref{thm:exist:unique:MFFBSDE:general}, and also under the quadratic reward in Assumption \ref{ass:lq}.  

    \begin{lemma}
    \label{lem:EX:Z:inequality}
        Let ${\bf X}$ be a square integrable $\mathbb{R}^d$-valued random vector, and ${\bf Z} := \mathbb{E}[{\bf X}]$. Then, for any $d\times d$ matrix $\mathcal{M}$, and any positive definite matrix $\mathcal{Q}$ such that $\lambda_{\min}(\mathcal{Q}-\mathcal{M})>0$, we have 
            \begin{equation*}
                \mathbb{E}[\langle {\bf X},\mathcal{Q} {\bf X}\rangle] - \langle \mathcal{M}{\bf Z},{\bf Z}\rangle > \min\left\{ \lambda_{\min}(\mathcal{Q}), \lambda_{\min}(\mathcal{Q}-\mathcal{M}) \right\} \mathbb{E}[|{\bf X}|^2].
            \end{equation*}
    \end{lemma}
    \begin{proof}
        Using the identity  $\mathbb{E}[\langle {\bf X},\mathcal{Q} {\bf X}\rangle] =   \mathbb{E}[\langle ({\bf X}-{\bf Z}),\mathcal{Q} ({\bf X}-{\bf Z})\rangle] + \langle {\bf Z},\mathcal{Q}{\bf Z} \rangle$, we have 
            \begin{align*}
                  \mathbb{E}[\langle {\bf X},\mathcal{Q} {\bf X}\rangle] - \langle \mathcal{M}{\bf Z},{\bf Z}\rangle &= \mathbb{E}[\langle ({\bf X}-{\bf Z}),\mathcal{Q} ({\bf X}-{\bf Z})\rangle] + \langle {\bf Z},(\mathcal{Q}-\mathcal{M}){\bf Z} \rangle  \\
                  &\geq \mathbb{E}[\langle ({\bf X}-{\bf Z}),\mathcal{Q} ({\bf X}-{\bf Z})\rangle] +\lambda_{\min}(\mathcal{Q}-\mathcal{M}) |{\bf Z}|^2 \\
                  &\geq \lambda_{\min}(\mathcal{Q}) \mathbb{E}[|{\bf X}-{\bf Z}|^2] + \lambda_{\min}(\mathcal{Q}-\mathcal{M}) |{\bf Z}|^2 \\
                  &=  \lambda_{\min}(\mathcal{Q}) \mathbb{E}[|{\bf X}|^2]   + \left( \lambda_{\min}(\mathcal{Q}-\mathcal{M}) - \lambda_{\min}(\mathcal{Q}) \right) |{\bf Z}|^2. 
            \end{align*}
        If $\lambda_{\min}(\mathcal{Q}-\mathcal{M})>\lambda_{\min}(\mathcal{Q})$, then we immediately have 
        $$\mathbb{E}[\langle {\bf X},\mathcal{Q} {\bf X}\rangle] - \langle \mathcal{M}{\bf Z},{\bf Z}\rangle \geq \lambda_{\min}(\mathcal{Q}) \mathbb{E}[|{\bf X}|^2].$$
        Otherwise, by Jensen's inequality, 
            \begin{align*}
              & \ \ \ \    \mathbb{E}[\langle {\bf X},\mathcal{Q} {\bf X}\rangle] - \langle \mathcal{M}{\bf Z},{\bf Z}\rangle\\
              &\geq \lambda_{\min}(\mathcal{Q}-\mathcal{M}) \mathbb{E}[|{\bf X}|^2]   + \left( \lambda_{\min}(\mathcal{Q})-\lambda_{\min}(\mathcal{Q}-\mathcal{M})  \right)\left( \mathbb{E}[|{\bf X}|^2]- |{\bf Z}|^2\right) \\
              &\geq \lambda_{\min}(\mathcal{Q}-\mathcal{M}) \mathbb{E}[|{\bf X}|^2] . 
            \end{align*}
        The result then follows by combining the two cases. 
    \end{proof}



\section{Proofs and Extensions for Section \ref{sec:mf:formulation}}
    This section contains the proof of Theorem \ref{thm:epsilon-Nash} and discusses an extension of the model incorporating member survivorship.
 
    \subsection{Proof of Theorem \ref{thm:epsilon-Nash}}
    \label{sec:app:e-nash}
    This section is devoted to proving Theorem \ref{thm:epsilon-Nash}. The entire proof is decomposed into four steps. To begin, for each $h\in[H]$ and $i\in[N^h]$, let $v^{i,h}$ be the optimal strategy obtained in Problems \ref{p:mf}-\ref{p:fixed:point}, except that the Brownian motion $W^h$ in the wealth process is replaced by $W^{i,h}$. We also let $\hat{y}^{i,h}$ and $\hat{x}^{i,h}$ be the dynamics \eqref{eq:y:empirical} under the $N$-player game and the mean field dynamics \eqref{eq:xh} when the strategy $v^{i,h}$ is taken, respectively. Since the Brownian motions $W^{i,h}$ and $W^{j,h}$ are independent for $i\neq j$, the controls $v^{i,h}$ and $v^{j,h}$ are  independent and identically distributed (i.i.d.), so does the associated wealth processes $\hat{x}^{i,h}$ and $\hat{x}^{j,h}$. However, $\hat{y}^{i,h}$ and $\hat{y}^{j,h}$ are in general dependent due to the presence of the idiosyncratic component. 

    The first result manifests that the difference between $\hat{y}^{i,h}$ and $\hat{x}^{i,h}$ decreases with the class sizes in the order of $1/2$.  
        \begin{lemma}
        \label{lem:e-Nash:1}
            For any $t\in[0,T]$, $h\in[H]$ and $i\in[N^h]$, we have 
                \begin{equation*}
                    \sup_{h\in[H]}\sup_{i\in[N^h]} \mathbb{E}\left[ \sup_{s\leq t}\left|\hat{x}^{i,h}_s - \hat{y}^{i,h}_s \right|^2 \right] =  \sum_{k=1}^H O\left(\frac{1}{N^k}\right). 
                \end{equation*}
        \end{lemma}
    \begin{proof}
        By \eqref{eq:y:empirical} and \eqref{eq:xh}, we have 
            \begin{align*}
                \hat{x}^{i,h}_t - \hat{y}^{i,h}_t =&\ \int_0^t\left(r\left(\hat{x}^{i,h}_s - \hat{y}^{i,h}_s\right) + \pi^h \sum_{k=1}^H \omega^k(\kappa^k-d^k)\frac{\sum_{j=1}^{N^k}\left( \mathbb{E}[v^{1,k}_s] -v^{j,k}_s \right)}{N^k}  \right)ds \\
                &\ - \int_0^t \pi^h\sum_{k=1}^H \frac{\sigma^k\omega^k}{N^k}\sum_{j=1}^{N^k}v^{j,k}_sdW^{j,k}_s. 
            \end{align*}
        Hence, there exists $K>0$ independent of $(N^k)_{k=1}^H$ such that 
            \begin{align*}
               &\ \mathbb{E}\left[\sup_{s\leq t}\left|\hat{x}^{i,h}_s - \hat{y}^{i,h}_s \right|^2  \right]\\
               \leq&\ K\int_0^t   \mathbb{E}\left[ \left|\hat{x}^{i,h}_s - \hat{y}^{i,h}_s \right|^2\right] ds + K \sum_{k=1}^H \int_0^t  \mathbb{E}\left[\left( \frac{1}{N^k}\sum_{j=1}^{N^k}\left( \mathbb{E}[v^{1,k}_s] -v^{j,k}_s  \right) \right)^2  \right]  ds \\
                &\  + K\sum_{k=1}^H   \mathbb{E}\left[\sup_{s\leq t} \left( \frac{1}{N^k}\sum_{j=1}^{N^k} \int_0^s v^{j,k}_l dW^{j,k}_l  \right)^2 \right] .
            \end{align*}
        Since $v^{j,k}$ and $v^{i,k}$ are i.i.d.~for $i\neq j$, we have  
            \begin{align*}
             \int_0^t   \mathbb{E}\left[\left( \frac{1}{N^k}\sum_{j=1}^{N^k}\left( \mathbb{E}[v^{1,k}_s] -v^{j,k}_s  \right) \right)^2\right]ds &= \frac{1}{(N^k)^2} \int_0^t\sum_{j=1}^{N^k} \mathbb{E}\left[\left( \mathbb{E}[v^{1,k}_s] -v^{j,k}_s  \right)^2\right]ds = O\left(\frac{1}{N^k} \right),
            \end{align*}
        as $N^k\to\infty$. Similarly, by the Burkholder-Davis-Gundy inequality,
            \begin{align*}
                 \mathbb{E}\left[\sup_{s\leq t} \left( \frac{1}{N^k}\sum_{j=1}^{N^k} \int_0^s v^{j,k}_l dW^{j,k}_l  \right)^2 \right]  ds &\leq  \frac{1}{N^k}  \mathbb{E}\left[ \int_0^t \left(v^{1,k}_s\right)^2 ds   \right] = O\left(\frac{1}{N^k}\right). 
            \end{align*}
        Therefore, we have 
            \begin{align*}
                 \mathbb{E}\left[\sup_{s\leq t}\left|\hat{x}^{i,h}_s - \hat{y}^{i,h}_s \right|^2  \right] \leq  K\int_0^t   \mathbb{E}\left[\sup_{l\leq s}\left|\hat{x}^{i,h}_l - \hat{y}^{i,h}_l \right|^2\right] ds + \sum_{k=1}^H O\left(\frac{1}{N^k}\right),
            \end{align*}
        and the result follows from Gr\"onwall's inequality. 
    \end{proof}

    The next result depicts that the discrepancy between the objective function under the $N$-player game, and the mean field objective function under the mean field optimal strategy, exhibits a square-root decay with respect to the class sizes. 

    \begin{lemma}
    \label{lem:e-Nash:2}
        For $h\in[H]$ and $i\in[N^h]$, we have 
            \begin{equation*}
                \left| \mathcal{J}^{i,h}\left(v^{i,h};\hat{{\bf y}}^{-i,h},{\bf v}^{-i,h} \right) - J^{i,h}\left(v^{i,h};z^h, \bar{v}^h \right) \right| =  \sum_{k=1}^H O\left(\frac{1}{\sqrt{N^k}}\right).
            \end{equation*}
    \end{lemma}
    \begin{proof}
        By \eqref{eq:empirical:objective}, \eqref{eq:Jh} and Assumption \ref{ass:concave:monotonicity:h}, we have 
             \begin{align}
            \label{eq:lem2:nash:proof:1}
               &\ \left| \mathcal{J}^{i,h}\left(v^{i,h};\hat{{\bf y}}^{-i,h},{\bf v}^{-i,h}\right) - J^{i,h}\left(v^{i,h}; {\bf z]},\bar{v}^h\right) \right|\nonumber \\
               \leq&\ \mathbb{E}\left[\int_0^T \left|f^h\left(t,\hat{y}^{i,h}_t,\frac{\sum_{j\neq i}^{N^h}\hat{y}^{j,h}_t}{N^h-1},v^{i,h}_t,\frac{\sum_{j\neq i}^{N^h}v^{j,h}_t}{N^h-1}  \right) - f^h(t,\hat{x}^{i,h}_t,z^h_t,v^{i,h}_t,\bar{v}^h_t) \right|dt \right]\nonumber \\
               &\ + \mathbb{E}\left[\left|g^h\left(\hat{y}^{i,h}_t,\frac{\sum_{j\neq i}^{N^h} \hat{y}^{j,h}_T }{N^h-1} \right) - g^h(x^{i,h}_T,z^h_T) \right|\right] \nonumber\\
               \leq&\ L\mathbb{E}\Bigg[\int_0^T \left(1 + |\hat{y}^{i,h}_t| + |\hat{x}^{i,h}_t| + \left| \frac{\sum_{j\neq i}^{N^h}\hat{y}^{j,h}_t}{N^h-1}\right| + |z^h_t| + 2|v^{i,h}_t| +\left| \frac{\sum_{j\neq i}^{N^h}\hat{v}^{j,h}_t}{N^h-1}\right| + |\bar{v}^h_t|  \right)\nonumber \\
               &\quad \cdot \left(\left|\hat{y}^{i,h}_t - \hat{x}^{i,h}_t \right| + \left|\frac{\sum_{j\neq i}^{N^h}\hat{y}^{j,h}_t}{N^h-1} - z^h_t \right| + \left|\frac{\sum_{j\neq i}^{N^h}v^{j,h}_t}{N^h-1} - \hat{v}^h_t \right| \right) dt
               \Bigg]\nonumber \\
               &\ + L\mathbb{E}\left[\left(1+ |\hat{y}^{i,h}_T| + |\hat{x}^{i,h}_T| +\left| \frac{\sum_{j\neq i}^{N^h}\hat{y}^{j,h}_T}{N^h-1}\right| + |z^h_T|  \right)\left(\left|\hat{y}^{i,h}_T - \hat{x}^{i,h}_T \right| + \left|\frac{\sum_{j\neq i}^{N^h}\hat{y}^{j,h}_T}{N^h-1} - z^h_T \right| \right)\right] \nonumber \\
               \leq&\  L\mathbb{E}\Bigg[\int_0^T \left(1 + |\hat{y}^{i,h}_t| + |\hat{x}^{i,h}_t| + \left| \frac{\sum_{j\neq i}^{N^h}\hat{y}^{j,h}_t}{N^h-1}\right| + |z^h_t| + 2|v^{i,h}_t| +\left| \frac{\sum_{j\neq i}^{N^h}\hat{v}^{j,h}_t}{N^h-1}\right| + |\bar{v}^h_t|  \right)\nonumber \\
               &\quad \cdot \left(\left|\hat{y}^{i,h}_t - \hat{x}^{i,h}_t \right| + \left|\frac{\sum_{j\neq i}^{N^h}(\hat{y}^{j,h}_t-\hat{x}^{j,h}_t)}{N^h-1}  \right| + \left| \frac{\sum_{j\neq i}^{N^h} (\hat{x}^{j,h}_t - z^h_t  ) }{N^h-1} \right| +\left|\frac{\sum_{j\neq i}^{N^h}v^{j,h}_t}{N^h-1} - \hat{v}^h_t \right| \right) dt
               \Bigg]\nonumber \\
               &\ + L\mathbb{E}\Bigg[\left(1+ |\hat{y}^{i,h}_T| + |\hat{x}^{i,h}_T| +\left| \frac{\sum_{j\neq i}^{N^h}\hat{y}^{j,h}_T}{N^h-1}\right| + |z^h_T|  \right) \nonumber\\
               &\qquad \cdot \left(\left|\hat{y}^{i,h}_T - \hat{x}^{i,h}_T \right| +\left|\frac{\sum_{j\neq i}^{N^h}(\hat{y}^{j,h}_T-\hat{x}^{j,h}_T)}{N^h-1}  \right| + \left| \frac{\sum_{j\neq i}^{N^h} (\hat{x}^{j,h}_T - z^h_T  ) }{N^h-1} \right| \right)\Bigg]. 
            \end{align}
        By applying the Cauchy-Schwarz inequality to \eqref{eq:lem2:nash:proof:1}, along with the fact that the processes $\hat{x}^{i,h},\hat{y}^{i,h}$ are square-integrable, we have
        \begin{align}
                    \label{eq:lem2:nash:proof:2}
              &\ \left| \mathcal{J}^{i,h}\left(v^{i,h}; \hat{{\bf y}}^{-i,h}, {\bf v}^{-i,h}\right) - J^{i,h}\left(v^h;z^h,\bar{v}^h\right) \right|\nonumber \\
              \leq&\ K^h_T \Bigg(\mathbb{E}\bigg[\int_0^T \bigg(\left|\hat{y}^{i,h}_t - \hat{x}^{i,h}_t \right|^2 + \left|\frac{\sum_{j\neq i}^{N^h}(\hat{y}^{j,h}_t-\hat{x}^{j,h}_t)}{N^h-1}  \right|^2 + \left| \frac{\sum_{j\neq i}^{N^h} (\hat{x}^{j,h}_t - \mathbb{E}[\hat{x}^{1,h}_t]  ) }{N^h-1} \right|^2 \nonumber \\
              &\  
              + \left|\frac{\sum_{j\neq i}^{N^h}(v^{j,h}_t -\mathbb{E}[\hat{v}^{1,h}_t])}{N^h-1}  \right|^2 \bigg) dt  \bigg]\Bigg)^{\frac{1}{2}}\nonumber \\
              &\ + K^h_T \Bigg(\mathbb{E}\Bigg[\left|\hat{y}^{i,h}_T - \hat{x}^{i,h}_T \right|^2 +\left|\frac{\sum_{j\neq i}^{N^h}(\hat{y}^{j,h}_T-\hat{x}^{j,h}_T)}{N^h-1}  \right|^2 + \left| \frac{\sum_{j\neq i}^{N^h} (\hat{x}^{j,h}_T - \mathbb{E}[\hat{x}^{1,h}_T]  ) }{N^h-1} \right|^2 
              \Bigg]\Bigg)^{\frac{1}{2}},
        \end{align}
where $K^h_T>0$ is a generic constant independent of $N^k$, $k\in[H]$, which may change from line to line. To proceed, by Lemma \ref{lem:e-Nash:1}, we have 
            \begin{equation}
                      \label{eq:lem2:nash:proof:3}
                \mathbb{E}\left[\int_0^T \left|\hat{y}^{i,h}_t - \hat{x}^{i,h}_t \right|^2 dt \right] = \sum_{k=1}^H O\left(\frac{1}{N^k}\right), \mathbb{E}\left[  \left|\hat{y}^{i,h}_T - \hat{x}^{i,h}_T \right|^2 \right] = \sum_{k=1}^H O\left(\frac{1}{N^k}\right). 
            \end{equation}
 Next, using the i.i.d.~property of $(v^{i,h})_{i\in[N^h]}$, we have 
    \begin{equation}
                \label{eq:lem2:nash:proof:4}
        \mathbb{E}\left[\int_0^T \left|\frac{\sum_{j\neq i}^{N^h}\left(v^{j,h}_t -\mathbb{E}[v^{1,h}_t]\right)}{N^h-1}  \right|^2  dt\right] = \frac{1}{N^h-1}  \mathbb{E}\left[\int_0^T \left(v^{j,h}_t -\mathbb{E}[v^{1,h}_t]\right)^2   dt\right] = O\left(\frac{1}{N^h}\right). 
    \end{equation}
In addition, by the i.i.d.~property of $(\hat{x}^{i,h})_{i\in[H]}$, we have  
    \begin{equation}
        \begin{aligned}
                            \label{eq:lem2:nash:proof:5}
            \mathbb{E}\left[\int_0^T  \left| \frac{\sum_{j\neq i}^{N^h} (\hat{x}^{j,h}_t - \mathbb{E}[\hat{x}^{1,h}_t]  ) }{N^h-1} \right|^2dt \right] &= O\left( \frac{1}{N^h} \right),  \ 
              \mathbb{E}\left[ \left| \frac{\sum_{j\neq i}^{N^h} (\hat{x}^{j,h}_T - \mathbb{E}[\hat{x}^{1,h}_T]  ) }{N^h-1} \right|^2 \right] &= O\left( \frac{1}{N^h} \right).
        \end{aligned}
    \end{equation}
Therefore, the claim follows by substituting   \eqref{eq:lem2:nash:proof:3}-\eqref{eq:lem2:nash:proof:5} into \eqref{eq:lem2:nash:proof:2}.

    \end{proof}

  In the next step, we fix an arbitrary Class $h_0\in[H]$ and a representative member $i_0\in[N^h]$. Suppose that this member takes an arbitrary admissible strategy $u^{i_0,h_0}$, while all the other members within the MIC adopt the mean field equilibrium strategy. In that case, the wealth process $\check{y}^{i_0,h_0}$ of that member is governed by 
    \begin{align*}
       d\check{y}^{i_0,h_0}_t &= \Bigg(r\check{y}^{i_0,h_0}_t+ l^{h_0} -\kappa^{h_0} u^{i_0,h_0}_t + \pi^{h_0}\bigg(\sum_{h_0\neq k\in[H]}\omega^k (\kappa^k-d^k)\frac{\sum_{j=1}^{N^k}v^{j,k}_t}{N^k} \\
       &\qquad +\omega^{h_0} (\kappa^{h_0}-d^{h_0})\frac{u^{i_0,h_0}_t + \sum_{j\neq i_0}^{N^{h_0}}v^{j,h_0}_t}{N^{h_0}} \bigg) \Bigg)dt  + \sigma^{h_0}(1-u^{i_0,h_0}_t)dW^{i_0,h_0}_t \\
          &\quad       + \pi^{h_0}\left(\sum_{h_0\neq k\in[H]} \frac{\sigma^k\omega^k}{N^k}\sum_{j=1}^{N^k}v^{j,k}_tdW^{j,k}_t + \frac{u^{i_0,h_0}_t + \sum_{j\neq i_0}^{N^{h_0}}v^{j,h_0}_t }{N^{h_0}} \sigma^{h_0}\omega^{h_0}dW^{i_0,h_0}_t \right).
    \end{align*}
Let also $\check{y}^{i,h}$ be the wealth process for the member $i$ from Class $h$, where   $(i,h)\neq (i_0,h_0)$. Then, $\check{y}^{i,h}$  is governed by 
 \begin{align*}
       d\check{y}^{i,h}_t &= \Bigg(r\check{y}^{i,h}_t+ l^{h} -\kappa^{h} v^{i,h}_t + \pi^{h}\bigg(\sum_{h_0\neq k\in[H]} \omega^k(\kappa^k-d^k)\frac{\sum_{j=1}^{N^k}v^{j,k}_t}{N^k}\\
       &\qquad + \omega^{h_0}(\kappa^{h_0}-d^{h_0})\frac{u^{i_0,h_0}_t + \sum_{j\neq i_0}^{N^{h_0}}v^{j,h_0}_t}{N^{h_0}} \bigg) \Bigg)dt  + \sigma^{h}(1-v^{i,h}_t)dW^{i,h}_t \\
          &\quad       + \pi^{h}\left(\sum_{h_0\neq k\in[H]} \frac{\sigma^k\omega^k}{N^k}\sum_{j=1}^{N^k}v^{j,k}_tdW^{j,k}_t + \frac{u^{i_0,h_0}_t + \sum_{j\neq i_0}^{N^{h_0}}v^{j,h_0}_t }{N^{h_0}} \sigma^{h_0}\omega^{h_0}dW^{i_0,h_0}_t \right).
    \end{align*}
We also define the process $\check{x}^{i_0,h_0}$ by 
    \begin{equation*}
             d\check{x}^{i_0,h_0}_t = \left(r\check{x}^{i_0,h_0}_t +l^{h_0} - \kappa^{h_0} u^{i_0,h_0}_t + \pi^{h_0}\sum_{k=1}^H \omega^k (\kappa^k-d^k)\mathbb{E}[v^{1,k}_t] \right)dt + \sigma^h(1-u^{i_0,h_0}_t)dW^{i_0,h_0}_t. 
    \end{equation*}

The following result indicates that, when the class sizes $N^k$, $k\in[H]$, are sufficiently large, the deviation from the mean field equilibrium wealth caused the member $i_0$ of Class $h_0$ would decline with the class sizes.   

\begin{lemma}
\label{lem:e-Nash:3}
    For $t\leq T$, $h\in[H]$, $i\in[N^h]$ with $(i,h)\neq (i_0,h_0)$,    we have 
        \begin{equation*}
            \mathbb{E}\left[\sup_{s\leq t}\left|\check{x}^{i_0,h_0}_s - \check{y}^{i_0,h_0}_s  \right|^2 \right] +  \mathbb{E}\left[\sup_{s\leq t}\left|\hat{x}^{i,h_0}_s - \check{y}^{i,h_0}_s  \right|^2 \right] +\mathbb{E}\left[\sup_{s\leq t}\left|\hat{x}^{i,h}_s - \check{y}^{i,h}_s  \right|^2 \right] = \sum_{k=1}^h O\left(\frac{1}{N^k}\right). 
        \end{equation*}
\end{lemma}
\begin{proof}
    Notice that 
     \begin{align*}
                \check{x}^{i_0,h_0}_t - \check{y}^{i_0,h_0}_t &= \int_0^t\Bigg[r\left(\check{x}^{i_0,h_0}_s - \check{y}^{i_0,h_0}_s\right)  + \pi^{h_0}\Bigg(\sum_{h_0\neq k\in[H]} \omega^k(\kappa^k-d^k)\frac{\sum_{j=1}^{N^k}\left( \mathbb{E}[v^{1,k}_s] -v^{j,k}_s \right)}{N^k}  \\
                &\qquad +\omega^{h_0}(\kappa^{h_0}-d^{h_0})\frac{ \sum_{j\neq i_0}^{N^{h_0}}\left(\mathbb{E}[v^{1,h_0}_s] -v^{j,h_0}_s \right) + \mathbb{E}[v_s^{1,h_0}] - u^{i_0,h_0}_s}{N^{h_0}}\Bigg) \Bigg]ds \\
                &\quad - \pi^{h_0}\left(\int_0^t \sum_{h_0\neq k\in[H]} \frac{\sigma^k\omega^k}{N^k}\sum_{j=1}^{N^k}v^{j,k}_tdW^{j,k}_s +\int_0^t \frac{u^{i_0,h_0}_s + \sum_{j\neq i_0}^{N^{h_0}}v^{j,h_0}_s }{N^{h_0}} \sigma^{h_0}\omega^{h_0}dW^{i_0,h_0}_s \right).
            \end{align*}
    Hence, there exists $K_T>0$ independent of $N^k$, $k\in[H]$, such that  
        \begin{align}
        \label{eq:proof:lem:eNash3:1}
            &\  \mathbb{E}\left[\sup_{s\leq t}\left|\check{x}^{i_0,h_0}_s - \check{y}^{i_0,h_0}_s \right|^2  \right]\nonumber\\
            \leq&\ K_T\int_0^t   \mathbb{E}\left[ \left|\check{x}^{i_0,h_0}_s - \check{y}^{i_0,h_0}_s \right|^2\right] ds + K_T \sum_{h_0\neq k\in[H]} \int_0^t  \mathbb{E}\left[\left( \frac{1}{N^k}\sum_{j=1}^{N^k}\left( \mathbb{E}[v^{1,k}_s] -v^{j,k}_s  \right) \right)^2  \right]  ds\nonumber \\
              &\  +  K_T \int_0^t  \mathbb{E}\left[\left( \frac{1}{N^{h_0}} \left( \sum_{j\neq i_0}^{N^{h_0}}\left(\mathbb{E}[v^{1,h_0}_s] -v^{j,h_0}_s \right) + \mathbb{E}[v_s^{1,h_0}] - u^{i_0,h_0}_s   \right)\right)^2  \right]  ds\nonumber \\
                &\ + K_T\sum_{k\neq h_0}   \mathbb{E}\left[\sup_{s\leq t} \left( \frac{1}{N^k}\sum_{j=1}^{N^k} \int_0^s v^{j,k}_l dW^{j,k}_l  \right)^2 \right]  ds \nonumber\\
                &\ + K_T   \mathbb{E}\left[\sup_{s\leq t} \left( \frac{1}{N^{h_0}}\sum_{j=1}^{N^{h_0}} \int_0^s\left( u^{i_0,h_0}_s + \sum_{j\neq i_0}v^{j,h_0}_s \right) dW^{i_0,h_0}_s \right)^2 \right].  
        \end{align}
    By i.i.d.~property of $(v^{i,h})_{i\in[N^h]}$, we have
    \begin{equation}
                \label{eq:proof:lem:eNash3:2}
        \begin{aligned}
            \sum_{k\neq h_0} \int_0^t  \mathbb{E}\left[\left( \frac{1}{N^k}\sum_{j=1}^{N^k}\left( \mathbb{E}[v^{1,k}_s] -v^{j,k}_s  \right) \right)^2  \right]  ds &= \sum_{k\neq h_0} O\left(\frac{1}{N^k}\right), \\
            \sum_{k\neq h_0}   \mathbb{E}\left[\sup_{s\leq t} \left( \frac{1}{N^k}\sum_{j=1}^{N^k} \int_0^s v^{j,k}_l dW^{j,k}_l  \right)^2 \right]  ds &= \sum_{k\neq h_0} O\left(\frac{1}{N^k}\right),
        \end{aligned}
    \end{equation}
 and that
 \begin{align}
         \label{eq:proof:lem:eNash3:3}
          & \ \    \int_0^t  \mathbb{E}\left[\left( \frac{1}{N^{h_0}} \left( \sum_{j\neq i_0}^{N^{h_0}}\left(\mathbb{E}[v^{1,h_0}_s] -v^{j,h_0}_s \right) + \mathbb{E}[v_s^{1,h_0}] - u^{i_0,h_0}_s   \right)\right)^2  \right]  ds\nonumber\\
          &\leq K_T\int_0^t  \left( \frac{1}{N^{h_0}} \mathbb{E}\left[\left(v^{1,h_0}_s - \mathbb{E}[v^{1,h_0}_s] \right)^2 \right] + \frac{1}{(N^{h_0})^2}\sum_{j\neq i_0}\mathbb{E}\left[\left|\mathbb{E}[v^{1,h_0}_s] -v^{j,h_0}_s \right|\left|\mathbb{E}[v_s^{1,h_0}] - u^{i_0,h_0}_s \right|  \right] \right) ds \nonumber\\
          &= O\left(\frac{1}{N^{h_0}}\right),
        \end{align}
    as $N^{h_0}\to\infty$. Likewise, using the Burkholder-Davis-Gundy inequality, one can show that 
        \begin{equation}
                \label{eq:proof:lem:eNash3:4}
             \mathbb{E}\left[\sup_{s\leq t} \left( \frac{1}{N^{h_0}}\sum_{j=1}^{N^{h_0}} \int_0^s\left( u^{i_0,h_0}_s + \sum_{j\neq i_0}v^{j,h_0}_s \right) dW^{i_0,h_0}_s \right)^2ds \right]   = O\left(\frac{1}{N^{h_0}}\right).
        \end{equation}
    By substituting \eqref{eq:proof:lem:eNash3:2}, \eqref{eq:proof:lem:eNash3:3} and \eqref{eq:proof:lem:eNash3:4} into \eqref{eq:proof:lem:eNash3:1}, we obtain 
        \begin{equation*}
             \mathbb{E}\left[\sup_{s\leq t}\left|\check{x}^{i_0,h_0}_s - \check{y}^{i_0,h_0}_s \right|^2  \right] =\sum_{k=1}^H O\left(\frac{1}{N^k}\right).
        \end{equation*}
    The fact that 
    \begin{equation*}
        \mathbb{E}\left[\sup_{s\leq t}\left|\hat{x}^{i,h_0}_s - \check{y}^{i,h_0}_s  \right|^2 \right] +\mathbb{E}\left[\sup_{s\leq t}\left|\hat{x}^{i,h}_s - \check{y}^{i,h}_s  \right|^2 \right] =\sum_{k=1}^H O\left(\frac{1}{N^k}\right)
    \end{equation*}
can be shown by a similar argument, henceforth the calculations are omitted. 
\end{proof}

The following result is a consequence of Lemma \ref{lem:e-Nash:3}, which can be shown by following the proof of Lemma \ref{lem:e-Nash:2}. 
\begin{lemma}
\label{lem:eNash-4}
    For $h_0\in[H]$, we have
      \begin{equation*}
                \left| \mathcal{J}^{i_0,h_0}\left(u^{i_0,h_0}; \check{{\bf y}}^{-i_0,h_0}, {\bf v}^{-i_0,h_0}\right) - J^{i_0,h_0}\left(u^{i_0,h_0}; z^{h_0}, \bar{v}^{h_0} \right) \right| = \sum_{k=1}^H O\left(\frac{1}{\sqrt{N^k}}\right),
            \end{equation*}
     where $\check{{\bf y}}^{-i_0,h_0} = (\check{y}^{i,h_0})_{i\in [N^{h_0}], i\neq i_0}$.
\end{lemma}

As a result of Lemmas \ref{lem:e-Nash:2} and \ref{lem:eNash-4}, we obtain 
    \begin{align*}
     \mathcal{J}^{i_0,h_0}\left(u^{i_0,h_0};  \check{{\bf y}}^{-i_0,h_0},{\bf v}^{-i_0,h_0} \right)&  \leq   J^{i_0,h_0}\left(u^{i_0,h_0}; z^{h_0},\bar{v}^{h_0}  \right) + \sum_{k=1}^H O\left(\frac{1}{\sqrt{N^k}}\right) \\
         &\leq  J^{i_0,h_0}\left(v^{i_0,h_0}; z^{h_0},\bar{v}^{h_0}\right) +\sum_{k=1}^H O\left(\frac{1}{\sqrt{N^k}}\right) \\
         &\leq  \mathcal{J}^{i_0,h_0}\left(v^{i_0,h_0};  \hat{{\bf y}}^{-i_0,h_0},{\bf v}^{-i_0,h_0}  \right) +\sum_{k=1}^H O\left(\frac{1}{\sqrt{N^k}}\right) ,
    \end{align*}
where the second inequality follows from the optimality of $v^{i_0,h_0}$. The desired $\varepsilon$-Nash equilibrium is thus established.

\section{Proofs of Assertions in Section \ref{sec:eqm:strategy}}
\label{sec:app:sec:eqm:strategy}

This section contains the proofs of statements in Section \ref{sec:eqm:strategy}.
\subsection{Proof of Lemma \ref{lem:cocercive}}
    \label{sec:app:pf:lem:cocercive}
     We first show that  $v^h \in L^2_{\mathbb{F}^h}([0,T];\mathbb{R}) \mapsto J^h(v^h)$ is continuous. Let $(z^h)_{h\in[H]}$ and $(\bar{v}^h)_{h\in[H]}$ be exogeneously given. Fix $\check{v}^h\in L^2_{\mathbb{F}^h}([0,T];\mathbb{R})$, and let $\check{x}^h$ be the associated wealth process when $\check{v}^h$ is adopted. For any $v^h \in L^2_{\mathbb{F}^h}([0,T];\mathbb{R})$ with the corresponding wealth process $x^h$, consider  
        \begin{align*}
            &\ \left|J^h(v^h) - J^h(\hat{v}^h) \right| \\
            \leq&\ \mathbb{E}\left[\int_0^T \left|f^h(t,x^h_t,z^h_t,v^h_t,\bar{v}^h_t) -f^h(t,\check{x}^h_t,z^h_t,\check{v}^h_t,\bar{v}^h_t)  \right|dt + \left|g^h(x^h_T,z^h_T) - g^h(\check{x}^h_T,z^h_T)\right| \right] \\
            \leq&\ L\mathbb{E}\Bigg[\int_0^T \left(1+|x^h_t|+|\check{x}^h_t| + |v^h_t| + |\check{v}^h_t| +  2|z^h_t| + 2|\bar{v}^h_t| \right)\left(\left|x^h_t-\check{x}^h_t\right| + \left|v^h_t-\check{v}^h_t\right|\right)dt \\
            &\quad + \left(1 +|x^h_T|+|\check{x}^h_T| + 2|z^h_T| \right)\left|x^h_T - \check{x}^h_T\right| \Bigg] \\
            \leq&\  L\mathbb{E}\Bigg[\int_0^T \left(1+|x^h_t-\check{x}^h_t|+2|\check{x}^h_t| + |v^h_t-\check{v}^h_t| + 2|\check{v}^h_t| +  2|z^h_t| + 2|\bar{v}^h_t| \right) \\
            &\quad \cdot \left(\left|x^h_t-\check{x}^h_t\right| + \left|v^h_t-\check{v}^h_t\right|\right)dt   + \left(1 +|x^h_T-\check{x}^h_T|+2|\check{x}^h_T| + 2|z^h_T| \right)\left|x^h_T - \check{x}^h_T\right| \Bigg],
        \end{align*}
where the second inequality follows from Assumption \ref{ass:concave:monotonicity:h}.A. By simple applications of  Young's and the Cauchy-Schwarz inequality, and noticing that 
     \begin{align*}
        \check{x}^h_t &= \int_0^t e^{r(t-s)}\left( l^h -\kappa^h\check{v}_s + \pi^h\sum_{k=1}^H\omega^k(\kappa^k-d^k)\bar{v}^k_t \right) ds + \sigma^h\int_0^t e^{r(t-s)} (1-\check{v}^h_s) dW^h_s, \\  
        x^h_t-\check{x}^h_t &= -\kappa^h\int_0^t e^{r(t-s)}(v^h_s-\check{v}^h_s)ds - \sigma^h\int_0^t e^{r(t-s)} (v^h_s-\check{v}^h_s)dW^h_s, 
    \end{align*}
we infer the existence of a constant $K^h_T>0$ independent of $x^h,v^h$,  such that 
  \begin{align*}
             \left|J^h(v^h) - J^h(\check{v}^h) \right|\leq K^h_T \mathbb{E}\left[\int_0^T\left|v^h_t - \check{v}^h_t\right|^2  dt \right]. 
        \end{align*}
Therefore, the continuity is established. 
   
    Next, we show that $v^h\mapsto J^h(v^h)$ is coercive and strictly concave in $L^2_{\mathbb{F}^h}([0,T];\mathbb{R})$.  To this end, let $\theta\in \mathbb{R}$ and $\hat{v}^h \in L^2_{\mathbb{F}^h}([0,T];\mathbb{R})$, and define $v^{h,\theta}:=v^h+\theta\hat{v}^h$. By linearity, the associated wealth process under the control $v^{h,\theta}$ is given by $x^{h,\theta} = x^h + \theta \hat{x}^h$, where $\hat{x}^h$ satisfies the following SDE:
        \begin{equation*}
            d\hat{x}^h_t = (r\hat{x}^h_t - \kappa^h\hat{v}^h_t)dt - \sigma^h\hat{v}^h_tdW^h_t, \ \hat{x}^h_0 = 0. 
        \end{equation*}
        
    To proceed, we shall first deduce an expression for the G\^ateaux derivative of $J^h(v^h)$. Notice that
        \begin{align}
        \label{eq:proof:coercive:1}
            \frac{d}{d\theta}J^h(v^{h,\theta}) &= \mathbb{E}\left[\int_0^T\left[f_x^h\left(t,x^{h,\theta}_t,z^h_t,v^{h,\theta}_t,\bar{v}^h_t\right)\hat{x}^h_t + f_v^h\left(t,x^{h,\theta}_t,z^h_t,v^{h,\theta}_t,\bar{v}^h_t\right) \hat{v}^h_t   \right] dt\right] \nonumber \\
            &\quad + \mathbb{E}[g_x^h(x^{h,\theta}_T,z^h_T)\hat{x}^h_T]. 
        \end{align}
Consider the following BSDE:
    \begin{equation*}
        \left\{ 
            \begin{aligned}
                -dp^{h,\theta}_t &= \left[rp^{h,\theta}_t -f_x^h\left(t,x^{h,\theta}_t,z^h_t,v^{h,\theta}_t,\bar{v}^h_t\right) \right]dt - \eta^{h,\theta}_tdW^h_t, \\
                p^{h,\theta}_T &= -g_x^h(x^{h,\theta}_T,z^h_T).
            \end{aligned}
        \right.
    \end{equation*}
Notice that the forward equations of $x^h$ and $\hat{x}^h$ are decoupled from $p^{h,\theta}$, and thus the latter admits a unique solution, thanks to Assumption \ref{ass:concave:monotonicity:h}. By applying It\^o's lemma on $p^{h,\theta}_t\hat{x}^h_t$, we obtain 
    \begin{align}
        \label{eq:proof:coercive:2}
        \mathbb{E}[g_x^h(x^{h,\theta}_T,z^h_T)\hat{x}^h_T] &= \mathbb{E}\left[\int_0^T \left[ -f_x^h\left(t,x^{h,\theta}_t,z^h_t,v^{h,\theta}_t,\bar{v}^h_t\right)\hat{x}^h_t + \left( \kappa^hp^{h,\theta}_t + \sigma^h\eta^{h,\theta}_t \right)\hat{v}^h_t \right]dt  \right]. 
    \end{align}
Substituting \eqref{eq:proof:coercive:2} into \eqref{eq:proof:coercive:1}, we have     
    \begin{align}
        \frac{d}{d\theta}J^h(v^{h,\theta}) &= \mathbb{E}\left[\int_0^T \left[f_v^h\left(t,x^{h,\theta}_t,z^h_t,v^{h,\theta}_t,\bar{v}^h_t\right) + \kappa^hp^{h,\theta}_t + \sigma^h\eta^{h,\theta}_t \right] \hat{v}^h_t    dt\right] \nonumber\\
        &= \mathbb{E}\left[\int_0^T q^{h,v^{h,\theta}}_t\hat{v}^h_tdt\right],
    \end{align}
where  
    \begin{equation*}
        q^{h,v^{h,\theta}}_t := f_v^h\left(t,x^{h,\theta}_t,z^h_t,v^{h,\theta}_t,\bar{v}^h_t\right) + \kappa^hp^{h,\theta}_t + \sigma^h\eta^{h,\theta}_t, \ t\in[0,T]. 
    \end{equation*}

Next, for any $\theta,\phi \in \mathbb{R}$, we have 
    \begin{align*}
      &\  \mathbb{E}\left[\int_0^T  \left(  q^{h,v^{h,\theta}}_t -  q^{h,v^{h,\phi}}_t\right)\left(v^{h,\theta}_t-v^{h,\phi}_t \right) \bigg) dt    \right] \\
        =&\ \mathbb{E}\left[\int_0^T\left[f_v^h(t,x^{h,\theta}_t,z^h_t,v^{h,\theta}_t,\bar{v}^h_t)- f_v^h(t,x^{h,\phi}_t,z^h_t,v^{h,\phi}_t,\bar{v}^h_t)\right]\left(v^{h,\theta}_t-v^{h,\phi}_t \right)dt \right] \\
        &\ + \mathbb{E}\left[\int_0^T \left[\kappa^h(p^{h,\theta}_t-p^{h,\phi}_t) + \sigma^h(\eta^{h,\theta}_t - \eta^{h,\phi}_t) \right] \left(v^{h,\theta}_t-v^{h,\phi}_t \right)dt\right] \\
        =&\  \mathbb{E}\left[\int_0^T\left[f_v^h(t,x^{h,\theta}_t,z^h_t,v^{h,\theta}_t,\bar{v}^h_t)- f_v^h(t,x^{h,\phi}_t,z^h_t,v^{h,\phi}_t,\bar{v}^h_t)\right]\left(v^{h,\theta}_t-v^{h,\phi}_t \right)dt \right] \\
        &\ + \mathbb{E}\left[\int_0^T\left[f_x^h(t,x^{h,\theta}_t,z^h_t,v^{h,\theta}_t,\bar{v}^h_t)- f_x^h(t,x^{h,\phi}_t,z^h_t,v^{h,\phi}_t,\bar{v}^h_t)\right]\left(x^{h,\theta}_t-x^{h,\phi}_t \right)dt \right] \\
        &\ + \mathbb{E}\left[ \left(g_x^h(x^{h,\theta}_T,z^h_T) - g_x^h(x^{h,\phi}_T,z^h_T) \right)\left( x^{h,\theta}_T-x^{h,\phi}_T\right)  \right] \\ 
        \leq&\  - \alpha^V_1  \mathbb{E}\left[\int_0^T\left| v^{h,\theta}_t-v^{h,\phi}_t \right|^2dt \right],
    \end{align*} 
where the second equality follows from applying It\^o's lemma on $(p^{h,\theta}_t-p^{h,\phi}_t)(x^{h,\theta}_t-x^{h,\phi}_t)$; and the last line follows from Assumption \ref{ass:concave:monotonicity:h}.B.  

Finally, for any fixed $\hat{v}^h\in L^2_{\mathbb{F}^h}([0,T];\mathbb{R})$ and any $v^h\in L^2_{\mathbb{F}^h}([0,T];\mathbb{R})$, we have 
    \begin{align*}
        J^h(v^h) - J^h(\hat{v}^h) &= -\int_0^1 \frac{d}{d\theta}J^h(v^h + \theta(\hat{v}^h-v^h)) d\theta \\
        &= - \int_0^1 \mathbb{E}\left[\int_0^T q^{h,v^h+\theta(\hat{v}^h-v^h)}_t
 (\hat{v}^h_t-v^h_t) dt\right] d\theta \\
        &= \int_0^1 \frac{1}{1-\theta} \mathbb{E}\left[\int_0^T q^{h,v^h+\theta(\hat{v}^h-v^h)}_t
 \left(v^h_t+\theta(\hat{v}^h_t-v^h_t)-\hat{v}^h_t\right) dt\right] d\theta \\
 &= \int_0^1 \mathbb{E}\left[\int_0^T q^{h,\hat{v}^h}_t\left(v^h_t-\hat{v}^h_t \right)dt\right] d\theta  \\
 &\ +   \int_0^1 \frac{1}{1-\theta} \mathbb{E}\left[\int_0^T \left(q^{h,v^h+\theta(\hat{v}^h-v^h)}_t -q^{h,\hat{v}^h}_t \right)
 \left(v^h_t+\theta(\hat{v}^h_t-v^h_t)-\hat{v}^h_t\right) dt\right] d\theta \\
 &\leq \mathbb{E}\left[\int_0^T q^{h,\hat{v}^h}_t\left(v^h_t-\hat{v}^h_t \right)dt\right] - \int_0^1\alpha^V_1(1-\theta) \mathbb{E}\left[\int_0^T \left|v^h_t-\hat{v}^h_t\right|^2 dt \right] d\theta \\
 &=  \mathbb{E}\left[\int_0^T q^{h,\hat{v}^h}_t\left(v^h_t-\hat{v}^h_t \right)dt\right] - \frac{\alpha^V_1}{2}\mathbb{E}\left[\int_0^T \left|v^h_t-\hat{v}^h_t\right|^2 dt \right],
    \end{align*}
thereby establishing the strict concavity (more precisely, the $\alpha^V_1$-concavity) of the objective function.

Finally, for any fixed $\hat{v}^h$, using the square integrability of $q^{h,\hat{v}^h}$ and Young's inequality, 
    \begin{equation*}
          J^h(v^h) - J^h(\hat{v}^h) \leq \frac{1}{\alpha^V_1}\mathbb{E}\left[\int_0^T \left(q^{h,\hat{v}^h}_t \right)^2dt \right] - \frac{\alpha^V_1}{4}\mathbb{E}\left[\int_0^T \left| v^h_t-\hat{v}^h_t \right|^2dt \right] \to  -\infty
    \end{equation*}
as $\mathbb{E}[\int_0^T|v^h_t|^2dt] \to \infty$. Therefore, the objective function is coercive.

\subsection{Proof of Theorem \ref{thm:sol:p2}}
\label{sec:app:pf:thm:sol:p2}

Let $\hat{v}^h \in \mathcal{A}_{\mathbb{F}^h}(I)$ be an arbitrary strategy, and $\hat{x}^h$ be its associated wealth process. Given $(z^h)_{h=1}^H$ and $(\bar{v}^h)_{h=1}^H$, consider
    \begin{align}
    \label{eq:optimal:proof:1}
     &\  J^h(\hat{v}^h)-J^h(v^h) \nonumber \\
     = &\ \mathbb{E}\left[\int_0^T \left(f^h(t,\hat{x}^h_t,z^h_t,\hat{v}^h_t,\bar{v}^h_t)-f^h(t,x^h_t,z^h_t,v^h_t,\bar{v}^h_t) \right)dt + \left( g^h(\hat{x}^h_T,z^h_T)-g^h(x^h_T,z^h_T)\right)\right]\nonumber \\
     \leq&\ \mathbb{E}\Bigg[\int_0^T \left(f^h_x(t,x^h_t,z^h_t,v^h_t,\bar{v}^h_t)(\hat{x}^h_t-x^h_t) + f^h_v(t,x^h_t,z^h_t,v^h_t,\bar{v}^h_t)(\hat{v}^h_t-v^h_t) \right)dt \nonumber \\ 
     &\quad + g^h_x(x^h_T,z^h_T)(\hat{x}^h_T - x^h_T) \Bigg]\nonumber \\
      =&\ \mathbb{E}\Bigg[\int_0^T \left(f^h_x(t,x^h_t,z^h_t,v^h_t,\bar{v}^h_t)(\hat{x}^h_t-x^h_t) + f^h_v(t,x^h_t,z^h_t,v^h_t,\bar{v}^h_t)(\hat{v}^h_t-v^h_t) \right)dt -p^h_T(\hat{x}^h_T - x^h_T) \Bigg],
    \end{align}
where the inequality follows from the Assumption \ref{ass:concave:monotonicity:h}.B and Remark \ref{remark:concavity}. Notice that $\hat{x}^h-x^h$ satisfies the following SDE:
    \begin{equation*}
        d(\hat{x}^h_t-x^h_t) = \left(r(\hat{x}^h_t-x^h_t) - \kappa^h(\hat{v}^h_t-v^h_t) \right)dt - \sigma^h(\hat{v}^h_t-v^h_t)dW^h_t, \ \hat{x}^h_0 - x^h_0 = 0. 
    \end{equation*}
By applying It\^o's lemma to $p^h_t(\hat{x}^h_t-x^h_t)$, we obtain 
    \begin{align*}
        \mathbb{E}\left[p^h_T(\hat{x}^h_T - x^h_T)\right] &= \mathbb{E}\Bigg[\int_0^T \bigg[p^h_t\left(r(\hat{x}^h_t-x^h_t) - \kappa^h(\hat{v}^h_t-v^h_t) \right) \\
        &\qquad - (\hat{x}^h_t-x^h_t)\left( rp^h_t - f^h_x(t,x^h_t,z^h_t,v^h_t,\bar{v}^h_t)\right)  -\sigma^h\eta^h_t(\hat{v}^h_t-v^h_t) \bigg]dt       \Bigg] \\
        &= \mathbb{E}\left[\int_0^T\left( f^h_x(t,x^h_t,z^h_t,v^h_t,\bar{v}^h_t)(\hat{x}^h_t-x^h_t) -\left(\kappa^hp^h_t + \sigma^h\eta^h_t \right)\left(\hat{v}^h_t-v^h_t \right) \right)dt\right].
    \end{align*}
Substituting this into \eqref{eq:optimal:proof:1}, we obtain 
    \begin{align*}
         J^h(\hat{v}^h)-J^h(v^h)  &\leq  \mathbb{E}\Bigg[\int_0^T\left(   f^h_v(t,x^h_t,z^h_t,v^h_t,\bar{v}^h_t) + \kappa^hp^h_t + \sigma^h\eta^h_t  \right)\left(\hat{v}^h_t-v^h_t \right)  dt\Bigg].
    \end{align*}
Hence, we derive the variational inequality 
    \begin{equation}
    \label{eq:VI}
        \mathbb{E}\Bigg[\int_0^T\left(    f^h_v(t,x^h_t,z^h_t,v^h_t,\bar{v}^h_t) +\kappa^hp^h_t + \sigma^h\eta^h_t \right)\left(\hat{v}^h_t-v^h_t \right)  dt\Bigg] \leq 0,
    \end{equation}
which implies $J^h(\hat{v}^h) < J^h(v^h)$. The arbitrariness of $\hat{v}^h$ then suggests that $v^h$ is indeed the optimal control, whose existence is warranted by Lemma \ref{lem:cocercive}. 

Notice that \eqref{eq:VI} holds if 
    \begin{equation}
         \left( f^h_v(t,x^h_t,z^h_t,v^h_t,\bar{v}^h_t) +\kappa^hp^h_t + \sigma^h\eta^h_t \right)\left(\hat{v}^h_t-v^h_t \right)  \leq 0
    \end{equation}
for all $t\in[0,T]$. Since $v\mapsto f^h_v(t,x,z,\cdot,\bar{v})$ is strictly decreasing, the above inequality holds iff 
    \begin{equation}
         \left[ (f^h_v)^{-1}\left(-\left(\kappa^hp^h_t + \sigma^h\eta^h_t\right);t,x^h_t,z^h_t,\bar{v}^h_t\right) - v^h_t \right]\left(\hat{v}^h_t-v^h_t \right)  \leq 0.
    \end{equation}
By Lemma \ref{lem:convex}, we conclude that the solution of the inequality is given by  \eqref{eq:v*}. 

\section{Proofs of Assertions in Section \ref{sec:MFFBSDE:wellpose}}
\label{sec:app:pf:wellpose}
This section contains the proofs of statements in Section \ref{sec:MFFBSDE:wellpose}.


\subsection{Proof of Proposition \ref{pp:I-M>0}}
\label{app:lem:I-M>0}
We shall need the following lemmas:

	\begin{lemma}\label{lem:ab+ba}
		For any ${\bf a}, {\bf b} \in \mathbb{R}^d$,  
		\begin{equation*}
			\lambda_{\min}({\bf a}{\bf b}^{\top}+{\bf b}{\bf a}^{\top})= {\bf b}^{\top}{\bf a} - |{\bf a}||{\bf b}| \text{ and } \lambda_{\max}({\bf a}{\bf b}^{\top}+{\bf b}{\bf a}^{\top})= {\bf b}^{\top}{\bf a} + |{\bf a}||{\bf b}|.
		\end{equation*}
	\end{lemma}
	\begin{proof}
		Let ${\bf U}:={\bf a}{\bf b}^{\top}+{\bf b}{\bf a}^{\top}$. 	If either ${\bf a}$ or ${\bf b}$ is the zero vector, the claim is clearly true. Henceforth, we assume that both ${\bf a}$ and ${\bf b}$ are non-zero column vectors. \\

	\noindent\underline{Case 1: ${\bf a}$ and ${\bf b}$ are linearly dependent}\\
    In this case, there exists a non-zero constant $c$ such that ${\bf b} = c {\bf a}$. Hence, ${\bf U}=2c{\bf a}{\bf a}^{\top}$ and ${\bf U}$ has at most one non-zero eigenvalue, $2c|{\bf a}|^2$. If $c > 0$,   
		\begin{align*}
			&\lambda_{\min}({\bf U}) =0 = {\bf b}^{\top}{\bf a} - |{\bf a}||{\bf b}|,  \	\lambda_{\max}({\bf U}) =2c|{\bf a}|^2 = c|{\bf a}|^2 + |c||{\bf a}|^2 = {\bf b}^{\top}{\bf a} + |{\bf a}||{\bf b}|.
		\end{align*}
		If $c < 0$,
		\begin{align*}
			\lambda_{\min}({\bf U}) = 2c|{\bf a}|^2 = c|{\bf a}|^2 - |c||{\bf a}|^2 = {\bf b}^{\top}{\bf a} - |{\bf a}||{\bf b}|, \ \lambda_{\max}({\bf U}) = 0 = {\bf b}^{\top}{\bf a} + |{\bf a}||{\bf b}|.
		\end{align*}

	\noindent\underline{Case 2: ${\bf a}$ and ${\bf b}$ are linearly independent}\\
    Let $\mathcal{S}:=\text{span}\{{\bf a},{\bf b}\}$ and $\mathcal{S}^{\perp}$ be its orthogonal complement. Since ${\bf a}$ and ${\bf b}$ are column vectors, we have $\text{rank}({\bf U}) \leq 2$ which implies that ${\bf U}$ has at most two non-zero eigenvalues. Note that, for any ${\bf x} \in \mathcal{S}^{\perp}$, we have ${\bf U}{\bf x} = 0$. Therefore, the eigenvectors corresponding to the non-zero eigenvalues of ${\bf U}$ belong to $\mathcal{S}$. Since
		\begin{equation*}
			{\bf U} {\bf a} = {\bf a}{\bf b}^{\top}{\bf a}+{\bf b}|{\bf a}|^2  \text{ and } {\bf U} {\bf b} = {\bf a}|{\bf b}|^2+{\bf b}{\bf a}^{\top}{\bf b},
		\end{equation*}
		the linear transform ${\bf U}$ in the basis $\{{\bf a},{\bf b}\}$ can be represented as a $2\times 2$ matrix  ${\bf U}_{{\bf a},{\bf b}}$, where 
		\begin{equation*}
		{\bf U}_{{\bf a},{\bf b}}  = \begin{pmatrix}
				{\bf b}^{\top}{\bf a} & |{\bf b}|^2\\
				|{\bf a}|^2 & {\bf a}^{\top}{\bf b}
			\end{pmatrix}.
		\end{equation*}
	
        The characteristic equation for   ${\bf U}_{{\bf a},{\bf b}}$  is $({\bf b}^{\top}{\bf a} - \lambda)^2 - |{\bf a}|^2|{\bf b}|^2 = 0$, which has solutions $\lambda_{\pm} = {\bf b}^{\top}{\bf a} \pm |{\bf a}||{\bf b}|$. By the Cauchy-Schwarz inequality, we have $\lambda_{-} \leq 0 \leq \lambda_{+}$. Therefore, $\lambda_{\min}({\bf U}) = \lambda_{-} = {\bf b}^{\top}{\bf a} - |{\bf a}||{\bf b}|$ and $\lambda_{\max}({\bf U}) = \lambda_{+} = {\bf b}^{\top}{\bf a} + |{\bf a}||{\bf b}|$.
		
		The desired result follows by combining the two cases. 
	\end{proof}

\begin{lemma}\label{lem:piw<2}
	Conditions 1-3 in Proposition \ref{pp:I-M>0} are equivalent.	
\end{lemma}
\begin{proof}
       We first prove the equivalence of Conditions 1 and 3. 
        Using the identity ${\bf I}-{\bf M}^\top = {\bf K}({\bf K}-{\bf \Pi})^{-1}$ and \eqref{eq:Pi-K:inv}, we have 
        \begin{equation}
        \label{eq:I-M:top}
            {\bf I} - {\bf M}^{\top} = {\bf I} + \frac{\boldsymbol{\pi} \boldsymbol{\upsilon}^{\top}{\bf K}^{-1}}{1-\boldsymbol{\upsilon}^{\top}{\bf K}^{-1}\boldsymbol{\pi} },
        \end{equation}
        where $\boldsymbol{\pi}$ and $\boldsymbol{\upsilon}$ are defined as in \eqref{eq:pi:v}. By considering the symmetrization of ${\bf I}-{\bf M}^\top$, we have 
            \begin{equation*}
                \lambda_{\min}({\bf I}-{\bf M}) = \lambda_{\min}({\bf I}-{\bf M}^\top) = 1 + \frac{\lambda_{\min}(\boldsymbol{\pi} \boldsymbol{\upsilon}^{\top}{\bf K}^{-1}+{\bf K}^{-1} \boldsymbol{\upsilon}\boldsymbol{\pi}^{\top})}{2-2\boldsymbol{\upsilon}^{\top}{\bf K}^{-1}\boldsymbol{\pi}} . 
            \end{equation*}
        By Lemma \ref{lem:ab+ba}, we have
        \begin{align*}
        	\lambda_{\min}(\boldsymbol{\pi} \boldsymbol{\upsilon}^{\top}{\bf K}^{-1}+{\bf K}^{-1} \boldsymbol{\upsilon}\boldsymbol{\pi}^{\top})&= \boldsymbol{\upsilon}^{\top}{\bf K}^{-1}\boldsymbol{\pi} - |\boldsymbol{\pi}||{\bf K}^{-1}\boldsymbol{\upsilon}| \\
        	&= \sum_{h=1}^H \pi^h\omega^h\frac{\kappa^h-d}{\kappa^h} - \sqrt{\left(\sum_{h=1}^H (\pi^h)^2\right)\left(\sum_{h=1}^H \left(\omega^h\frac{\kappa^h-d}{\kappa^h}\right)^2\right)}. 
        \end{align*}
        Thus
        \begin{equation}
        \label{eq:lambda:min:I-M}
        	\lambda_{\min}\left({\bf I} - {\bf M}^{\top}\right) = \frac{2-\sum_{h=1}^H \pi^h\omega^h\frac{\kappa^h-d}{\kappa^h}-\sqrt{\left(\sum_{h=1}^H (\pi^h)^2\right)\left(\sum_{h=1}^H \left(\omega^h\frac{\kappa^h-d}{\kappa^h}\right)^2\right)}}{2-2\sum_{h=1}^H \pi^h\omega^h\frac{\kappa^h-d}{\kappa^h}},
        \end{equation}
        where the denominator $2-2\sum_{h=1}^H \pi^h\omega^h\frac{\kappa^h-d}{\kappa^h} > 0$, since $\kappa^h>d\geq 0$ for all $h\in[H]$, and $\sum_{h=1}^H\pi^h\omega^h = 1$. It is then easy to see $\lambda_{\min}\left({\bf I} - {\bf M}^{\top}\right)  > 0$ if and only if Condition 3 holds. 
        
         Next, we prove the equivalence of Conditions 2 and 3. 
         Using Lemma \ref{lem:ab+ba}, we have
        \begin{align*}
		& 2\lambda_{\max}({\bf \Pi}{\bf K}^{-1}) = \lambda_{\max}(\boldsymbol{\pi} \boldsymbol{\upsilon}^{\top}{\bf K}^{-1} +  {\bf K}^{-1}\boldsymbol{\upsilon}\boldsymbol{\pi}^{\top}) =   \boldsymbol{\upsilon}^{\top}{\bf K}^{-1}\boldsymbol{\pi} +  |{\bf K}^{-1}\boldsymbol{\upsilon}||\boldsymbol{\pi}|\\
		=& \sum_{h=1}^H \pi^h\omega^h\frac{\kappa^h-d}{\kappa^h} + \sqrt{\left(\sum_{h=1}^H (\pi^h)^2\right)\left( \sum_{h=1}^H \left(\omega^h\frac{\kappa^h-d}{\kappa^h}\right)^2\right)}.
	\end{align*}
        Therefore,  $\lambda_{\max}({\bf \Pi}{\bf K}^{-1}) < 1$ if and only if Condition 3 holds.
\end{proof}

\begin{lemma}\label{lem:uniformofpi/w}
	Condition 4 of Proposition \ref{pp:I-M>0} implies Conditions 1-3 of the same proposition. 
\end{lemma}
\begin{proof}
    Given Condition 4 of Proposition \ref{pp:I-M>0}, there exists a $c>0$ such that $\frac{\pi^h}{\omega^h} < c < \frac{\pi^h}{\omega^h} \frac{\kappa^h}{\kappa^h-d}$ for all $h\in[H]$. Let $\delta^h  := \frac{\pi^h}{c \omega^h}$, which satisfies $\frac{\kappa^h-d}{\kappa^h} < \delta^h < 1$. Hence, we have 
\begin{align*}
	& \ \ \ \ \sum_{h=1}^H \pi^h\omega^h\frac{\kappa^h-d}{\kappa^h} + \sqrt{\left(\sum_{h=1}^H (\pi^h)^2\right)\left( \sum_{h=1}^H \left(\omega^h\frac{\kappa^h-d}{\kappa^h}\right)^2\right)} \\
    &<  \sum_{h=1}^H \pi^h\omega^h\delta^h + \sqrt{\left(\sum_{h=1}^H (\pi^h)^2\right)\left(\sum_{h=1}^H \left(\omega^h\delta^h\right)^2\right)}.
\end{align*}
By the Cauchy–Schwarz inequality and the fact that $\pi^h = c \omega^h \delta^h $, we have
    \begin{align*}
    	\sum_{h=1}^H \pi^h\omega^h\delta^h +  \sqrt{\left(\sum_{h=1}^H (\pi^h)^2 \right)\left(\sum_{h=1}^H \left(\omega^h\delta^h\right)^2\right)} = 2\sum_{h=1}^H \pi^h\omega^h\delta^h< 2 \sum_{h=1}^H \pi^h\omega^h = 2.
    \end{align*}
    Therefore, Condition 4 implies Condition 3. By Lemma \ref{lem:piw<2}, the proof is complete.
\end{proof}

\subsection{Proof of Lemma \ref{lem:exist:general}}
\label{sec:app:pf:lem:exist}
For $\mu \in[0,1]$, we define the operator $\Psi_{\mu+\delta}({\bf x},{\bf p}, \boldsymbol{\eta}) = (\hat{\bf x},\hat{\bf p}, \hat{\boldsymbol{\eta}})$, where the latter is the solution of the parametrized system \eqref{eq:MFFBSDE:general:parametrized} with 
    \begin{align*}
        \boldsymbol{\phi}_t &= \delta\left( \hide{\beta_2} {\bf p}_t + (r{\bf x}_t + {\bf l} - {\bf K}{\bf v}_t + {\bf \Pi}\mathbb{E}[{\bf v}_t] \right) + \hat{\boldsymbol{\phi}}_t, \\
        \boldsymbol{\psi}_t &= \delta\left( \hide{\beta_2}\text{diag}(\boldsymbol{\eta}_t) + {\bf \Sigma}({\bf I}-\text{diag}({\bf v}_t)) \right) + \hat{\boldsymbol{\psi}}_t, \\
        \boldsymbol{\xi}_t &= \delta\left( -\hide{\beta_1}{\bf x}_t + r{\bf p}_t - \partial_{\bf x}{\bf F}(t,{\bf x}_t,{\bf z}_t,{\bf v}_t,\mathbb{E}[{\bf v}_t])   \right) + \hat{\boldsymbol{\xi}}_t, \\
        \boldsymbol{\zeta}_T &= -\delta \left[\partial_{\bf x}{\bf G}({\bf x}_T,{\bf z}_T) + {\bf x}_T \right] + \hat{\boldsymbol{\zeta}}_T,
    \end{align*}
Here, $\delta>0$ is a small positive constant to be chosen independently of $\mu$, $\hat{\boldsymbol{\phi}}, \hat{\boldsymbol{\xi}} \in L^2_{\mathbb{F}^{[H]}}([0,T];\mathbb{R}^H)$, $\hat{\boldsymbol{\psi}}\in L^2_{\mathbb{F}^h}([0,T];\mathbb{R}^H\times \mathbb{R}^H)$, $\hat{\boldsymbol{\zeta}}_T\in L^2(\Omega,\mathcal{F}_T,\mathbb{P})$, and $({\bf z}_t)_{t\in[0,T]}$, $({\bf v}_t)_{t\in[0,T]}$ are given by 
    \begin{equation*}
     {\bf z}_t = \mathbb{E}[{\bf x}_t], \    {\bf v}_t = \text{Proj}_{I^H}\left[ \left(\partial_{\bf v} {\bf F}\right)^{-1}\left(-\left({\bf K}{\bf p}_t + {\bf \Sigma} \boldsymbol{\eta}_t\right) ;t, {\bf x}_t,{\bf z}_t,\mathbb{E}[{\bf v}_t]\right) \right]. 
    \end{equation*}

Suppose that the system \eqref{eq:MFFBSDE:general:parametrized} admits a solution for some $\mu_0\in[0,1)$. Let $\Psi_{\mu_0+\delta}({\bf x}^i,{\bf p}^i, \boldsymbol{\eta}^i) = (\hat{\bf x}^i,\hat{\bf p}^i, \hat{\boldsymbol{\eta}}^i)$, $i=1,2$. Let also $\tilde{{\bf x}} := {\bf x}^1-{\bf x}^2$, $\tilde{{\bf p}} := {\bf p}^1-{\bf p}^2$, $\tilde{\boldsymbol{\eta}} := \boldsymbol{\eta}^1-\boldsymbol{\eta}^2$, $\tilde{{\bf v}} := {\bf v}^1-{\bf v}^2$; $\tilde{\hat{{\bf x}}} := \hat{{\bf x}}^1-\hat{{\bf x}}^2$, $\tilde{\hat{{\bf p}}} := \hat{{\bf p}}^1-\hat{{\bf p}}^2$, $\tilde{\hat{\boldsymbol{\eta}}} := \hat{\boldsymbol{\eta}}^1-\hat{\boldsymbol{\eta}}^2$, $\tilde{\hat{{\bf v}}} := \hat{{\bf v}}^1-\hat{{\bf v}}^2$. We shall show that $\Psi_{\mu_0+\delta}$ is a contraction for any sufficiently small $\delta>0$ independent of $\mu_0$. Consequently, by the Banach fixed point theorem,  one can deduce that the operator $\Psi_{\mu_0+\delta}$ admits a fixed point, which is indeed the solution of  \eqref{eq:MFFBSDE:general:parametrized} with $\mu=\mu_0+\delta$.

By applying It\^o's lemma to $\langle \tilde{\hat{{\bf x}}}_t, \tilde{\hat{{\bf p}}}_t \rangle$ and using Assumption \ref{ass:concave:monotonicity:h}, we have 
 \begin{align}
    \label{eq:exist:general:proof:1}
       & \mu_0 \mathbb{E}\left[ \left\langle  \tilde{\hat{{\bf x}}}_T, -\left( \partial_{\bf x}{\bf G}(\hat{{\bf x}}^1_T,\hat{{\bf z}}^1_T) - \partial_{\bf x}{\bf G}(\hat{{\bf x}}^2_T,\hat{{\bf z}}^2_T) \right)  \right\rangle  \right] + (1-\mu_0) \mathbb{E}\left[\left|  \tilde{\hat{{\bf x}}}_T\right|^2\right] \nonumber \\
       &\quad - \delta \mathbb{E}\left[\left\langle  \tilde{\hat{{\bf x}}}_T, \left( \partial_{\bf x}{\bf G}({\bf x}^1_T,{\bf z}^1_T) - \partial_{\bf x}{\bf G}({\bf x}^2_T,{\bf z}^2_T)  \right) + \tilde{{\bf x}}_T \right\rangle \right] \nonumber \\
       &= \mu_0\mathbb{E}\left[\int_0^T \left\langle  \tilde{\hat{{\bf x}}}_t , \partial_{\bf x} {\bf F}(\hat{{\bf x}}^1_t,\hat{{\bf z}}^1_t, \hat{{\bf v}}^1_t, \mathbb{E}[\hat{{\bf v}}^1_t]) - \partial_{\bf x} {\bf F}(\hat{{\bf x}}^2_t,\hat{{\bf z}}^2_t, \hat{{\bf v}}^2_t, \mathbb{E}[\hat{{\bf v}}^2_t]) \right\rangle dt \right] \nonumber \\
       &\ - \mu_0 \mathbb{E}\left[\int_0^T \left\langle  \tilde{\hat{{\bf v}}}_t ,  {\bf K} \tilde{\hat{{\bf p}}}_t +{\bf \Sigma} \tilde{\hat{\boldsymbol{\eta}}}_t \right\rangle dt \right]  + \mu_0 \mathbb{E}\left[\int_0^T \left\langle  \tilde{\hat{{\bf p}}}_t , {\bf \Pi} \mathbb{E}[ \tilde{\hat{{\bf v}}}_t] \right\rangle dt\right]\nonumber \\
       &\ - (1-\mu_0) \mathbb{E}\left[\int_0^T\left( \hide{\beta_1} \left| \tilde{\hat{{\bf x}}}_t \right|^2 + \hide{\beta_2}\left|  \tilde{\hat{{\bf p}}}_t\right|^2 + \hide{\beta_2}\left|  \tilde{\hat{\boldsymbol{\eta}}}_t\right|^2\right) dt\right] \nonumber \\
       &\ -\delta \mathbb{E}\left[\int_0^T \left\langle  \tilde{\hat{{\bf x}}}_t, -\hide{\beta_1}\tilde{{\bf x}}_t + r\tilde{{\bf p}}_t - \left(\partial_{\bf x} {\bf F}({\bf x}^1_t,{\bf z}^1_t,{\bf v}^1_t,\mathbb{E}[{\bf v}^1_t]) - \partial_{\bf x} {\bf F}({\bf x}^1_t,{\bf z}^1_t,{\bf v}^1_t,\mathbb{E}[{\bf v}^1_t])\right) \right\rangle dt \right] \nonumber \\
       &  \  + \delta  \mathbb{E}\left[\int_0^T \left\langle  \tilde{\hat{{\bf p}}}_t, \hide{\beta_2} \tilde{{\bf p}}_t + r\tilde{{\bf x}}_t -{\bf K}\tilde{{\bf v}}_t + {\bf \Pi} \mathbb{E}[\tilde{{\bf v}}_t]  \right\rangle dt \right] +  \delta  \mathbb{E}\left[\int_0^T \left\langle  \tilde{\hat{\boldsymbol{\eta}}}_t, \hide{\beta_2} \tilde{\boldsymbol{\eta}}_t - {\bf \Sigma}\Tilde{{\bf v}}_t \right\rangle dt \right] \nonumber \\
       &\leq \mu_0\mathbb{E}\left[\int_0^T \left\langle  \tilde{\hat{{\bf x}}}_t , \nabla {\bf F}_{\bf x}(\hat{{\bf x}}^1_t,\hat{{\bf z}}^1_t, \hat{{\bf v}}^1_t, \mathbb{E}[\hat{{\bf v}}^1_t]) - \nabla {\bf F}_{\bf x}(\hat{{\bf x}}^2_t,\hat{{\bf z}}^2_t, \hat{{\bf v}}^2_t, \mathbb{E}[\hat{{\bf v}}^2_t]) \right\rangle dt \right] \nonumber \\
       &\ - \mu_0 \mathbb{E}\left[\int_0^T \left\langle  \tilde{\hat{{\bf v}}}_t ,  {\bf K} \tilde{\hat{{\bf p}}}_t +{\bf \Sigma} \tilde{\hat{\boldsymbol{\eta}}}_t \right\rangle dt \right]  + \mu_0 \mathbb{E}\left[\int_0^T \left\langle  \tilde{\hat{{\bf p}}}_t , {\bf \Pi} \mathbb{E}[ \tilde{\hat{{\bf v}}}_t] \right\rangle dt\right]\nonumber \\
       &\ - (1-\mu_0) \mathbb{E}\left[\int_0^T\left( \hide{\beta_1} \left| \tilde{\hat{{\bf x}}}_t \right|^2 + \hide{\beta_2}\left|  \tilde{\hat{{\bf p}}}_t\right|^2 + \hide{\beta_2}\left|  \tilde{\hat{\boldsymbol{\eta}}}_t\right|^2\right) dt\right] \nonumber \\
       &\ + K_T\delta\left( \mathbb{E}\left[\int_0^T \left(\left|\tilde{\hat{{\bf x}}}_t \right|^2 + \left|\tilde{\hat{{\bf p}}}_t \right|^2 + \left|\tilde{\hat{\boldsymbol{\eta}}}_t \right|^2 + \left|\tilde{{\bf x}}_t \right|^2 + \left|\tilde{{\bf p}}_t \right|^2 + \left|\tilde{\boldsymbol{\eta}}_t \right|^2 \right)dt \right] \right),
    \end{align}
where $K_T>0$ is a generic constant depending solely on $T\hide{\beta_1}$, which changes from line to line in the subsequent calculations. 

We estimate the terms on the right-hand side of \eqref{eq:exist:general:proof:1}. Following the proof of \eqref{eq:bad:1} in Theorem \ref{thm:unique:general}, one can show that 
    \begin{equation}
    \label{eq:bad:1:proof:exist}
         \mathbb{E}\left[\int_0^T \left\langle  \tilde{\hat{{\bf v}}}_t ,  {\bf K} \tilde{\hat{{\bf p}}}_t +{\bf \Sigma} \tilde{\hat{\boldsymbol{\eta}}}_t \right\rangle dt \right]  \geq 0. 
    \end{equation}
Next, we  estimate the term
    \begin{equation*}
        \mathbb{E}\left[\int_0^T \left\langle  \tilde{\hat{{\bf p}}}_t , {\bf \Pi} \mathbb{E}[ \tilde{\hat{{\bf v}}}_t] \right\rangle dt\right]. 
    \end{equation*}
By considering the dynamics of $\langle {\bf M} \tilde{\hat{\bf z}}_t , \mathbb{E}[\tilde{\hat{{\bf p}}}_t] \rangle$, we obtain 
 \begin{align*}
      &\  \mu_0 \mathbb{E}\left[\left\langle {\bf M} \Tilde{\hat{{\bf z}}}_T, -\left(\partial_{\bf x}{\bf G}(\hat{{\bf x}}^1_T,\hat{{\bf x}}^1_T)-\partial_{\bf x}{\bf G}(\hat{{\bf x}}^2_T,\hat{{\bf x}}^2_T) \right)  \right\rangle\right] +(1-\mu_0)\left\langle {\bf M} \Tilde{\hat{{\bf z}}}_T,  \Tilde{\hat{{\bf z}}}_T  \right\rangle\nonumber \\
      &\quad + \delta \mathbb{E}\left[\left\langle {\bf M}\Tilde{\hat{{\bf z}}}_T,  -\left(\partial_{{\bf x}}{\bf G}({\bf x}^1_T,{\bf z}^1_T)-\partial_{{\bf x}}{\bf G}({\bf x}^2_T,{\bf z}^2_T) + \tilde{{\bf z}}_T \right) \right\rangle \right]\nonumber \\
      =&\ -(1-\mu_0)  \int_0^T \left( \hide{\beta_2}\left\langle  {\bf M}\mathbb{E}[\Tilde{\hat{{\bf p}}}_t],\mathbb{E}[\Tilde{\hat{{\bf p}}}_t]   \right\rangle + \hide{\beta_1} \left\langle {\bf M}\tilde{\hat{{\bf z}}}_t  , \tilde{\hat{{\bf z}}}_t\right\rangle  \right) dt \nonumber\\
      &\quad + \mu_0 \int_0^T \left\langle \mathbb{E}[\tilde{\hat{{\bf p}}}_t], {\bf \Pi} \mathbb{E}[\tilde{\hat{\bf v}}_t]  \right\rangle dt \nonumber \\
      &\ -\mu_0 \int_0^T   \left\langle {\bf M}\tilde{\hat{{\bf z}}}_t ,  -\left( \partial_{{\bf x}}{\bf F}(t,\hat{{\bf x}}^1_t,\hat{{\bf z}}^1_t,\hat{{\bf v}}^1_t,\mathbb{E}[\hat{{\bf v}}^1_t]) -  \partial_{{\bf x}}{\bf F}(t,\hat{{\bf x}}^2_t,\hat{{\bf z}}^2_t,\hat{{\bf v}}^2_t,\mathbb{E}[\hat{{\bf v}}^2_t]) \right) \right\rangle  dt \nonumber\\
      &\ + \delta \int_0^T \bigg(\left\langle  \mathbb{E}[\Tilde{\hat{{\bf p}}}_t], {\bf M} \left(\hide{\beta_2}\mathbb{E}[\Tilde{{\bf p}}_t] + r \tilde{{\bf z}}_t + {\bf \Pi} \mathbb{E}[\tilde{{\bf v}}_t] \right)\right\rangle  \nonumber\\
      &\qquad + \left\langle {\bf M}\tilde{\hat{{\bf z}}}_t, \hide{\beta_1} \tilde{{\bf z}}_t - r\mathbb{E}[\tilde{\bf p}_t]  + \partial_{\bf x}{\bf F}({\bf x}^1_t,{\bf z}^1_t,{\bf v}^1_t,\mathbb{E}[{\bf v}^1_t])-\partial_{\bf x}{\bf F}({\bf x}^1_t,{\bf z}^1_t,{\bf v}^1_t,\mathbb{E}[{\bf v}^1_t]) \right\rangle \bigg) dt \nonumber \\
      \geq&\ -(1-\mu_0)  \int_0^T \left( \hide{\beta_2}\left\langle  {\bf M}\mathbb{E}[\Tilde{\hat{{\bf p}}}_t],\mathbb{E}[\Tilde{\hat{{\bf p}}}_t]   \right\rangle + \hide{\beta_1} \left\langle {\bf M}\tilde{\hat{{\bf z}}}_t  , \tilde{\hat{{\bf z}}}_t\right\rangle  \right) dt \nonumber\\
      &\ + \mu_0 \int_0^T \left\langle \mathbb{E}[\tilde{\hat{{\bf p}}}_t], {\bf \Pi} \mathbb{E}[\tilde{\hat{\bf v}}_t]  \right\rangle dt \nonumber \\
      &\ -\mu_0 \int_0^T   \left\langle {\bf M}\tilde{\hat{{\bf z}}}_t ,  -\mathbb{E}\left[ \partial_{{\bf x}}{\bf F}(t,\hat{{\bf x}}^1_t,\hat{{\bf z}}^1_t,\hat{{\bf v}}^1_t,\mathbb{E}[\hat{{\bf v}}^1_t]) -  \partial_{{\bf x}}{\bf F}(t,\hat{{\bf x}}^2_t,\hat{{\bf z}}^2_t,\hat{{\bf v}}^2_t,\mathbb{E}[\hat{{\bf v}}^2_t]) \right] \right\rangle  dt \nonumber\\
      &\ - K_T\delta   \mathbb{E}\left[\int_0^T \left(\left|\tilde{\hat{{\bf x}}}_t \right|^2 + \left|\tilde{\hat{{\bf p}}}_t \right|^2+ \left|\tilde{{\bf x}}_t \right|^2 + \left|\tilde{{\bf p}}_t \right|^2 + \left|\tilde{\boldsymbol{\eta}}_t \right|^2 \right)dt \right].
    \end{align*}
Rearranging yields 
    \begin{align}
         \label{eq:bad:2:proof:exist}
       &\   \mu_0 \int_0^T \left\langle \mathbb{E}[\tilde{\hat{{\bf p}}}_t], {\bf \Pi} \mathbb{E}[\tilde{\hat{\bf v}}_t]  \right\rangle dt \nonumber \\
       \leq &\ \mu_0 \mathbb{E}\left[\left\langle {\bf M} \Tilde{\hat{{\bf z}}}_T, -\left(\partial_{\bf x}{\bf G}(\hat{{\bf x}}^1_T,\hat{{\bf x}}^1_T)-\partial_{\bf x}{\bf G}(\hat{{\bf x}}^2_T,\hat{{\bf x}}^2_T) \right)  \right\rangle\right] +(1-\mu_0)\left\langle {\bf M} \Tilde{\hat{{\bf z}}}_T,  \Tilde{\hat{{\bf z}}}_T  \right\rangle\nonumber \\
      &\ + \delta \mathbb{E}\left[\left\langle {\bf M}\Tilde{\hat{{\bf z}}}_T,  -\left(\partial_{{\bf x}}{\bf G}({\bf x}^1_T,{\bf z}^1_T)-\partial_{{\bf x}}{\bf G}({\bf x}^2_T,{\bf z}^2_T) +\tilde{{\bf z}}_T \right) \right\rangle \right] \nonumber \\
      &\ + (1-\mu_0)  \int_0^T \left( \hide{\beta_2}\left\langle  {\bf M}\mathbb{E}[\Tilde{\hat{{\bf p}}}_t],\mathbb{E}[\Tilde{\hat{{\bf p}}}_t]   \right\rangle + \hide{\beta_1} \left\langle {\bf M}\tilde{\hat{{\bf z}}}_t  , \tilde{\hat{{\bf z}}}_t\right\rangle  \right) dt \nonumber \\
      &\ -\mu_0 \int_0^T   \left\langle {\bf M}\tilde{\hat{{\bf z}}}_t ,  \mathbb{E}\left[ \partial_{{\bf x}}{\bf F}(t,\hat{{\bf x}}^1_t,\hat{{\bf z}}^1_t,\hat{{\bf v}}^1_t,\mathbb{E}[\hat{{\bf v}}^1_t]) -  \partial_{{\bf x}}{\bf F}(t,\hat{{\bf x}}^2_t,\hat{{\bf z}}^2_t,\hat{{\bf v}}^2_t,\mathbb{E}[\hat{{\bf v}}^2_t]) \right] \right\rangle  dt \nonumber \\
      &\ +  K_T\delta   \mathbb{E}\left[\int_0^T \left(\left|\tilde{\hat{{\bf x}}}_t \right|^2 + \left|\tilde{\hat{{\bf p}}}_t \right|^2+ \left|\tilde{{\bf x}}_t \right|^2 + \left|\tilde{{\bf p}}_t \right|^2 + \left|\tilde{\boldsymbol{\eta}}_t \right|^2 \right)dt \right]. 
    \end{align}

  Following the derivation of \eqref{eq:proof:matrix:concavity}, using  Assumptions \ref{ass:concave:monotonicity:h}, \ref{ass:M}, and substituting \eqref{eq:bad:1:proof:exist}-\eqref{eq:bad:2:proof:exist} into \eqref{eq:exist:general:proof:1},  we arrive at
    \begin{align}
    \label{eq:exist:general:pf}
  &\   K_T\delta  \mathbb{E}\left[ |\tilde{\hat{{\bf x}}}_T|^2 + |\tilde{{\bf x}}|_T^2+   \int_0^T \left(\left|\tilde{\hat{{\bf x}}}_t \right|^2 + \left|\tilde{\hat{{\bf p}}}_t \right|^2 + \left|\tilde{\hat{\boldsymbol{\eta}}}_t \right|^2 + \left|\tilde{{\bf x}}_t \right|^2 + \left|\tilde{{\bf p}}_t \right|^2 + \left|\tilde{\boldsymbol{\eta}}_t \right|^2 \right)dt \right]  \nonumber \\
      \geq&\  K_T\delta\left( \mathbb{E}\left[\int_0^T \left(\left|\tilde{\hat{{\bf x}}}_t \right|^2 + \left|\tilde{\hat{{\bf p}}}_t \right|^2 + \left|\tilde{\hat{\boldsymbol{\eta}}}_t \right|^2 + \left|\tilde{{\bf x}}_t \right|^2 + \left|\tilde{{\bf p}}_t \right|^2 + \left|\tilde{\boldsymbol{\eta}}_t \right|^2 \right)dt \right] \right) \nonumber \\
      &\ + \delta \mathbb{E}\left[\left\langle  \tilde{\hat{{\bf x}}}_T-{\bf M} \tilde{\hat{{\bf z}}}_T, \partial_{\bf x}{\bf G}({\bf x}^1_T,{\bf z}^1_T) - \partial_{\bf x}{\bf G}({\bf x}^2_T,{\bf z}^2_T)  \right\rangle \right] \nonumber \\
       \geq&\ \mu_0 \mathbb{E}\left[ \left\langle  \tilde{\hat{{\bf x}}}_T- {\bf M}\tilde{\hat{{\bf z}}}_T , -\left( \partial_{\bf x}{\bf G}(\hat{{\bf x}}^1_T,\hat{{\bf z}}^1_T) - \partial_{\bf x}{\bf G}(\hat{{\bf x}}^2_T,\hat{{\bf z}}^2_T) \right)  \right\rangle  \right] \nonumber\\
       &\ + (1-\mu_0)\left(\mathbb{E}\left[\left|\tilde{\hat{{\bf x}}}_T\right|^2\right] - \left\langle {\bf M}\tilde{\hat{{\bf z}}}_T,  \tilde{\hat{{\bf z}}}_T\right\rangle \right)\nonumber \\
       &\ - \mu_0\mathbb{E}\left[\int_0^T \left\langle  \tilde{\hat{{\bf x}}}_t -{\bf M}\tilde{\hat{{\bf z}}}_t, \partial_{\bf x} {\bf F}(t,\hat{{\bf x}}^1_t,\hat{{\bf z}}^1_t, \hat{{\bf v}}^1_t, \mathbb{E}[\hat{{\bf v}}^1_t]) - \partial_{\bf x} {\bf F}(t,\hat{{\bf x}}^2_t,\hat{{\bf z}}^2_t, \hat{{\bf v}}^2_t, \mathbb{E}[\hat{{\bf v}}^2_t]) \right\rangle dt \right] \nonumber\\
       &\ + (1-\mu_0) \mathbb{E}\left[\int_0^T\left( \hide{\beta_1} \left| \tilde{\hat{{\bf x}}}_t \right|^2 + \hide{\beta_2}\left|  \tilde{\hat{{\bf p}}}_t\right|^2 + \hide{\beta_2}\left|  \tilde{\hat{\boldsymbol{\eta}}}_t\right|^2\right) dt\right]\nonumber\\ 
       &\ - (1-\mu_0)  \int_0^T \left( \hide{\beta_2}\left\langle  {\bf M}\mathbb{E}[\Tilde{\hat{{\bf p}}}_t],\mathbb{E}[\Tilde{\hat{{\bf p}}}_t]   \right\rangle + \hide{\beta_1} \left\langle {\bf M}\tilde{\hat{{\bf z}}}_t  , \tilde{\hat{{\bf z}}}_t\right\rangle  \right) dt \nonumber \\
       \geq&\ \left[\mu_0 \alpha_{\bf M}^{\bf G} + (1-\mu_0)\min\{\lambda_{\min}({\bf I}-{\bf M}),1\}\right] \mathbb{E}\left[\left|\tilde{\hat{{\bf x}}}_T\right|^2\right] \nonumber\\
       &\ + \hide{\beta_1}\left[\mu_0 {\alpha}_{\bf M} + (1-\mu_0)\min\{\lambda_{\min}({\bf I}-{\bf M}),1\}\right]
        \mathbb{E}\left[\int_0^T \left|\tilde{\hat{{\bf x}}}_t\right|^2 dt  \right]\nonumber \\
       &\ +(1-\mu_0) \mathbb{E}\left[\int_0^T\left(    \hide{\beta_2} \min \{\lambda_{\min}({\bf I}-{\bf M}),1\}\left|  \tilde{\hat{{\bf p}}}_t\right|^2 + \hide{\beta_2}\left|  \tilde{\hat{\boldsymbol{\eta}}}_t\right|^2\right) dt\right],
    \end{align}
Note that the last inequality is a consequence of Lemma \ref{lem:EX:Z:inequality}.  
    
Next, we estimate 
    \begin{equation*}
        \mathbb{E}\left[\int_0^T \left(\left|\tilde{\hat{{\bf p}}}_t \right|^2 + \left|\tilde{\hat{\boldsymbol{\eta}}}_t \right|^2 \right)dt   \right]. 
    \end{equation*}
By applying It\^o's lemma to $|\tilde{\hat{{\bf p}}}_t|^2$, we obtain 
    \begin{align}
    \label{eq:exist:general:proof:7}
      & \  \mathbb{E}\bigg[\bigg|-\mu_0\left(\partial_{\bf x}{\bf G}(\hat{{\bf x}}^1_T, \hat{{\bf z}}^1_T) - \partial_{\bf x}{\bf G}(\hat{{\bf x}}^2_T, \hat{{\bf z}}^2_T) \right)  + (1-\mu_0) \tilde{\hat{{\bf x}}}_T \nonumber \\
      &\quad - \delta \left(\partial_{\bf x}{\bf G}({\bf x}^1_T, {\bf z}^1_T) - \partial_{\bf x}{\bf G}({\bf x}^2_T, {\bf z}^2_T) + \tilde{{\bf x}}_T \right)    \bigg|^2\bigg] - \mathbb{E}\left[\left|\tilde{\hat{{\bf p}}}_t\right|^2\right] \nonumber \\
      =&\ 2\mu_0 \mathbb{E}\left[\int_t^T \left( r\left|\tilde{\hat{{\bf p}}}_s\right|^2 + \left\langle \tilde{\hat{{\bf p}}}_s , \partial_{\bf x}{\bf F}(s,{\hat{{\bf x}}}^1_s,{\hat{{\bf z}}}^1_s,{\hat{{\bf v}}}^1_s,\mathbb{E}[{\hat{{\bf v}}}^1_s]) - \partial_{\bf x}{\bf F}(s,{\hat{{\bf x}}}^2_s,{\hat{{\bf z}}}^2_s,{\hat{{\bf v}}}^2_s,\mathbb{E}[{\hat{{\bf v}}}^2_s]) \right\rangle \right)ds  \right]\nonumber \\
      &\quad - 2(1-\mu_0)\hide{\beta_1} \mathbb{E}\left[\int_t^T \left\langle \tilde{\hat{{\bf p}}}_s , \tilde{\hat{{\bf x}}}_s \right\rangle ds\right] +  \mathbb{E}\left[\int_t^T \left|\tilde{\hat{\boldsymbol{\eta}}}_s \right|^2 ds \right] \nonumber \\
      &\quad  + \delta \mathbb{E}\Bigg[\int_t^T \bigg\langle \tilde{\hat{{\bf p}}}_s, -\hide{\beta_1} \tilde{{\bf x}}_s +r  \tilde{{\bf p}}_s   -\big(\partial_{\bf x}{\bf F}(s,{\bf x}^1_s,{{\bf z}}^1_s,{{\bf v}}^1_s,\mathbb{E}[{{\bf v}}^1_s]) \nonumber\\
      &\qquad\quad - \partial_{\bf x}{\bf F}(s,{{\bf x}}^2_s,{{\bf z}}^2_s,{{\bf v}}^2_s,\mathbb{E}[{{\bf v}}^2_s])  \big)  \bigg\rangle ds \Bigg] .
    \end{align}
By  Assumption \ref{ass:fg:sep}, we further obtain  
    \begin{align}
    \label{eq:exist:general:proof:8}
       &\ \mathbb{E}\left[\left|\tilde{\hat{{\bf p}}}_t\right|^2\right] +  \mathbb{E}\left[\int_t^T \left|\tilde{\hat{\boldsymbol{\eta}}}_s \right|^2 ds \right]  \nonumber\\
       \leq &\ K_T \mathbb{E}\left[\left|\tilde{{\hat{{\bf x}}}}_T\right|^2\right] + K_T\mathbb{E}\left[\int_t^T \left(  \left|\tilde{\hat{{\bf p}}}_s\right|^2 + \left|\tilde{\hat{{\bf x}}}_s\right|^2 \right) ds \right] \nonumber\\
       &\  + \delta K_T\left(   \mathbb{E}[|\tilde{{\bf x}}_T|^2]  + \mathbb{E}\left[\int_t^T \left(  \left|\tilde{\hat{{\bf p}}}_s\right|^2+  \left|\tilde{{\bf p}}_s\right|^2 + \left|\tilde{{\bf x}}_s\right|^2 \right) ds \right]\right).
    \end{align}
By Gr\"onwall's inequality, we infer from \eqref{eq:exist:general:proof:8} that 
    \begin{align}
    \label{eq:exist:general:proof:9}
        \mathbb{E}\left[\left|\tilde{\hat{{\bf p}}}_t\right|^2\right] 
       &\leq  K_T   \mathbb{E}\left[\left|\tilde{{\hat{{\bf x}}}}_T\right|^2\right] + K_T\mathbb{E}\left[\int_0^T     \left|\tilde{\hat{{\bf x}}}_s\right|^2   ds \right] \nonumber \\
       &\ + \delta K_T \left( \mathbb{E}[|\tilde{{\bf x}}_T|^2]  + \mathbb{E}\left[\int_0^T \left(   \left|\tilde{{\bf p}}_s\right|^2 + \left|\tilde{{\bf x}}_s\right|^2 \right) ds \right]\right).
    \end{align}
  Substituting \eqref{eq:exist:general:proof:9} into the right-hand side of \eqref{eq:exist:general:proof:8}, followed by integrating both sides over $t=0$ to $t=T$, we obtain 
    \begin{align}
    \label{eq:exist:general:proof:10}
         \mathbb{E}\left[\int_0^T \left(\left|\tilde{\hat{{\bf p}}}_t\right|^2 + \left|\tilde{\hat{\boldsymbol{\eta}}}_t \right|^2 \right)dt\right] &\leq  K_T \mathbb{E}\left[\left|\tilde{{\hat{{\bf x}}}}_T\right|^2\right] + K_T\mathbb{E}\left[\int_0^T     \left|\tilde{\hat{{\bf x}}}_t\right|^2   dt \right] \nonumber \\
       &\ + \delta K_T \left( \mathbb{E}[|\tilde{{\bf x}}_T|^2]  + \mathbb{E}\left[\int_0^T \left(   \left|\tilde{{\bf p}}_t\right|^2 + \left|\tilde{{\bf x}}_t\right|^2 \right) dt \right]\right).
    \end{align}
If $\alpha^{\bf G}_{{\bf M}} >0$, by combining \eqref{eq:exist:general:proof:10} and    \eqref{eq:exist:general:pf}, there exists $K_T>0$ such that for any $\mu_0\in[0,1]$ and sufficiently small $\delta>0$, 
 \begin{align*}
       &\  \mathbb{E}\left[\left|  \tilde{\hat{{\bf x}}}_T\right|^2\right]  + \mathbb{E}\left[\int_0^T \left|\tilde{\hat{{\bf x}}}_t  \right|^2 dt \right] +  K_T\mathbb{E}\left[\int_0^T \left(\left|\tilde{\hat{{\bf p}}}_t\right|^2 + \left|\tilde{\hat{\boldsymbol{\eta}}}_t \right|^2 \right)dt\right] \\
      \leq&\ \delta K_T\mathbb{E}\left[   \left|\tilde{\hat{{\bf x}}}_T \right|^2  +  \left|\tilde{{\bf x}}_T \right|^2  +  \int_0^T \left(\left|\tilde{\hat{{\bf x}}}_t \right|^2 + \left|\tilde{\hat{{\bf p}}}_t \right|^2 + \left|\tilde{\hat{\boldsymbol{\eta}}}_t \right|^2 + \left|\tilde{{\bf x}}_t \right|^2 + \left|\tilde{{\bf p}}_t \right|^2 + \left|\tilde{\boldsymbol{\eta}}_t \right|^2 \right)dt  \right] \\
      &\ + \delta^2K_T \left( \mathbb{E}[|\tilde{{\bf x}}_T|^2]  + \mathbb{E}\left[\int_0^T \left(   \left|\tilde{{\bf p}}_t\right|^2 + \left|\tilde{{\bf x}}_t\right|^2 \right) dt \right]\right). 
    \end{align*}
Therefore, one can pick $\delta>0$  such that $\Psi_{\mu_0+\delta}$ is a contraction for any $\mu_0\in[0,1]$ and the proof is complete.

\section{Proofs of Assertions in Section \ref{sec:lq} }	
\label{sec:app:proof:LQ}
This section contains the proofs of statements in Section \ref{sec:lq}.

\subsection{Proof of Theorem \ref{thm:riccati}}
\label{sec:app:pf:thm:riccati}

We consider an \textit{ansatz} of the adjoint process $p^h_t$ with the following feedback form:
    \begin{equation}
    \label{eq:p:ansatz}
        p^h_t = \Gamma^h_t (x^h_t - z^h_t) + \bar{p}^h_t.
    \end{equation}
Using \eqref{eq:FBSDE:lq:insurance:constrained}, applying It\^o's lemma on the right hand side of \eqref{eq:p:ansatz} yields 
    \begin{align}
    \label{eq:rhs:ansatz}
       &\ d\left(\Gamma^h_t (x^h_t - z^h_t) + \bar{p}^h_t \right) \nonumber\\
       =&\ \left(\frac{d\Gamma^h_t}{dt} (x^h_t - z^h_t) - \kappa^h\Gamma^h_t(v^h_t-\bar{v}^h_t) + r\left(\Gamma^h_t\left(x^h_t - z^h_t\right) -\bar{p}^h_t\right) -Q^h_tz^h_t(1-S^h_t)  \right)dt \nonumber \\
        &\quad + \Gamma^h_t \sigma^h(1-v^h_t)dW^h_t. 
    \end{align}
By comparing \eqref{eq:rhs:ansatz} with the equation satisfied by $p^h$ in          \eqref{eq:FBSDE:lq:insurance:constrained}
, we find that 
    \begin{equation}
    \label{eq:eta:ansatz}
        \eta^h_t = \Gamma^h_t \sigma^h(1-v^h_t), \text{ and thus } \mathbb{E}[\eta^h_t]  = \Gamma^h_t \sigma^h(1-\bar{v}^h_t).
    \end{equation}
Substituting \eqref{eq:eta:ansatz} into \eqref{eq:v:ansatz}, we obtain 
    \begin{equation}
        \bar{v}^h_t = \frac{\kappa^h\bar{p}^h_t + (\sigma^h)^2\Gamma^h_t}{P^h_t(1-R^h_t) + (\sigma^h)^2\Gamma^h_t} \text{ and } v^h_t = \frac{\kappa^h p^h_t +(\sigma^h)^2\Gamma^h_t}{P^h_t +(\sigma^h)^2\Gamma^h_t} + \frac{R^h_tP^h_t}{P^h_t+(\sigma^h)^2\Gamma^h_t}\bar{v}^h_t. 
    \end{equation}
By further substituting this into \eqref{eq:rhs:ansatz}, and using the \textit{ansatz} \eqref{eq:p:ansatz}, we have 
     \begin{align}
    \label{eq:rhs:ansatz:2}
       &\ d\left(\Gamma^h_t (x^h_t - z^h_t) + \bar{p}^h_t \right)\nonumber\\ =&\ \left(\frac{d\Gamma^h_t}{dt} (x^h_t - z^h_t) -\frac{ (\kappa^h)^2\Gamma^h_t\left(p^h_t - \bar{p}^h_t \right)}{P^h_t +(\sigma^h)^2\Gamma^h_t} + r\left(\Gamma^h_t\left(x^h_t - z^h_t\right) -\bar{p}^h_t\right) -Q^h_tz^h_t(1-S^h_t)  \right)dt \nonumber\\ 
       &\ + \Gamma^h_t \sigma^h(1-v^h_t)dW^h_t \nonumber \\
        =&\ \Bigg( (x^h_t - z^h_t)\left( \frac{d\Gamma^h_t}{dt}  -\frac{ (\kappa^h)^2(\Gamma^h_t)^2}{P^h_t +(\sigma^h)^2\Gamma^h_t} +2r\Gamma^h_t + Q^h_t \right) -r\left( \Gamma^h_t(x^h_t-z^h_t) + \bar{p}^h_t\right)\nonumber \\
        &\quad -Q^h_t(x^h_t-S^h_tz^h_t)  \Bigg)dt  + \Gamma^h_t \sigma^h(1-v^h_t)dW^h_t
    \end{align}
Using \eqref{eq:p:ansatz}, \eqref{eq:rhs:ansatz:2}, and comparing with the equation satisfied by $p^h$ in \eqref{eq:FBSDE:lq:insurance:constrained}, we deduce that $\Gamma^h$ has to satisfy \eqref{eq:Gamma:mf}. The proof that the solution of \eqref{eq:FBODE:unconstraint} can be expressed as \eqref{eq:fbode:ansatz} and \eqref{eq:FBODE:riccati} can be proven similarly, and thus we omit the proof.

\subsection{Well-posedness of \eqref{eq:FBODE:unconstraint}}
\label{sec:well-posed:unconstraint}
In this section, we provide a global existence condition for the equation \eqref{eq:FBODE:unconstraint}. To this end, we need the following assumption. 

\begin{assumption} \hfill 
\label{ass:global:unconstraint}
    \begin{enumerate}[label=(\alph*)]
        \item $\lambda_1:=\inf_{t\in[0,T]}\lambda_{\min}({\bf I}-{\bf S}_t)>0$;
        \item $\lambda_2:=\inf_{t\in[0,T]} \lambda_{\min}({\bf \Lambda}{\bf A}_t ) >0$, 
    \end{enumerate}
where ${\bf \Lambda} := {\bf K} - {\bf \Pi}$. 
\end{assumption}

\begin{theorem}
    Under Assumption \ref{ass:global:unconstraint}, there is at most one solution for the equation \eqref{eq:FBODE:unconstraint}. 
\end{theorem}

\begin{proof}
    Let $({\bf z}^1,{\bf p}^1)$ and $({\bf z}^2,{\bf p}^2)$ be two solutions of \eqref{eq:FBODE:unconstraint}. Then, the functions $\tilde{{\bf z}}_t := {\bf z}^1_t - {\bf z}^2_t$ and $\tilde{\bar{{\bf p}}}_t := \tilde{\bar{{\bf p}}}^1_t - \tilde{\bar{{\bf p}}}^2_t$ satisfies 
    \begin{equation*}
    \begin{dcases}
    d\tilde{{\bf z}}_t = \left(r{\bf z}_t - {\bf \Lambda}{\bf A}_t\tilde{\bar{{\bf p}}}_t     \right)dt, \\
    -d\tilde{\bar{{\bf p}}}_t = \left(r\tilde{\bar{{\bf p}}}_t + {\bf Q}_t({\bf I}-{\bf S}_t)\tilde{{\bf z}}_t \right)dt, \\
    \tilde{{\bf z}}_0 = {\bf 0},\\
    \tilde{\bar{{\bf p}}}_T = {\bf Q}_T({\bf I}-{\bf S}_T)\tilde{{\bf z}}_T.
    \end{dcases}
\end{equation*}
    Using this, by considering the differential of $\tilde{{\bf z}}_t\tilde{\bar{{\bf p}}}_t$, we arrive at 
    \begin{align*}
        \tilde{{\bf z}}_T{\bf Q}_T({\bf I}-{\bf S}_T)\tilde{{\bf z}}_T + \int_0^T   \tilde{{\bf z}}_t{\bf Q}_t({\bf I}-{\bf S}_T)\tilde{{\bf z}}_t dt &= - \int_0^T  \tilde{\bar{{\bf p}}}_t {\bf \Lambda}{\bf A}_t  \tilde{\bar{{\bf p}}}_tdt \\
        &\leq - \frac{1}{2}\inf_{t\in[0,T]} \lambda_{\min}({\bf \Lambda}{\bf A}_t + {\bf A}_t{\bf \Lambda}^\top) \int_0^T | \tilde{\bar{{\bf p}}}_t|^2dt \leq 0. 
    \end{align*}
    This implies $\tilde{{\bf z}}$ and thus $\tilde{{\bf p}}$ must be identical to zero. 
\end{proof}

To show that the FBODE indeed admits a solution, we again employ the continuation approach. To this end, we let $\hat{{\bf z}}$ and $\hat{{\bf p}}$ be the solution of the FBODE parametrized by $\mu_0\in[0,1]$:
    \begin{equation}
    \label{eq:FBODE:exist}
    \begin{dcases}
        d\hat{{\bf z}}_t =\left( -(1-\mu)\lambda_2\hat{{\bf p}}_t + \mu\left(r \hat{{\bf z}}_t - {\bf \Lambda}{\bf A}_t \hat{{\bf p}}_t + {\bf l}  \right) + \phi_t \right)dt, \\
        -d\hat{{\bf p}}_t = \left( (1-\mu)\lambda_1\hat{{\bf z}}_t+ \mu\left(r\hat{{\bf p}}_t + {\bf Q}_t({\bf I}-{\bf S}_t) \right) +\psi_t\right)dt, \\
        \hat{{\bf z}}_0 = (\mathbb{E}[\xi^h])_{h=1}^H,\\
        \hat{{\bf p}}_T = (1-\mu)\lambda_1{\bf z}_T + \mu {\bf Q}_T({\bf I}-{\bf S}_T)\hat{\bf z}_T - \boldsymbol{\gamma},
        \end{dcases}
    \end{equation}
where  $\phi_t,\xi_t$ are square integrable functions over $[0,T]$. The spirit of the approach is in line with the proof of Lemma \ref{lem:exist:general}: if the system \eqref{eq:FBODE:exist} has a solution for $\mu=\mu_0$, and for any square-integrable functions  $\phi_t,\xi_t$, then the operator defined by \eqref{eq:FBODE:exist} is a contraction for any $\mu\in[\mu_0,\mu_0+\delta]$, where $\delta>0$ is independent of $\mu_0$. Hence, the system \eqref{eq:FBODE:exist} admits a solution whenever $\mu\in[\mu_0,\mu_0+\delta]$. Using the fact that the solution \eqref{eq:FBODE:exist} admits a solution when $\mu=0$, we can conclude the existence of solution for any $\mu\in[0,1]$.  The details of the calculations are omitted. 
\begin{theorem}
    Under Assumption \ref{ass:global:unconstraint}, Equation \eqref{eq:FBODE:unconstraint} admits a solution. 
\end{theorem}

\section{Supplementary Tables for Section \ref{sec:mf:numerical}}
\label{sec:app:Supp:tables}
This section presents tables summarizing the training errors and computational efficiency of the neural network algorithm used in Section \ref{sec:mf:numerical}.

\begin{table}[htbp]
    \caption{Computation errors of neural network approach with respect to the ODE approach under non-constrained cases.} 
	\smallskip
	\centering 
	\begin{tabular}{c c c c c c}
        \multicolumn{6}{c}{Relative Error (\%)}\\
        \hline
          Penalty coefficient $\lambda$   &0.1 & 1.0 & 10.0 & 100.0 & 1000.0 \\
        \hline
        Case 1(a)  &9.844555	&0.915430	&\textbf{0.678508}	&7.141356	&20.914042\\
        Case 1(b)  &9.739147	&1.407824	&\textbf{0.608529}	&7.287133	&19.793709\\
        Case 1(c)  &9.704202	&1.084904	&\textbf{0.732753}	&7.099290	&23.721090\\
        \hline
        Case 2(a)  &5.520915	&\textbf{1.421929}	&1.565124	&2.647546	&8.175174\\
        Case 2(b)  &5.209600	&\textbf{1.402305}	&1.574674	&2.877253	&11.329953\\
        Case 2(c)  &5.610542	&\textbf{1.438909}	&1.556929	&2.462988	&11.801251\\
        \hline
        Case 3(a)  &3.701866	&\textbf{1.549085}	&2.105542	&3.224022	&7.313468\\
        Case 3(b)  &5.162252	&\textbf{1.134077}	&1.410400	&3.693166	&6.850523\\
        \hline
        Case 4(a)  &9.575447	&\textbf{0.908286}	&1.445545	&3.657386	&11.387693\\
        Case 4(b)  &9.883666	&1.643441	&\textbf{1.464042}	&3.789543	&11.719880\\
        Case 4(c)  &9.904698	&1.665042	&\textbf{1.383423}	&3.927721	&12.774165\\   
        \hline
	\end{tabular}

    \begin{flushleft}
        The average time to compute each total error is 1886.94 seconds. 
    \end{flushleft}
    \label{app:sec:tables}

 \end{table} 

\begin{table}[htbp]
    	 \caption{Final loss functions for unconstrained cases. }
        \smallskip
	
	\centering 
    
        	\begin{tabular}{c c c c }
            \hline
Unconstrained Cases & Case 1(a) & Case 1(b)  & Case 1(c) \\
        \hline
        Terminal Condition Error &$1.548988 \times 10^{-3}$ & $1.667531 \times 10^{-3}$ & $1.450754 \times 10^{-3}$\\
        Mean Field Term Error & $3.235802 \times 10^{-5}$ & $4.012196 \times 10^{-5}$ & $2.563629 \times 10^{-5}$\\
        Time elapsed (secs)& 1893.98 & 1881.24 & 1931.42\\
        \hline
         & Case 2(a) & Case 2(b)  & Case 2(c)\\
        \hline
        Terminal Condition Error &$6.176409 \times 10^{-4}$ & $6.400801 \times 10^{-4}$ & $6.024750 \times 10^{-4}$\\
        Mean Field Term Error    & $1.265802 \times 10^{-5}$ & $2.451912 \times 10^{-5}$ & $6.414652 \times 10^{-6}$\\
        Time elapsed (secs)& 1896.25 & 1869.49 &1868.24\\
        \hline
         & Case 3(a) & Case 3(b)  & Case 4(a) \\
        \hline
        Terminal Condition Error &$2.784362 \times 10^{-4}$ &$5.350168 \times 10^{-4}$ & $8.624881 \times 10^{-3}$
\\
        Mean Field Term Error    & $2.395581 \times 10^{-5}$ &$5.165220 \times 10^{-5}$& $1.511587 \times 10^{-4}$ \\
        Time elapsed (secs)& 1865.63 &1867.42 &1879.71 \\
        \hline 
         & Case 4(b)  & Case 4(c)  & Case 5\\
         \hline
         Terminal Condition Error  &$1.638430 \times 10^{-3}$ & $1.698711 \times 10^{-3}$ & $1.663634 \times 10^{-5}$\\
         Mean Field Term Error    &$5.030000 \times 10^{-7}$ & $1.427944 \times 10^{-6}$ & $4.262555\times 10^{-6}$\\
         Time elapsed (secs)  &1865.49 & 1863.38 &2124.92\\
         \hline 
        \end{tabular}
 	 \begin{flushleft}
        The terminal condition error and the mean field term error refers to the term $ \sum \limits_{h=1}^2 \mathbb{E}\left[(p^h_T + g_x(x^h_T,z^h_T)\right]$ and $\frac{1}{M} \sum \limits_{i=0}^{M-1} \sum \limits_{h=1}^2 (\mathbb{E}[v^h_{t_i}]-\bar{v}^h_{t_i})^2$, respectively.
 	 \end{flushleft}
    \label{loss_penalty_table1}
\end{table} 

\begin{table}[htbp]
    \caption{Final loss and penalty values for constrained cases}
    \smallskip
	\centering 
	\begin{tabular}{c c c c }
        \hline
          Constrained Cases    & Case 1(a) & Case 1(b)  & Case 1(c) \\
        \hline
        Terminal Condition Error & $2.009435 \times 10^{-3}$ & $2.245595 \times 10^{-3}$ & $1.819428 \times 10^{-3}$ \\
        Mean Field Term Error & $6.938315 \times 10^{-5}$ & $8.738402 \times 10^{-5}$ & $5.548106 \times 10^{-5}$ \\
        Time elapsed (secs)&1918.70 &1945.75 &1930.34\\
        \hline  
           & Case 2(a) & Case 2(b) & Case 2(c) \\
        \hline
        Terminal Condition Error & $2.155019 \times 10^{-3}$ & $2.155019 \times 10^{-3}$ & $2.118284 \times 10^{-3}$\\
        Mean Field Term Error & $2.286199 \times 10^{-4}$ & $2.286199 \times 10^{-4}$ & $2.108407 \times 10^{-4}$\\
        Time elapsed (secs)&1927.40 & 1936.75 & 1898.35\\
        \hline  
           &  Case 3(a)  & Case 3(b) & Case 4(a)\\
        \hline
        Terminal Condition Error & $4.444087 \times 10^{-3}$ & $6.950257 \times 10^{-4}$& $9.873541\times 10^{-4}$
\\
        Mean Field Term Error &$2.919305 \times 10^{-3}$ & $8.796966 \times 10^{-5}$ &$1.538552\times 10^{-4}$\\
        Time elapsed (secs)& 1957.89 &1907.67 & 1965.20 \\
        \hline        
         & Case 4(b) &  Case 4(c) & Case 5 \\
         \hline
         Terminal Condition Error & $2.209101 \times 10^{-3}$ &  $2.348184 \times 10^{-3}$ & $1.382007\times 10^{-5}$\\
         Mean Field Term Error & $2.228195 \times 10^{-5}$ & $2.664459 \times 10^{-5}$ & $4.025126 \times 10^{-6}$\\
         Time elapsed (secs) & 1910.48 & 1914.44 & 2214.13\\        
        \hline 
	\end{tabular} \\
 	\begin{flushleft}
		{\small
        }
	\end{flushleft}
    \label{loss_penalty_table2}
\end{table} 

\begin{table}[h!]
	\centering 
    \caption{Equilibrium insurance strategies with and without constraints for each case}
    \smallskip
        \begin{tabular}{c c c c c c c }
        \hline
        Case & Parameter & Constraint & $\bar{v}^1_0$ & $\bar{v}^2_0$ & $\bar{v}^1_{T-\Delta t}$ & $\bar{v}^2_{T-\Delta t}$  \\ \hline \hline
        \multirow{2}{*}{1(a)} & \multirow{2}{*}{\makecell{$\sigma^1=0.1$,  $\sigma^2=0.3$} }  &  No &0.359091	&0.440495	&-0.066468	&0.015177 \\ 
                          &  & Yes & 0.355441	&0.435981	&0.000000	&0.033949\\
       \hline
        \multirow{2}{*}{1(b)} & \multirow{2}{*}{\makecell{1(a) with \\  $\omega^1=0.8$} } & No & 0.352799	&0.437634	&-0.069131	&0.012958\\
                            &  & Yes & 0.347622	&0.432593	&0.000000	&0.032654\\ 
        \hline
        \multirow{2}{*}{1(c)} & \multirow{2}{*}{\makecell{1(a) with \\  $\omega^1=0.2$} } & No & 0.365121	&0.442901	&-0.063986	&0.017695\\
                         &     & Yes &0.362339	&0.439052	&0.000000	&0.035201\\ 
        \hline
        \multirow{2}{*}{2(a)}  & \multirow{2}{*}{\makecell{$\gamma^1=1$, $\gamma^2=1.6$} }   & No & 0.430947	&0.175347	&0.009570	&-0.265701\\
                          &    & Yes & 0.431735	&0.133601	&0.035397	&0.000000\\ 
        \hline
        \multirow{2}{*}{2(b)} & \multirow{2}{*}{\makecell{2(a) with \\  $\omega^1=0.8$} }   & No & 0.442197	 &0.182480	&0.017060	&-0.254958\\
                          &    & Yes & 0.440488	&0.137245 &0.038818	&0.000000\\ 
        \hline
        \multirow{2}{*}{2(c)}  & \multirow{2}{*}{\makecell{2(a) with \\  $\omega^1=0.2$} }   & No & 0.419688 &0.167624	&0.001187	&-0.275922\\
                         &     & Yes &0.422559	&0.129154	&0.028175	&0.000000\\ 
        \hline
        \multirow{2}{*}{3(a)} & \multirow{2}{*}{\makecell{$\kappa^1=0.1$, $\kappa^2=0.5$ \\
        $\gamma^1=\gamma^2=1.6$} } & No & 0.165144	&0.167612	&0.024968	&-0.280722\\
                           &   & Yes & 0.174865	&0.152342	&0.020742	&0.000000\\ 
        \hline
        \multirow{2}{*}{3(b)} &  \multirow{2}{*}{\makecell{3(a) with \\ $\gamma^1=\gamma^2=1$ } }  & No & 0.220572	&0.431938	&0.092434	&0.009920\\
                           &   & Yes &0.220646	&0.426106	&0.092528	&0.029771\\ 
        \hline
        \multirow{2}{*}{4(a)} &  \multirow{2}{*}{\makecell{$\tilde{l}^1-\mu^1=0.02$\\  $\tilde{l}^2-\mu^2=0.1$}}  & No &0.437253	&0.449271	&0.029694	&0.050107\\
                            &  & Yes & 0.433595	&0.448720	&0.047438	&0.063334\\ 
        \hline
        \multirow{2}{*}{4(b)} &  \multirow{2}{*}{\makecell{4(a) with \\ $e^1=0.1$, $e^2=0.01$}}  & No & 0.473885	 &0.443457	&0.038497	&0.019084\\
                            &  & Yes & 0.471770	&0.438603 &0.053599	&0.036133 \\ 
        \hline
        \multirow{2}{*}{4(c)} &  \multirow{2}{*}{\makecell{4(a) with \\ $e^1=0.01$, $e^2=0.1$}}  & No &0.421126	&0.495810	&0.006337	&0.052741\\
                            &  & Yes & 0.414905	&0.495798 &0.028501	&0.065609 \\  
        \hline
        \multirow{2}{*}{5} &  \multirow{2}{*}{\makecell{$\gamma^1=0.5$, $\gamma^2=3.0$}}  & No &0.057585	&0.103351	&0.037535	&0.057978\\
                            &  & Yes &0.058370	&0.104163	&0.038957	&0.058851 \\  
        \hline
        
        \end{tabular}

    \label{tab:v:values}
\end{table}

\end{appendices}
 
\end{document}